\documentclass[12pt]{article}
\usepackage{appendix}
\usepackage{anysize}
\usepackage{graphics}
\usepackage{color}
\usepackage{amssymb}
\usepackage{graphicx}
\usepackage{epstopdf}
\usepackage[hang,small,bf]{caption}
\usepackage{hyperref}
\usepackage{amsmath}
\usepackage{latexsym}
\usepackage{setspace}
\usepackage{sectsty}
\usepackage[square,comma,numbers,sort&compress,merge]{natbib}
\newcommand{\be}{\begin{equation}}
\newcommand{\ee}{\end{equation}}
\newcommand{\bea}{\begin{eqnarray}}
\newcommand{\eea}{\end{eqnarray}}

\def\4vol{{\int d^4x \sqrt{-g}}}

\def\simlt{\stackrel{<}{{}_\sim}}
\def\simgt{\stackrel{>}{{}_\sim}}
\newcommand{\nc}{\newcommand}

\DeclareMathOperator{\Tr}{Tr}

\nc{\nt}{\tilde{N}}
\nc{\ra}{\rightarrow}
\nc{\lsim}{\begin{array}{c}\,\sim\vspace{-21pt}\\< \end{array}}
\nc{\gsim}{\begin{array}{c}\sim\vspace{-21pt}\\> \end{array}}
\nc{\tnt}{\tilde{N}}
\nc{\tst}{\tilde{t}}
\nc{\LL}{L}
\nc{\vv}{\tilde{v}}

\renewcommand{\theequation}{\thesection.\arabic{equation}}

\sectionfont{\large\bf}
\subsectionfont{\normalsize\bf}
\subsubsectionfont{\small\bf}
\title{
\vspace*{-1.3cm}
\begin{flushright}
\normalsize{
ANL-HEP-PR-10-9\\
EFI-10-05\\
FERMILAB-PUB-10-049-T}
\end{flushright}
\vspace{0.5cm}
\Large
\textbf{Determining the Structure of Supersymmetry-Breaking with Renormalization Group Invariants}
\author{\textbf{Marcela Carena$^{a,b}$, Patrick Draper$^{b}$,}\\
\textbf{Nausheen R.~Shah$^{a}$, and Carlos E.~M.~Wagner$^{b,c,d}$}\\
[1.5cm]
\normalsize\emph{$^a$~Fermi National Accelerator Laboratory, P.~O.~Box 500, Batavia, IL 60510, USA}\\
\normalsize\emph{$^b$~Enrico Fermi Institute and $^c$~Kavli Institute for Cosmological Physics,}\\
\normalsize\emph{University of Chicago, Chicago, IL 60637, USA} \\
\normalsize\emph{$^d$~HEP Division, Argonne National Laboratory, 9700 Cass Ave., Argonne, IL 60439, USA}}}
\begin{document}
\nocite{*}
\setcounter{page}{0}
\maketitle
\thispagestyle{empty}
\begin{abstract}

If collider experiments demonstrate that the Minimal Supersymmetric Standard Model (MSSM) is a good description of nature at the weak scale, the experimental priority will be the precise determination of superpartner masses. These masses are governed by the weak scale values of the soft supersymmetry (SUSY)-breaking parameters, which are in turn highly dependent on the SUSY-breaking scheme present at high scales. It is therefore of great interest to find patterns in the soft parameters that can distinguish different high scale SUSY-breaking structures, identify the scale at which the breaking is communicated to the visible sector, and determine the soft breaking parameters at that scale. In this work, we demonstrate that 1-loop Renormalization Group~(RG) invariant quantities present in the MSSM may be used to answer each of these questions. We apply our method first to generic flavor-blind models of SUSY-breaking, and then examine in detail the subset of these models described by General Gauge Mediation and the constrained MSSM with non-universal Higgs masses. As RG invariance generally does not hold beyond leading-log order, we investigate the magnitude and direction of the 2-loop corrections. We find that with superpartners at the TeV scale, these 2-loop effects are either negligible, or they are of the order of optimistic experimental uncertainties and have definite signs, which allows them to be easily accounted for in the overall uncertainty.

\end{abstract}
\thispagestyle{empty}

\newpage

\setcounter{page}{1}

\onehalfspacing


\section*{Introduction}

The Standard Model~(SM) provides an excellent description of the experimental data collected at high energy physics facilities. However, it has a series of shortcomings and is believed to provide only an effective theory at energies below a scale of the order of 1~TeV. The Minimal Supersymmetric Standard Model~(MSSM) addresses most of the SM shortcomings: it provides a solution to the hierarchy problem, is consistent with the unification of couplings at high energies, leads to an understanding of the origin of the negative Higgs mass parameters generating the breakdown of the electroweak symmetry, and includes a natural Dark Matter candidate~\cite{Nilles:1983ge,Haber:1984rc,Martin:1997ns}. Furthermore, the MSSM provides a weakly interacting extension of the SM in which the SM-like Higgs mass is bounded from above by about 130~GeV~\cite{Haber:1990aw,Okada:1990vk,Ellis:1991zd,Carena:1996,Haber:1996fp,Heinemeyer:1998np,Espinosa:2000df,Carena:2000npb,Degrassi2,Martin:2003} and will be within observational reach of colliders within a few years~\cite{Abazov:2008eb,Aaltonen:2009dh,:2009je,Aad:2009wy,Ball:2007zza,Stange:1993ya,Hahn:2006my,Spira:1998wh,Berger:2003pd,Dittmaier:2003ej,Dawson:2005vi,Balazs:1998nt,Carena:1998gk,Carena:1999bh,Plehn:1999nw,Draper:2009fh,Draper:2009au}. Finally, MSSM sparticle effects on the precision electroweak observables efficiently decouple for sparticle masses above a few hundred GeV, in agreement with the best fit to these measurements~(see, for example,~\cite{Barbieri:1989dc,Chankowski:1993eu,Erler:1998ur}).

If sparticles corresponding to the MSSM are discovered at the LHC or any future collider, resolving the properties of the supersymmetry~(SUSY)-breaking mechanism will become an important theoretical and experimental problem. Since most well-motivated scenarios place the ``messenger scale'' at which SUSY-breaking is communicated to the MSSM far beyond direct experimental access, it is likely that the renormalization group~(RG) will be our primary tool for investigating the nature of this mechanism.

The standard approach is ``top-down,'' where a high scale model with a limited number of parameters~(for instance the CMSSM or mSUGRA) is assumed, and the parameters are fit to the low scale data using RG evolution. This method is well studied, both in the context of present constraints from dark matter searches and low-energy observables as well as for future LHC projections~(for a sample of references, see~\cite{Baer:2003wx,Chattopadhyay:2003xi,Ellis:2004bx,Battaglia:2003ab,Djouadi:2006be,Baltz:2006fm,Allanach:2007qk,Berger:2008cq}).
It is now quite a sophisticated technique~(see for instance~\cite{Lafaye:2004cn}), but it has a few limitations.  Primarily, the reliability of the fit may be reduced as the number of high scale parameters is increased. It is also subject to the behavior of uncertainties under RG flow: if they tend to spread as the scale increases, the fit will become weaker. The top-down approach is also sensitive to the experimental uncertainties in gauge couplings and SM particle masses, and multiple iterations of RG evolution from high to low scales are required to consistently incorporate SUSY corrections to the Yukawa couplings~\cite{deltamb1,deltamb2,deltamb3,AguilarSaavedra:2005pw}.

Alternatively, a direct ``bottom-up'' approach has been studied~\cite{Kneur:2008ur,Kneur:1998gy,Blair:2002pg,Blair:2005ui,Carena:1996km} and found to possess utility complementary to the top-down method. The bottom-up reconstruction has the advantage of being easy and quite transparent: simple analytic formulae are used to convert measured pole masses into low scale values of the Lagrangian parameters, including dominant radiative effects. The running parameters are then evolved up to the high scale at which the mediation of SUSY-breaking takes place and their structure can be analyzed. This method is relatively fast and attractive, and it is less sensitive to the total number of free parameters, but it also presents challenges: foremost, one does not know the scale that should be used as the ultraviolet boundary. Additionally, uncertainties in the running parameters must again be propagated carefully under RG evolution. As in the top-down approach, the $\beta$-functions  for each sfermion mass are subject to the present experimental uncertainties in the gauge and Yukawa couplings. Furthermore, these $\beta$-function depend on many MSSM parameters even at 1-loop order, coupling all of the soft scalar masses to each other through the hypercharge $D$-term, as well as to the gaugino masses and the soft trilinear couplings. In Ref.~\cite{Kane:2006hd}, it was demonstrated that the failure to infer even one soft mass experimentally may be sufficient to drastically alter the high-scale prediction for other masses.

Here we would like to suggest an alternative technique for the bottom-up reconstruction program. Within the parameters of the MSSM, neglecting first and second generation Yukawa couplings, there exists a significant set of combinations of the parameters that are RG invariant~(RGI) at 1-loop. We propose that these RGIs may be used to efficiently test and extract parameters of high scale SUSY-breaking models, including in certain cases the messenger scale.

There are several reasons why RGIs are of particular interest. The first is that in some appropriate bases, subsets of the RGIs are predicted to vanish in different classes of SUSY-breaking models. A low scale measurement of any of these invariants may therefore provide immediate evidence for or against several attractive high scale models without requiring knowledge of the value of the high scale. The second reason is that for broad classes of SUSY-breaking models, most or all of the fundamental high scale parameters can be expressed uniquely in terms of the nonzero RGIs. If measurements of the first type of RGIs indicate that a particular class of models is favored, the second type may be used to constrain the parameter space of those models. Thirdly, none of the RGIs individually involve all of the parameters of the theory. This property limits the set of physical masses and couplings that must be measured before high-scale theories can be tested. Finally, from a practical point of view, the use of RGIs simplifies the standard bottom-up approach: it avoids the integration of RG equations and the complicated propagation of errors between scales.

Although RG invariance only holds strictly at leading-log order, 2-loop corrections are expected to be quite small and are likely to be negligible compared with experimental uncertainties. To confirm this expectation, we analyze the 2-loop effects and demonstrate that they can either be neglected, or absorbed to a good approximation into a simple shift of the measured value of the RGIs. For each RGI, we show that the shift can be computed from a linear function of a few dominant sparticle masses.

In this work, we will study RGIs within theories satisfying the following conditions:
\begin{itemize}
  \item The effective theory at the electroweak scale is the MSSM;
  \item No new physics alters the 1-loop MSSM $\beta$-functions below the messenger scale, at which SUSY-breaking is transmitted to the observable sector\footnote{Here we neglect the possibility that strong couplings in the hidden sector could affect the running of the soft scalar masses, as pointed out in~\cite{Cohen:2006qc}.}.
\end{itemize}

After deriving a set of invariants under these assumptions, we will apply our method to test the class of theories in which the SUSY-breaking mechanism is flavor-blind. In addition to the general case, we will consider two major subclasses of flavor-blind theories. The first is general gauge mediation~(GGM), where the interactions between the SUSY-breaking sector and the sfermion/gaugino fields of the MSSM are controlled entirely by the MSSM gauge couplings~\cite{Meade:2008wd}. Integrating out the hidden sector generates soft mass terms for these fields that can be parameterized by six constants. Soft masses are also generated for the Higgs doublets, but it is likely that additional, non-gauge couplings link the Higgs and hidden sectors. We demonstrate that certain RGIs may be used to test the GGM hypothesis, and if the data is found to be consistent, other RGIs may be used to extract the full set of GGM parameters, including the soft Higgs masses at the messenger scale. The second type of SUSY-breaking structure we examine is the Constrained MSSM~(CMSSM) with non-universal Higgs masses~(NUHM). In this case, flavor-blindness is maintained with fewer input parameters than in GGM. Consequently, all messenger scale parameters may be expressed in terms of a subset of the RGIs. Among the remaining invariants, several nontrivial relationships must be satisfied for the low energy mass spectrum to be consistent with the CMSSM~+~NUHM.

The RGI quantities discussed in this work have all appeared previously in the literature in various guises, but either as limited subsets or not in the context of a bottom-up determination for the high scale SUSY-breaking parameters. In Ref.~\cite{Kane:2006hd} several invariants were introduced for the purpose of testing high-scale flavor universality in the presence of extra intermediate-scale GUT multiplets. Ref.~\cite{Kane:2006hd} also commented on the use of two quantities, $D_Y$ and $D_{B-L}$~(defined in Section 2), for probing GUT embeddings of the MSSM. These functions are related to the RGIs but are not themselves invariant in general: we will discuss the use of the corresponding RGIs in greater detail for testing gauge mediation and the CMSSM. In Ref.~\cite{Demir:2004aq} most of the invariants we will use are listed, but in a basis without the properties discussed above, rendering them less relevant for probing high scale models with low, single-scale measurements. In Ref.~\cite{Meade:2008wd}, $D_Y$, $D_{B-L}$, and one other invariant combination of them were listed as potential tests of GGM; they will be included and expanded upon in our discussion. Finally, Ref.~\cite{Martin:1993ft,Ananthanarayan:2007fj,Ananthanarayan:2003ca,Ananthanarayan:2004dm,Balazs:2010ha} discuss in detail some of the invariants in the context of sum rules for mSUGRA and a variety of SUSY-GUTs.

We attempt to go beyond previous work on the subject of RGIs by providing a full set of invariants in a useful basis, a program for using them to explore the structure of the messenger scale effective theory, a discussion of experimental uncertainties, and an analysis of 2-loop effects. Our presentation is divided into four sections. In the first, we will derive the RGIs in a convenient basis from symmetry arguments. In the second we will discuss tests of flavor-blindness, and under that hypothesis we will express the high scale soft parameters as simple functions of RGIs and a single gauge coupling at that scale. We will then use the RGIs to test and determine the parameter spaces of GGM and of the CMSSM~+~NUHM. In the third section, we analyze experimental prospects and 2-loop corrections for the RGI method in the context of GGM. We discuss briefly the minimal set of physical observables necessary to determine the soft parameters in the RGIs, and estimate the precision with which the RGIs may be used to constrain the high scale parameter space given a set of uncertainties in the low scale soft parameters. We derive simple estimates that may be used to subtract out the bulk of the low scale 2-loop shifts in the RGIs, and demonstrate that residual 2-loop contributions are typically small compared with optimistic assumptions for experimental uncertainties in the soft masses. We reserve the final section for our summary and conclusions. An appendix collects some analytic approximations for the 2-loop $\beta$-functions.

\vspace{0.2 cm}
\indent

\section{1-loop RGI Combinations of Soft Masses and Gauge Couplings}

In this section we will derive fourteen RG invariants\footnote{A few more RGIs exist, such as $I_2$ and $I_4$ of Ref.~\cite{Demir:2004aq}. However, they are not needed in our investigation of flavor-blind theories, and they require more MSSM parameters than the set considered in this work, including the soft trilinear couplings and $\mu$.} and relate  several of them to symmetries of the MSSM Lagrangian. For reference, the RGIs are collected in Table~\ref{table.Inv}.

We begin with RGIs constructed solely out of soft scalar masses. We will assume that the soft sfermion masses are flavor diagonal in the super-weak basis and that the first and second generation masses are degenerate at the input scale, as is strongly motivated by low-energy flavor mixing constraints. Similarly, we assume that there are no new sources of CP-violation in the soft SUSY-breaking sector beyond those induced by the Yukawa couplings. We will also neglect first and second generation Yukawa and trilinear couplings since they give very small contributions to the evolution of the soft SUSY-breaking parameters, smaller than the 2-loop corrections associated with the gauge and third generation Yukawa couplings. In general, these approximations tend to be excellent for models in which there exists flavor universality at the messenger scale, and all flavor violation effects are due to radiative corrections induced by Yukawa couplings. Finally, we assume that the right-handed neutrino, if it exists, effectively decouples from the spectrum and the RGEs~(see discussion in Section 3).  With these conditions, the soft scalar masses $m_i$ obey the following RGEs at 1-loop:

\begin{equation}
16\pi^2\frac{dm^2_i}{dt}=\sum_{jk}y^{*}_{ijk}y^{ijk}(m^2_i+m^2_j+m^2_k+A^{*}_{ijk}A^{ijk})-8\sum_aC_a(i)g_a^2|M_a|^2+\frac{6}{5}Y_ig_1^2D_Y,
\label{RGE}
\end{equation}
where
\begin{eqnarray}
D_Y&\equiv&\Tr(Ym^2)\nonumber\\
&=&\sum_{gen}\left(m^2_{\tilde{Q}}-2m^2_{\tilde{u}}+m^2_{\tilde{d}}-m^2_{\tilde{L}}+m^2_{\tilde{e}}\right)+m^2_{H_u}-m^2_{H_d}.
\label{Dy}
\end{eqnarray}
The first sum in Eq.~(\ref{RGE}) is over all the degrees of freedom available to run in self-energy loops, the second sum is over the three gauge groups $a$ of the MSSM, $M_a$ and $A_{ijk}$ are the gaugino masses and soft trilinear parameters, $C_a(i)$ is the quadratic Casimir for the representation $i$ of $a$, and $t\equiv \log(\mu/M_Z)$. The trace in the definition of the hypercharge $D$-term, $D_Y$, runs over all chiral multiplets. A summary of the $B$, $L$ and the hypercharge representations for the particle content in the MSSM is given in Table~\ref{table.YBL}.

\begin{table}[!tbp]
\caption{U(1) Representations in the MSSM}
\centering
\begin{tabular}{c c c c}
\hline
\hline\\
[.1ex]
Particle & Y & B & L\\
[1ex]
\hline\\
[.1ex]
Q=$\left(\begin{array}{c}
u \\
d\end{array}\right)_L$ & 1/6 & 1/3 & 0\\ [3ex]

L=$\left(\begin{array}{c}
\nu \\
l\end{array}\right)_L$ & -1/2 & 0 & 1\\ [3ex]

u=$u_R^{\mbox{\tiny{\textit{C}}}}$ & -2/3 & -1/3 & 0\\ [3ex]

d=$d_R^{\mbox{\tiny{\textit{C}}}}$ & 1/3 & -1/3 & 0\\ [3ex]

e=$l_R^{\mbox{\tiny{\textit{C}}}}$ & 1 & 0 & -1\\ [3ex]

$H_u$ & 1/2 & 0 & 0\\ [3ex]
$H_d$ & -1/2 & 0 & 0\\ [3ex]
\hline
\end{tabular}
\label{table.YBL}
\end{table}

We first construct linear combinations, $D_i$, of the soft masses that evolve only with $D_Y$,
\begin{equation}
D_i\equiv\Tr (Q_{i} m^2),\qquad\qquad \frac{d D_i}{dt} = a_i D_Y,
\label{Didef}
\end{equation}
where $Q_{i}$ and $a_i$ are constants, and again the trace is defined over all chiral multiplets. With five soft SUSY-breaking masses contributed per generation~(for a total of ten soft masses if the first generation masses are identified with those of the second generation) and the two soft SUSY-breaking Higgs mass terms, twelve coefficients $Q_i$ are associated with each combination $D_i$. That there are six independent $D_i$ satisfying Eq.~(\ref{Didef}) can be seen from Eq.~(\ref{RGE}) as follows. For the Yukawa terms to vanish from the $\beta$-function, the $Q_{i}$ must correspond to charges of a global symmetry of the classical Yukawa potential. This affects only the third generation in our approximation, so it implies three independent constraints on the twelve $Q_{i}$. For the gaugino terms to cancel, the symmetry must have vanishing mixed anomalies with the SM gauge groups. This supplies three more independent constraints on the $Q_{i}$.

Furthermore, we can construct a basis in which five of these six combinations will also satisfy $\Tr Q_iY=0$. This condition eliminates the $D_Y$ term from the RGEs. These five combinations are then genuine RG invariants, independent of the vanishing of $D_Y$.

Let us first consider baryon number, $Q=B$, and lepton number, $Q=L$. Classically, these are symmetries of the action, but both are anomalous in the MSSM. Within our approximation, both the baryon and the lepton number associated with each generation is separately conserved at the classical level and their anomalies are flavor-independent. Hence the differences between the first~(or second) and third generation baryon and lepton number are anomaly-free. With this in mind, one can define RGIs associated with the new global symmetries: $B_{13}$ and $L_{13}$~\cite{Kane:2006hd},
\begin{eqnarray}
D_{B_{13}}&\equiv& D_{B_1}-D_{B_3}\nonumber\\
&=& 2m_{\tilde{Q}_1}^2-m_{\tilde{u}_1}^2-m_{\tilde{d}_1}^2-2m_{\tilde{Q}_3}^2+m_{\tilde{u}_3}^2+m_{\tilde{d}_3}^2\;,\nonumber\\
&&\nonumber\\
D_{L_{13}}&\equiv& D_{L_1}-D_{L_3}\nonumber\\
&=& 2m_{\tilde{L}_1}^2-m_{\tilde{e}_1}^2-2m_{\tilde{L}_3}^2+m_{\tilde{e}_3}^2\;,
\end{eqnarray}
where numbers in the subscripts indicate the generation.

We can also consider non-anomalous $U(1)$ symmetries. The obvious choices are hypercharge and $(B-L)$. However, since $\Tr Y^2$ and $\Tr Y(B-L)$ are non-vanishing, these combinations will evolve with $D_Y$.

In the hypercharge combination, the $D_Y$ dependence can be eliminated by using nearly the same trick with family non-universality as we did with $B$ and $L$. In this case, we must include the Higgs doublets with the third generation, since their soft mass evolution is linked by the Yukawa couplings. The proper RG invariant combination is given by
\begin{eqnarray}
D_{Y_{13H}}&\equiv& D_{Y_1}-\frac{10}{13}D_{Y_{3H}}\;,\nonumber\\
&=& m^2_{\tilde{Q}_1}-2m^2_{\tilde{u}_1}+m^2_{\tilde{d}_1}-m^2_{\tilde{L}_1}+m^2_{\tilde{e}_1}\nonumber\\
&&-\frac{10}{13}\left( m^2_{\tilde{Q}_3}-2m^2_{\tilde{u}_3}+m^2_{\tilde{d}_3}-m^2_{\tilde{L}_3}+m^2_{\tilde{e}_3}+m^2_{H_u}-m^2_{H_d}\right)\;.
\label{y13h}
\end{eqnarray}

For $(B-L)$, generation subtraction is redundant, since we can construct it out of $D_{B_{13}}$ and $D_{L_{13}}$. However, even restricting the trace to a single generation, $D_Y$ and $D_{B-L}$ evolve only with $D_Y$. Therefore we can construct a fourth RGI combination which depends only on the soft masses of the first generation~\cite{Meade:2008wd},
\begin{eqnarray}
D_{\chi_1}&\equiv&4D_{Y_1}-5D_{(B-L)_1}\nonumber\\
&=&-6m_{\tilde{Q}_1}^2-3m_{\tilde{u}_1}^2+9m_{\tilde{d}_1}^2+6m_{\tilde{L}_1}^2-m_{\tilde{e}_1}^2\;.
\end{eqnarray}
Here $\chi_1$ is used to indicate that this charge assignment is related by an overall re-scaling to the $U(1)_{\chi}$ symmetry generated by the spontaneous breaking of $E_6$.

The $E_6$ breaking generates an additional $U(1)$, $E_6\rightarrow SU(5)\times U(1)_{\chi} \times U(1)_{\psi}$, but the $U(1)_{\psi}$ symmetry is anomalous when restricted to the MSSM and involves the Higgs sector, which prevents a family non-universal invariant combination. Fortunately, by taking the linear combination~(the ``inert $U(1)'$''~\cite{Everett:2009cn}),
\begin{equation}
\frac{\sqrt{10}}{2}U(1)_{\chi}+\frac{\sqrt{6}}{2}U(1)_{\psi}\;,
\end{equation}
and setting the charge of the first generation left-handed sleptons to zero, we obtain an anomaly-free symmetry, $U(1)_Z$, and the corresponding RGI~\cite{Kane:2006hd}:
\begin{eqnarray}
D_{Z}&\equiv&3m_{\tilde{d}_3}^2+2m_{\tilde{L}_3}^2-2m_{H_d}^2-3m_{\tilde{d}_1}^2\;.
\end{eqnarray}
This new symmetry is evidently independent from those already discussed, as it depends only on $m^2_{H_d}$ and not on the combination $m^2_{H_u}-m^2_{H_d}$.

\begin{table}[!hp]
\caption{1-Loop RG Invariants in the MSSM }
\centering
\begin{tabular}{| c | c | c |}
\hline
\hline
&&\\
Invariant & Symmetry & Dependence on Soft Masses   \\ [0.8ex]
 & &  \\
\hline
\hline
&&\\
$D_{B_{13}}$&$B_1-B_3$&$2(m_{\tilde{Q}_1}^2-m_{\tilde{Q}_3}^2)-m_{\tilde{u}_1}^2+m_{\tilde{u}_3}^2-m_{\tilde{d}_1}^2+m_{\tilde{d}_3}^2$\\ [3ex]
\hline
&&\\
$D_{L_{13}}$&$L_1-L_3$&$2(m_{\tilde{L}_1}^2-m_{\tilde{L}_3}^2)-m_{\tilde{e}_1}^2+m_{\tilde{e}_3}^2$\\ [3ex]
\hline
&&\\
$D_{\chi_1}$&$\chi_1$&$3(3m_{\tilde{d}_1}^2-2(m_{\tilde{Q}_1}^2-m_{\tilde{L}_1}^2)-m_{\tilde{u}_1}^2)-m_{\tilde{e}_1}^2$\\ [3ex]
\hline&&\\
$D_{Y_{13H}}$&$Y_1-\frac{10}{13}Y_{3H}$&$\begin{array}{c} m^2_{\tilde{Q}_1}-2m^2_{\tilde{u}_1}+m^2_{\tilde{d}_1}-m^2_{\tilde{L}_1}+m^2_{\tilde{e}_1}\\-\frac{10}{13}\left(m^2_{\tilde{Q}_3}-2m^2_{\tilde{u}_3}+m^2_{\tilde{d}_3}-m^2_{\tilde{L}_3}+m^2_{\tilde{e}_3}+m^2_{H_u}-m^2_{H_d}\right)\end{array}$\\ [6ex]
\hline
&&\\
$D_{Z}$&$Z$&$3(m_{\tilde{d}_3}^2-m_{\tilde{d}_1}^2)+2(m_{\tilde{L}_3}^2-m_{H_d}^2)$\\ [3ex]
\hline
&&\\
$I_{Y\alpha}$&$Y$&$\left(m^2_{H_u}-m^2_{H_d}+\sum_{gen}(m^2_{\tilde{Q}}-2m^2_{\tilde{u}}+m^2_{\tilde{d}}-m^2_{\tilde{L}}+m^2_{\tilde{e}})\right)/g_1^2$\\ [3ex]
\hline
&&\\
$I_{B_r}$&&$M_r/g_r^2$\\ [3ex]
\hline
&&\\
$I_{M_1}$&&$M_1^2-\frac{33}{8}(m_{\tilde{d}_1}^2-m_{\tilde{u}_1}^2-m_{\tilde{e}_1}^2)$\\ [3ex]
\hline
&&\\
$I_{M_2}$&&$M_2^2+\frac{1}{24}\left(9(m_{\tilde{d}_1}^2-m_{\tilde{u}_1}^2)+16m_{\tilde{L}_1}^2-m_{\tilde{e}_1}^2\right)$\\ [3ex]
\hline
&&\\
$I_{M_3}$&&$M_3^2-\frac{3}{16}(5m_{\tilde{d}_1}^2+m_{\tilde{u}_1}^2-m_{\tilde{e}_1}^2)$\\ [3ex]
\hline
&&\\
$I_{g_2}$&&$ 1/g_1^2-33/(5g_2^2)$\\ [3ex]
\hline
&&\\
$I_{g_3}$&&$ 1/g_1^2+33/(15g_3^2)$\\ [3ex]
\hline
\end{tabular}
\label{table.Inv}
\end{table}

Of course, the sixth combination must now run with $D_Y$. The RGE for $D_Y$ is given by
\begin{equation}
16\pi^2\frac{ d D_Y}{dt}=\frac{66}{5}g_1^2D_Y,
\label{DYRGE}
\end{equation}
so this combination can be taken to be $D_Y$ itself. $D_{B-L}$ also evolves only with $D_Y$, and as mentioned previously, so do the restrictions of $D_Y$ and $D_{B-L}$ to any given generation~(plus Higgs in the case of the third generation $D_{Y}$). These quantities are then RGIs if $D_Y$ vanishes, although only those related to $D_Y$ are linearly independent of the invariants introduced above. However, one can construct a genuine independent invariant out of $D_Y$ that is nonlinear in running parameters. Recall that the gauge couplings obey homogeneous RGEs at one loop:
\begin{equation}
16\pi^2\frac{d g_r}{dt} = g_r^3\left(\Tr_nI_r(n)-3C_r(G)\right)\;,
\label{gaugeRGE}
\end{equation}
where $I_r(n)$ is the Dynkin index of the chiral multiplet $n$, and $C_r(G)$ is the quadratic Casimir invariant of the adjoint representation. The $g_1$ group theory factor is equal to $33/5$ in the MSSM, and so we obtain the RGI,
\begin{equation}I_{Y\alpha}\equiv \frac{D_Y}{g_1^2}\;.
\end{equation}

Additionally, because of the form of Eq.~(\ref{gaugeRGE}), we obtain two additional RGIs
\begin{equation}
I_{g_2}\equiv \frac{1}{g_1^2}-\frac{33}{5g_2^2} \;\;\;\;\;\; \mbox{and} \;\;\;\;\;\; I_{g_3}\equiv \frac{1}{g_1^2}+\frac{33}{15g_3^2}\;,
\label{gaugeRGI}
\end{equation}
whose values may be computed from measurements at the scale $M_Z$. In the $\overline{MS}$ scheme, for instance, they are approximately $-10.9$ and $6.2$, respectively.

We now construct RGIs with gaugino mass dependence. The three soft gaugino masses, $M_r$, evolve with 1-loop RGEs given by
\begin{equation}
16\pi^2\partial_t M_r = g_r^2M_r\left(2\Tr_nI_r(n)-6C_r(G)\right)\;.
\label{gauginoRGE}
\end{equation}
Again replacing $g_r^2$ in Eq.~(\ref{gauginoRGE}) with $\partial_t \log(g_r)$ and the group theory factor, we recover the well-known set of three RGIs,
\begin{equation}
I_{B_r}\equiv M_r/g_r^2\;.
\end{equation}

Finally, there are three RGIs that mix gaugino and sfermion soft mass parameters. Following the method of Ref.~\cite{Demir:2004aq}, these can be obtained easily from the 1st generation masses because of the absence of Yukawa couplings and the homogeneity of the $M_r$ and $D_Y$ RGEs,

\begin{eqnarray}
I_{M_1}&\equiv& M_1^2-\frac{33}{8}\left(m_{\tilde{d}_1}^2-m_{\tilde{u}_1}^2-m_{\tilde{e}_1}^2\right)\;,\nonumber\\
I_{M_2}&\equiv& M_2^2+\frac{1}{24}\left(9(m_{\tilde{d}_1}^2-m_{\tilde{u}_1}^2)+16m_{\tilde{L}_1}^2-m_{\tilde{e}_1}^2\right)\;,\nonumber\\
I_{M_3}&\equiv& M_3^2-\frac{3}{16}\left(5m_{\tilde{d}_1}^2+m_{\tilde{u}_1}^2-m_{\tilde{e}_1}^2\right)\;.
\end{eqnarray}

The invariant $I_{M_1}$ involves the subtraction of potentially large squark masses squared with a large coefficient. In models possessing a large mass hierarchy between strongly and weakly interacting particles, the splitting between $m_{\tilde{d}_1}$ and $m_{\tilde{u}_1}$ is likely to be much smaller than either mass, whereas the experimental uncertainty in $I_{M_1}$ grows approximately linearly with both $m_{\tilde{d}_1}$ and $m_{\tilde{u}_1}$. Therefore, the measured value of this RGI is likely to be consistent with zero within experimental errors. A similar effect may occur with $I_{M_2}$, but in this case the squarks coefficient is significantly smaller. One can form a combination of $I_{M_1}$ and $I_{M_2}$ that is independent of the squark masses, and therefore suffers lesser uncertainties:
\begin{eqnarray}
I_{M_{12}} & = & I_{M_1} + 11 I_{M_2}  \nonumber\\
&   = &  M_1^2 + \frac{11}{3}\left(3 M_2^2 + 2 m_{\tilde{L}_1}^2+  m_{\tilde{e}_1}^2 \right)\;.
\end{eqnarray}
In most cases, $I_{M_{12}}$ may be determined much more precisely than $I_{M_1}$ or $I_{M_2}$. Indeed, since all of the terms are positive, the percentage error in its determination is governed by the percentage error in the measurement of the weakly interacting sparticle masses.

With the exception of $I_{Y\alpha}$, the RGIs given in Table~1 of Ref.~\cite{Demir:2004aq} may be obtained\footnote{A few typos appear in Ref.~\cite{Demir:2004aq}: the coefficient of $M_2^2$ in $I_{12}$ should be $+3/2$, and similarly for $I_{17}$. The coefficient of $M_1^2$ should be $1/33$, and all coefficients of $S$ should be multiplied by $13/33$. Note that with these corrections, five invariants in the list $I_5$-$I_{17}$ of Ref.~\cite{Demir:2004aq} can be expressed as linear combinations of the others.} as linear combinations of those we have listed in Table~\ref{table.Inv}. However, the explicit relationship to symmetries in our basis allows a clean application of the invariants to the problem of reconstructing the theory at the messenger scale.

\section{Extraction of High Scale Parameters from RGIs}

We now examine the extraction of high scale SUSY-breaking parameters from the RGIs constructed in the previous section. The five soft masses contributed by the degenerate first and second sfermion generations, five soft masses from the third sfermion generation, two Higgs soft mass parameters, three gaugino masses, and the gauge couplings~(assuming no knowledge of the high scale) constitute eighteen unknowns~(more are introduced if the right-handed neutrino or other new physics is included). In the most generic case, assuming all RGIs are non-zero at the high scale, fourteen of these parameters may be obtained. Although the system is under-constrained, the high scale degrees of freedom are considerably reduced by the determination of the invariants.

As mentioned previously, a large class of SUSY-breaking models favor flavor-universal interactions. Even without the ability to extract all high scale parameters, using the RGIs to test flavor blindness would be a non-trivial result.  Furthermore, in the most generic flavor-blind models, the number of unknown parameters is reduced to thirteen, whereas the number of nonzero RGIs is twelve. We will consider the generic case below, as well as the specific models of GGM and the CMSSM with non-universal Higgs masses. The high scale values of the invariants listed in Table~\ref{table.Inv} are tabulated in Table~\ref{table.InvVals} for these three models.


\begin{table}[!hp]
\caption{Values of the 1-loop RGIs in terms of high scale soft parameters.}
\centering
\begin{tabular}{| c | c | c | c |}
\hline
\hline
&&&\\
Invariant & Generic Flavor Blind Model & GGM & CMSSM with NUHM  \\ [0.8ex]
 & & &\\
\hline
\hline
&&&\\
$D_{B_{13}}$&0&0&0\\ [3ex]
\hline
&&&\\
$D_{L_{13}}$&0&0&0\\ [3ex]
\hline
&&&\\
$D_{\chi_1}$&$3(3 m_{\tilde{d}}^2-2 (m_{\tilde{Q}}^2- m_{\tilde{L}}^2)- m_{\tilde{u}}^2)-m_{\tilde{e}}^2 $&0&$5m_0^2$\\ [3ex]
\hline
&&&\\
$D_{Y_{13H}}$&$\begin{array}{c}\frac{1}{13} \left(3 (m_{\tilde{Q}}^2-2 m_{\tilde{u}}^2+m_{\tilde{d}}^2- m_{\tilde{L}}^2+ m_{\tilde{e}}^2)\right.\\+\left.10( m_{H_d}^2- m_{H_u}^2)\right)\end{array}$&$-\frac{10}{13}(\delta_u-\delta_d)$&$-\frac{10}{13}(\delta_u-\delta_d)$\\ [6ex]
\hline
&&&\\
$D_{Z}$&$2 \left(m_{\tilde{L}}^2-m_{H_d}^2\right) $&$-2\delta_d$&$-2\delta_d$\\ [3ex]
\hline
&&&\\
$I_{Y\alpha}$&$ \begin{array}{c}\left(3 (m_{\tilde{Q}}^2-2 m_{\tilde{u}}^2+m_{\tilde{d}}^2- m_{\tilde{L}}^2+ m_{\tilde{e}}^2)\right.\\-\left.m_{H_d}^2+m_{H_u}^2\right)/g_1^2\end{array}$&$\left(\delta_u-\delta_d\right)/g_1^2$&$\left(\delta_u-\delta_d\right)/g_1^2$\\ [3ex]
\hline
&&&\\
$I_{B_r}$&$M_r/g_r^2$&$MB_r$& $m_{1/2}/g_r^2$\\ [3ex]
\hline
&&&\\
$I_{M_1}$&$M_1^2+\frac{33}{8} \left(m_{\tilde{u}}^2-m_{\tilde{d}}^2+m_{\tilde{e}}^2\right)$&$g_1^4\left((MB_1)^2+\frac{33}{10}A_1\right)$&$m_{1/2}^2+\frac{33}{8} m_0^2$\\ [3ex]
\hline
&&&\\
$I_{M_2}$&$M_2^2+\frac{1}{24} \left(9( m_{\tilde{d}}^2- m_{\tilde{u}}^2)+16 m_{\tilde{L}}^2-m_{\tilde{e}}^2\right)$&$g_2^4\left((MB_2)^2+\frac{1}{2}A_2\right)$&$m_{1/2}^2+\frac{5}{8} m_0^2$\\ [3ex]
\hline
&&&\\
$I_{M_3}$&$ M_3^2-\frac{3}{16} \left(5 m_{\tilde{d}}^2+m_{\tilde{u}}^2-m_{\tilde{e}}^2\right)$&$g_3^4\left((MB_3)^2-\frac{3}{2}A_3\right)$&$m_{1/2}^2-\frac{15}{16} m_0^2$\\ [3ex]
\hline
&&&\\
$I_{g_2}$&$\approx -10.9$&$\approx -10.9$&$\approx -10.9$\\ [3ex]
\hline
&&&\\
$I_{g_3}$&$\approx 6.2$&$\approx 6.2$&$\approx 6.2$\\ [3ex]
\hline
\end{tabular}
\label{table.InvVals}
\end{table}

\subsection{Generic Flavor-Blind Models}

The most immediate consequence of flavor blindness is the vanishing of $D_{B_{13}}$ and $D_{L_{13}}$. Therefore these invariants provide us with a direct test of the flavor-independent hypothesis with a minimal set of measurements. More precisely, they allow this hypothesis to be ruled out: measuring $D_{B_{13}}\neq 0$ or $D_{L_{13}}\neq 0$ at the low scale implies high-scale family non-universality; however, as noted in Ref.~\cite{Kane:2006hd}, measuring $D_{B_{13}}=0$ and $D_{L_{13}}=0$ at the low scale does not necessarily indicate high-scale universality.

Current experimental data from flavor physics strongly motivates a flavor-universal mediation mechanism for SUSY-breaking~(for a selection of Minimal Flavor Violation studies, see~\cite{Ellis:1981ts,Bertolini:1990if,Isidori:2001fv,Buras:2002vd,Babu:1999hn,Dedes:2002er,Demir:2003bv,Carena:2006ai,Lunghi:2006uf,Carena:2007aq,Ellis:2006jy,Ellis:2007ss,Isidori:2007jw,Barenboim:2007sk,Paradisi:2008qh,Ellis:2007kb,Carena:2008ue,Foster:2005wb}). Accordingly, if $D_{B_{13}}$ and $D_{L_{13}}$ are found to vanish, it is reasonable to proceed a step further and attempt to extract constraints on the high-scale values of flavor-blind MSSM soft parameters from the RGIs.

The ten free soft mass parameters can be expressed uniquely in terms of the ten invariants $D_{\chi_1}$ through $I_{M_3}$ listed in Table~\ref{table.Inv} and are given in Eq.~(\ref{flaveq1})-(\ref{flaveq2}). Note that these relations depend on the three gauge couplings at the high scale; with the invariants $I_{g_2}$ and $I_{g_3}$ defined in Eq.~(\ref{gaugeRGI}), they may be expressed entirely in terms of $g_1$. Equivalently, one can reduce the degrees of freedom at the high scale to a single parameter which can be taken to be the value of that scale. In particular this permits tests of more restrictive flavor-universal models such as mSUGRA, taking $g_1$ at the GUT scale. However, if the high scale is not known, one may take advantage of the fact that the gauge couplings do not vary too wildly with scale and simply guess values for them, at the cost of introducing a further uncertainty. We will discuss this possibility further below.

\begin{align}
M_1 &= g_1^2 I_{B_1}\;,\nonumber\\
M_2 &= g_2^2 I_{B_2}\;,\nonumber\\
M_3 &= g_3^2 I_{B_3}\;,
\label{flaveq1}
\end{align}

\begin{align}
m_{\tilde{L}}^2 &= -\frac{1}{440} \big(26 D_{Y_{13H}}+11 D_{\chi_1}+20 \big(\big(g_1^4I_{B_1}^2 +33 g_2^4 I_{B_2}^2\big)-\big(I_{M_1}+33 I_{M_2}\big)+g_1^2I_{Y\alpha} \big)\big)\;,\nonumber\\
m_{H_d}^2 &=  m_{\tilde{L}}^2 - \frac{1}{2}D_Z \; ,\nonumber\\
m_{H_u}^2 &= m_{\tilde{L}}^2 - \frac{1}{2}D_Z - \frac{13}{11} D_{Y_{13H}} + \frac{g_1^2}{11}  I_{Y\alpha} \; ,\nonumber\\
m_{\tilde{e}}^2 &= \frac{1}{220} \big(26 D_{Y_{13H}}+11 D_{\chi_1}-20 \big(2\big( g_1^4I_{B_1}^2  - I_{M_1}\big)-g_1^2I_{Y\alpha}\big)\big)\;,\nonumber\\
m_{\tilde{u}}^2 &= -\frac{1}{990} \big(78 D_{Y_{13H}}+33 D_{\chi_1}+20 \big(4\big(\big(g_1^4 I_{B_1}^2 -11 g_3^4 I_{B_3}^2\big) -\big( I_{M_1}-11 I_{M_3}\big)\big)+3 g_1^2I_{Y\alpha}\big)\big)\;,\nonumber\\
m_{\tilde{d}}^2 &= \frac{1}{1980}\big(78 D_{Y_{13H}}+33 D_{\chi_1}-20 \big(2 \big(\big(g_1^4I_{B_1}^2 -44 g_3^4 I_{B_3}^2\big) -\big(I_{M_1}-44 I_{M_3}\big)\big)-3 g_1^2I_{Y\alpha}\big)\big)\;,\nonumber\\
m_{\tilde{Q}_1}^2 &= \frac{1}{3960}\big( 78 D_{Y_{13H}}-627 D_{\chi_1}\nonumber\\
&-20 \big(\big(g_1^4 I_{B_1}^2 +297 g_2^4 I_{B_2}^2-176 g_3^4 I_{B_3}^2\big)
 -\big(I_{M_1}+297 I_{M_2}-176 I_{M_3}\big)-3 g_1^2I_{Y\alpha}\big)\big)\; .\label{flaveq2}
\end{align}
In the above, all couplings and soft parameters are assumed to be given at the messenger scale.

\subsection{General Gauge Mediation}

Gauge mediated SUSY-breaking encompasses a broad class of models in which flavor blindness is perhaps most natural~\cite{Giudice:1998bp,Affleck:1984xz,Dine:1995ag,Martin:1996zb,Wagner:1998vd,McGarrie:2010kh}. In Ref.~\cite{Meade:2008wd} General Gauge Mediation~(GGM) was defined as any theory in which all SUSY-breaking effects decouple from the MSSM in the limit of vanishing MSSM gauge couplings.

Let us first review the formulae for the soft masses in GGM at the messenger scale given in Ref.~\cite{Meade:2008wd}. The soft sfermion masses can be parameterized by a set of three parameters, $A_r$:\footnote{We neglect a possible explicit hypercharge Fayet-Iliopoulos~(FI) dependent term, $g_1^2Y_{\tilde{f}}\xi$, in Eq.~(\ref{softfm}). The FI term may drive some of the $m_{\tilde{f}}^2$ to negative values, and we assume it is fixed to zero by a $\mathbf{Z}_2$ symmetry in the hidden sector. For further discussion of difficulties associated with FI terms, see~\cite{Komargodski:2009pc,Dienes:2009td}. We comment again on this term below Eq.~(\ref{IYalphaeq}).}
\begin{equation}\label{softfm}
m_{\tilde{f}}^2=\sum_{r=1}^3g_r^4C_r(f)A_r\;,
\end{equation}
where the sum runs over the gauge groups of the MSSM, and the $A_r$ can be expressed as loop integrals over hidden sector current-current correlation functions,
\begin{align}
A_r\equiv-\int\frac{d^4p}{2\pi^4}\frac{1}{p^2}\bigg(-\frac{1}{p^2}\eta^{\mu\nu}\langle j^{(r)}_{\mu}(p)j^{(r)}_{\nu}(-p)\rangle+\frac{2p_{\nu}}{p^2}\bar{\sigma}^{\nu\dot{\beta}\alpha}\langle j_{\alpha}(p)\bar{j}_{\dot{\beta}}(-p)\rangle+\langle J(p)J(-p)\rangle\bigg)\;.
\end{align}

The gaugino masses are expressed in terms of three more parameters, $MB_r$, given by
\begin{equation}\label{softMm}
M_r=g_r^2MB_r\;,
\end{equation}
where $M$ is the scale of SUSY-breaking and
\begin{align}
MB_r=-\frac{1}{2}\epsilon^{\alpha\beta}\langle j_{\alpha}(0)j_{\beta}(0)\rangle\;.
\end{align}

In order to generate a Higgsino mass parameter, $\mu$, and soft term, $B_{\mu}$, of the correct order, gauge mediation may need to be supplemented by additional SUSY-breaking contributions in the Higgs sector. Therefore, we assume that in the case of the soft Higgs masses, the expression given in Eq.~(\ref{softfm}) may be modified,
\begin{align}
m_{H_u}^2&=m_{\tilde{L}_3}^2+\delta_u\;,\nonumber\\
m_{H_d}^2&=m_{\tilde{L}_3}^2+\delta_d\;.
\label{deltas}
\end{align}.

With these definitions for the soft masses, we can now calculate the high scale values of the RGIs in GGM, listed in Table~\ref{table.InvVals}. By inspection, we see that they can be separated into
\begin{itemize}
\item Three that test consistency of GGM;
\item Six that probe the high-scale mass parameters of pure GGM;
\item Five that are sensitive to extra structures linking the Higgs and SUSY-breaking sectors and to the values of the gauge couplings at the messenger scale.
\end{itemize}

The flavor-independence of gauge mediation is manifest in the formulae listed for the soft masses, and therefore, as before $D_{B_{13}}$ and $D_{L_{13}}$ should vanish at all scales.

The third invariant testing consistency of data with GGM is $D_{\chi_1}$~\cite{Meade:2008wd}. It should also remain approximately zero at all scales, which can be seen as follows. The condition for anomaly cancelation between a $U(1)$ symmetry $Q$ and another gauge group $G$ with generators $t^G_A(i)$ in the representation $i$ reads
\begin{equation}
\Tr Q t^G_A t^G_A = 0\;,
\end{equation}
where the trace runs over all degrees of freedom. If $Q$ commutes with all generators of $G$, this can be written equivalently as
\begin{align}
0&=\sum_i Q_i \Tr t^G_A(i) t^G_A(i)\;,\nonumber\\
&= \sum_i Q_i\;d(i) C_G(i)\;\nonumber\\
&=\Tr Q_iC_G(i)\;,
\end{align}
where the sums run over all representations $i$ of $G$ and $d(i)$ is the dimension of $i$. The sfermion masses in GGM depend on the gauge couplings and the Casimirs of the gauge groups; hence, if the mixed anomalies between $Q$ and the MSSM gauge groups cancel, we can multiply both sides of Eq.~(\ref{softfm}) by the charges $Q$ and take the trace to obtain
\begin{align}
\label{dq}
&D_Q=\Tr (Q m^2)=\sum_G g_G^4 A_G\Tr Q_i C_G(i)= 0\;.
\end{align}
The second equality in Eq.~(\ref{dq}) is only valid if the Higgs fields are neutral under $Q$; otherwise, the modifications of Eq.~(\ref{deltas}) imply $D_Q=2(Q_{H_u}\delta_u+Q_{H_d}\delta_d)$. Choosing $Q=\chi_1$, the Higgs fields are uncharged, and thus $D_{\chi_1}$ vanishes. This property may differentiate GGM from alternate theories with universal soft masses such as mSUGRA, where $D_{\chi_1}$ does not vanish unless a relevant right-handed neutrino is included.

This argument also suggests $D_Y$ and $D_{B-L}$ vanish and are RG invariant, since both are anomaly-free symmetries of the MSSM and both run with $D_Y$. However, as mentioned above the modifications to the Higgs sector in Eq.~(\ref{deltas}) may spoil the vanishing of $D_Y$. Furthermore, although $D_Y$ is zero in pure GGM, it also vanishes in other SUSY-breaking scenarios, including mSUGRA.

These first three RGIs are sufficient to either rule out GGM or demonstrate consistency of GGM with the data. However, they do not constrain the~($A_r$, $MB_r$, $\delta_u$, $\delta_d$, $M$) parameter space. Therefore, we now turn to the RGIs with non-zero values at the messenger scale.

In the most generic case where $D_Y$~(and thus the invariants $D_{Y_{13H}}$ and $I_{Y\alpha}$) is nonzero, we can determine the gauge couplings at the high scale. The following relations are satisfied at the input scale:
\begin{eqnarray}
D_{Y_{13H}}&=&-\frac{10}{13}(\delta_u-\delta_d)\;,\\
&&\nonumber\\
I_{Y\alpha}&=&\frac{(\delta_u-\delta_d)}{g_1^2}\;.
\label{IYalphaeq}
\end{eqnarray}
Therefore, if we can calculate the value of $D_{Y_{13H}}/I_{Y\alpha}$ at the low scale, we can infer the value of $g_1^2$ at the input scale via
\begin{align}
g_1^2(M)=-\frac{13}{10}\frac{D_{Y_{13H}}}{I_{Y\alpha}}.
\label{g1mess}
\end{align}
Then the $I_{g_r}$ RGIs given in Eq.~(\ref{gaugeRGI}) can be used with Eq.~(\ref{g1mess}) to compute $g_2$ and $g_3$ at the high scale.

The invariants $D_Z$ and $D_{Y_{13H}}$ can be used to extract the individual input-scale values of $\delta_u$ and $\delta_d$:
\begin{align}
\delta_u&=-\frac{1}{2}\left(D_{Z}+\frac{13}{5}D_{Y_{13H}}\right)\;,\nonumber\\
\delta_d&=-\frac{1}{2}D_{Z}\;.
\label{softhiggs}
\end{align}

Now we turn to the RGIs with explicit dependence on the gaugino mass parameters to extract information about the $A_r$ and $MB_r$. From the $I_{g_r}$ we immediately obtain
\begin{equation}
MB_r=I_{B_r}\;.
\label{Br}
\end{equation}

For the sfermions, using the $I_{M_r}$, we can obtain the $A_r$:
\begin{align}
A_1&=\;\frac{10}{33}\left(\frac{I_{M_1}}{g_1^4}-I_{B_1}^2\right)\;,\nonumber\\
A_2&=\quad2\left(\frac{I_{M_2}}{g_2^4}-I_{B_2}^2\right)\;,\nonumber\\
A_3&=-\frac{2}{3}\left(\frac{I_{M_3}}{g_3^4}-I_{B_3}^2\right)\;,
\label{Ar}
\end{align}
where the $g_r$ are the gauge couplings at the unknown scale, $M$, that can be deduced from Eq.~(\ref{g1mess})~(note that the $g_r$ used to compute the $I_{B_r}$ are at the low, measurement scale)\footnote{If there is an explicit hypercharge FI term at the input scale, the only effect on the RGIs is to change the input scale value of $I_{Y\alpha}$:
\begin{equation}
\frac{\delta_u-\delta_d}{g_1^2}\rightarrow\frac{\delta_u-\delta_d}{g_1^2}+11\xi.
\end{equation}
Eq.~(\ref{Ar}) for the $A_r$ remains unchanged; however, this shift is sufficient to spoil the extraction of $g_1$ at the input scale from the inhomogeneities in the Higgs sector.}.

If $\delta_u$ is found to be equal to $\delta_d$, and thus the gauge couplings at the high scale cannot be extracted from the RGIs, explicit RG evolution may be the only available method for determining $M$, choosing it to be the scale at which $m_{\tilde{u}_1}^2=m_{\tilde{u}_3}^2$. This would require knowledge not only of all up-type soft mass parameters at the low scale, but also of the soft trilinear coupling $A_t$ of the stop to the Higgs, which appears in the $m_{\tilde{u}_3}^2$ RGE\footnote{One could equally seek the scale where $m_{\tilde{d}_1}^2=m_{\tilde{d}_3}^2$, but it is likely that the soft trilinear coupling $A_b$ will be more difficult to determine experimentally than $A_t$.}. Moreover, as discussed previously, the evolution of the soft masses is linked to the evolution of all other parameters via the $D_Y$-terms. Therefore, it is expected that there will be large uncertainties in the determination of the messenger scale by this method.

As mentioned before and as we will see in Section 3 below, $I_{M_1}$ and $I_{M_2}$ could suffer from large uncertainties induced by cancelation between potentially large masses. If $I_{M_{12}}$ is used instead, this pitfall is avoided, but in general only the following correlation between $A_{1,2}$ and $g_{1,2}$ can be obtained,
\begin{equation}
g_1^4 \frac{33}{10} A_1 + g_2^4 \frac{11}{2} A_2 = I_{M_{12}} - g_1^4 I_{B_1}^2 - 11 g_2^4 I_{B_2}^2\;.
\end{equation}
In the above, the coefficients of the $I_{B_1}$ and $A_1$ terms are significantly smaller than those of $I_{B_2}$ and $A_2$, particularly for low values of the messenger scale where $g_2^2 \simeq 2 g_1^2$. Therefore, unless $A_1 \gg A_2$, for low messengers scales, the above equation then gives the approximate expression for $A_2$,
\begin{align}
A_2\approx\frac{2}{11}\left(\frac{I_{M_{12}}}{g_2^4}-11I^2_{B_2}\right)\;.
\end{align}

\subsection{Constrained MSSM with Non-Universal Higgs Masses}

The CMSSM~+~NUHM~\cite{Matalliotakis:1994ft,Olechowski:1994gm,Berezinsky:1995cj,Drees:2000he,Nath:1997qm,Ellis:2002iu,Ellis:2008eu} is another common model realizing flavor universality. The messenger scale is the GUT scale, however, here we will consider the more generic possibility of $M\neq M_{\rm{GUT}}$, as for example occurs in Mirage Mediation~\cite{Choi:2004sx,Choi:2005uz,Endo:2005uy}. The sfermions are given a common soft mass $m_0$ and the gauginos share a soft mass $m_{1/2}$ at the scale $M$. As in the GGM case, the Higgs masses are allowed to differ from the sfermion masses by the factors $\delta_u$ and $\delta_d$.

From the fourth column of Table~\ref{table.InvVals}, we see that the non-vanishing RGIs overconstrain the ($m_0$, $m_{1/2}$, $g_r(M)$, $\delta_u$, $\delta_d$) system, allowing multiple ways of testing consistency and extracting parameters. Both experimental errors and 2-loop contributions to the RGIs should be considered to decide which avenues to use for each purpose. The unification of gaugino mass parameters at some scale $M^*$ demands the following consistency relationships:
\begin{equation}
I_{B_1} - \frac{33}{5} I_{B_2} = I_{g_2} m_{1/2}\;,
\label{GauginoUnif1}
\end{equation}
\begin{equation}
I_{B_1} + \frac{33}{15} I_{B_3} = I_{g_3} m_{1/2}\;.
\label{GauginoUnif2}
\end{equation}
where $m_{1/2}$ is now a universal gaugino mass parameter at $M^*$, which is not necessarily constrained to be equal to $M$ by these relations.

In the MSSM the gauge couplings unify at the grand unification scale, and therefore at any arbitrary scale the following relationship is satisfied
\begin{equation}\label{g2g3}
I_{g_3} = - \frac{4}{7} I_{g_2}\;.
\end{equation}
Together Eqs.~(\ref{GauginoUnif1})--(\ref{g2g3}) indicate that gaugino mass unification requires the simple condition
\begin{equation}
\frac{5}{12} I_{B_1} -I_{B_2} + \frac{7}{12} I_{B_3} = 0\;.
\label{GauginoUnif3}
\end{equation}
This relation is necessary but not sufficient to ensure unification. In addition, a second requirement following from Eqs.~(\ref{GauginoUnif1})--(\ref{g2g3}) is
\begin{align}\label{GauginoUnif4}
I_{B_1}g^2_1(M^*)=\frac{5I_{B_1}-33I_{B_2}}{5I_{g_2}}.
\end{align}
If we demand that Eq.~(\ref{GauginoUnif4}) should be satisfied for a sensible value of $M^*$, between about $10^5$ and $10^{16}$ GeV, then the $I_{B_i}$ should obey the inequality
\begin{align}\label{GauginoIneq}
\frac{I_{B_1}}{4}   \lesssim \frac{5I_{B_1}-33I_{B_2}}{5I_{g_2}}   \lesssim   \frac{I_{B_1}}{2}.
\end{align}
In the particular case where the gaugino masses unify at $M^*=M_{\rm{GUT}}$, $I_{B_1}=I_{B_2}=I_{B_3}$ and Eq.~(\ref{GauginoUnif3}) holds directly, while the upper bound of Eq.~(\ref{GauginoIneq}) is approximately saturated.
Note that it is not unreasonable for a model to obey Eq.~(\ref{GauginoUnif3}) while still violating Eq.~(\ref{GauginoIneq}). This happens, for example, in anomaly mediated SUSY-breaking~\cite{Randall:1998uk,Giudice:1998xp}, where the $I_{B_i}$ are proportional to the $\beta$-function coefficients of the $g_i$, and so Eq.~(\ref{GauginoUnif3}) holds. The gaugino masses do not unify in anomaly mediation, however, spoiling Eq.~(\ref{GauginoIneq})~(and the extracted value of $m_{1/2}$ from Eq.~(\ref{GauginoUnif1}) or Eq.~(\ref{GauginoUnif2}) would be zero).  Note further that these relations involve invariants sensitive only to the gaugino sector, and therefore are expected to have small experimental uncertainties. 

Eqs.~(\ref{GauginoUnif1})--(\ref{GauginoIneq}) guarantee gaugino mass unification, but do not enforce $M=M^*$, which is a requirement in the CMSSM. In mirage mediation or minimal gauge mediation, the masses unify, but $M\neq M^*$ in general. Therefore, to differentiate the CMSSM from these other models, one may test the equality of $M$ with $M^*$, if $\delta_u\neq\delta_d$, by equating the messenger scale value of $g^2_1$ calculated via Eq.~(\ref{g1mess}) with the $M^*$ value of $g^2_1$ found from Eq.~(\ref{GauginoUnif4}). Additionally, the extracted value of $m_{1/2}$ from Eq.~(\ref{GauginoUnif1}) has to be consistent with the gaugino mass calculated from the other RGIs. These considerations enforce that the following set of relations, in addition to Eq.~(\ref{GauginoUnif3}), have to be satisfied for the spectrum to be consistent with the CMSSM:

\begin{align}\label{CMSSM.Con}
I_{Y\alpha}&=-\frac{13 D_{Y_{13H}} I_{B_1} I_{g_2}}{2(5 I_{B_1}-33 I_{B_2})}\;,\nonumber\\
I_{M_1}&=\left(\frac{5 I_{B_1}-33 I_{B_2}}{5I_{g_2}}\right)^2 +\frac{33 D_{\chi_1}}{40}\;,\nonumber\\
I_{M_2}&=\left(\frac{5 I_{B_1}-33 I_{B_2}}{5I_{g_2}}\right)^2 +\frac{D_{\chi_1}}{8}\;,\nonumber\\
I_{M_3}&=\left(\frac{5 I_{B_1}-33 I_{B_2}}{5I_{g_2}}\right)^2 -\frac{3 D_{\chi_1}}{16}\;.
\end{align}
If the data satisfies Eq.~(\ref{GauginoUnif3}) and Eq.~(\ref{CMSSM.Con}), one can proceed to the extraction of parameters. $D_{\chi_1}$ immediately yields the value of the soft sfermion mass $m_0$, and Eq.~(\ref{softhiggs}) can still be used to obtain $\delta_d$ and $\delta_u$. Turning to the 2-loop $\beta$-functions for the RGIs listed in Appendix A, we find that $I_{M_1}$ is a 2-loop invariant in the approximation that $g_1$, the soft trilinear couplings, and the lepton Yukawa couplings are set to zero. It may therefore be useful to extract $m_{1/2}$ from $I_{M_1}$. However, the advantage of 2-loop invariance can be mitigated by the consideration of experimental errors, which as noted previously may be large for $I_{M_1}$. In that event $m_{1/2}$ should be taken directly from either Eq.~(\ref{GauginoUnif1}) or Eq.~(\ref{GauginoUnif2}), for which the experimental errors are expected to be small:
\begin{align}
m_0^2&= \frac{D_{\chi_1}}{5}\;,\nonumber\\
m_{1/2}&=\frac{5 I_{B_1}-33 I_{B_2}}{5I_{g_2}}\;.
\end{align}
The $g_1^2$ at the messenger scale can then be obtained via Eq.~(\ref{GauginoUnif4}):
\begin{equation}
g_1^2=m_{1/2}/I_{B_1}.
\end{equation}

Taken together, the consistency relations, Eq.~(\ref{GauginoUnif3}) and Eq.~(\ref{CMSSM.Con}), provide strong constraints that make it highly unlikely, for example, for a generic flavor-blind or GGM spectrum to mimic the CMSSM~+~NUHM.

\section{Experimental Prospects for the MSSM RGIs}

\subsection{Extracting Soft Masses from Observables}

We now turn to the prospects for measuring the RGIs at the LHC. In practice it is necessary to convert the observed pole masses into the soft masses entering into the RGIs at the TeV scale. Here we comment only briefly on some relevant features of the analysis in the sfermion sector. For comprehensive studies, see Ref.~\cite{Kneur:2008ur,Kneur:1998gy,AguilarSaavedra:2005pw}.

For the first and second squark generations, the off-diagonal components of the mass matrices are proportional to the corresponding fermion masses and can therefore be neglected. Furthermore, the soft masses for the up-type and down-type left-chiral states are equal by gauge invariance. This allows one squark pole mass to be written in terms of the others and thus removed from the RGIs. For instance,
\begin{equation}
m_{\tilde{u}_L}^2 = m_{\tilde{d}_L}^2  + m_W^2 \cos 2 \beta\;,
\end{equation}
where we have ignored the quark masses and the subscript on the squarks refers to the predominantly left-handed~(and therefore chargino-interacting) mass eigenstate of the squarks in the spectrum.

For the third generation squarks, the off-diagonal components cannot be ignored. Therefore, in addition to pole masses, mixing angles must be measured in order to extract the soft masses. However, there is still one more observable than there are parameters in each sector, allowing the sbottom mixing angle to be removed
\begin{equation}
\cos^2\theta_{\tilde{b}} =
\frac{m_{\tilde{t}_1}^2 \cos^2\theta_{\tilde{t}} +m_{\tilde{t}_2}^2 \sin^2\theta_{\tilde{t}} - m_{\tilde{b}_2}^2  - m_t^2 + m_b^2
-m_W^2 \cos 2 \beta}{m_{\tilde{b}_1}^2 - m_{\tilde{b}_2}^2}\;.
\end{equation}
Here and below, the subscripts $1$ and $2$ denote the mass eigenstates. This leaves only the stop mixing angle to be determined experimentally.

For the slepton sector, the result depends on the treatment of the sneutrinos. In GGM models, the right handed sneutrinos, if present in the spectrum, do not receive SUSY-breaking masses at the messenger scale. Their masses are therefore controlled by the Majorana mass scale, $M_R$, of the right-handed neutrino partners, which only couple with the rest of the observable sector via the neutrino Yukawa couplings\footnote{In more general flavor-blind models, the SUSY-breaking masses for the right handed sneutrinos are naturally of the order of the soft scale, and the same conclusion holds.}. Since the neutrino masses
\begin{equation}
m_{\nu} \sim {\cal{O}}\left(\frac{h_\nu^2 \sin^2\beta v^2}{M_R} \right)
\end{equation}
are smaller or of the order of 1~eV, these Yukawa coupling effects are very small provided $M_R \simlt \mathcal{O}~(10^{10}~\mbox{GeV})$~. For values of $M_R > 10^{10}$~GeV, the Yukawa effects must be taken into account. However, they will only have an impact on the soft masses if the messenger scale is larger than $M_R$, since otherwise the right-handed sneutrinos decouple at the messenger scale. In this work we will make the assumption that either $ M_R\simlt \mathcal{O}~(10^{10}~\mbox{GeV})$ or $ M_R \simgt M $. This justifies our neglect of any right-handed sneutrino effects in the RGIs, as we have implicitly done in our treatment of the lepton sector in this article. With this assumption, the mixing angles in the sneutrino sector are negligibly small for all generations.

For the first two generations, as in the squark sector we can remove either a slepton or a sneutrino mass. For the third generation we can remove either $m^2_{\tilde{\nu}_{\tau}}$ or the stau mixing angle using the relation
\begin{equation}
m_{\tilde{\nu}_1}^2 = m_{\tilde{\tau}_1}^2 \cos^2\theta_{\tilde{\tau}} + m_{\tilde{\tau}_2}^2 \sin^2\theta_{\tilde{\tau}}
+m_W^2 \cos 2 \beta\;,
\end{equation}
where we have ignored the explicit dependence on the relatively small $\tau$ mass.

Even if a large number of masses can be determined at the LHC, a finite set of ambiguities will arise and affect the calculation of the RGIs. For the first and second generation sfermions, unless the mass eigenvalues are nearly degenerate, their measurement is not sufficient to compute the RGIs, because it must still be determined which eigenstate is the superpartner of the left chiral and the two right chiral quarks. An incorrect assignment leads~(in the limit of vanishing $D$-terms in the mass matrix), for instance, to a reversal of the soft masses $m^2_{\tilde{Q}}$ and $m^2_{\tilde{u}}$ for the case of the up-type first and second generation squarks, and these parameters do not enter into the RGIs in a symmetric way. Similar reversals may occur for the cases of the down-type squarks and the sleptons. Measurements in the chargino sector may be able to unravel this ambiguity, since only the left-handed first and second generation sparticles couple to the charginos.

Another problem is the difficulty in distinguishing up-type and down-type first and second generation sfermions at hadron colliders. Since the dominant production mechanism is through the strong force, the electromagnetic charges may be unmeasurable. However, if our intent is to test GGM, it is likely that $m_{\tilde{u}_1}^2>m_{\tilde{d}_1}^2$, since this relation holds at the input scale if $A_1\geq 0$~(which is necessary for a nontachyonic selectron mass in the absence of an FI term) and is maintained if $D_Y\geq0$~(although this condition is not necessary). Alternatively, one can test which, if any, of the two squark mass choices leads to a fulfillment of the necessary condition of $D_{B_{13}}=D_{\chi_1}=0$, and take that as the hypothetical correct assignment when using the other RGIs to test the GGM scenario further. The contribution of $m_{\tilde{u}_1}^2$ and $m_{\tilde{d}_1}^2$ to $D_{\chi_1}$ is quite different and of opposite sign, and therefore a cancelation of this RGI within errors for both choices of the squark masses is highly unlikely.

\subsection{Measurement Uncertainties in RGIs and GGM Parameters}

\begin{figure}[!htbp]
\begin{center}
\includegraphics[width=1.01\textwidth,trim=0in 0in 0 0,clip=true]{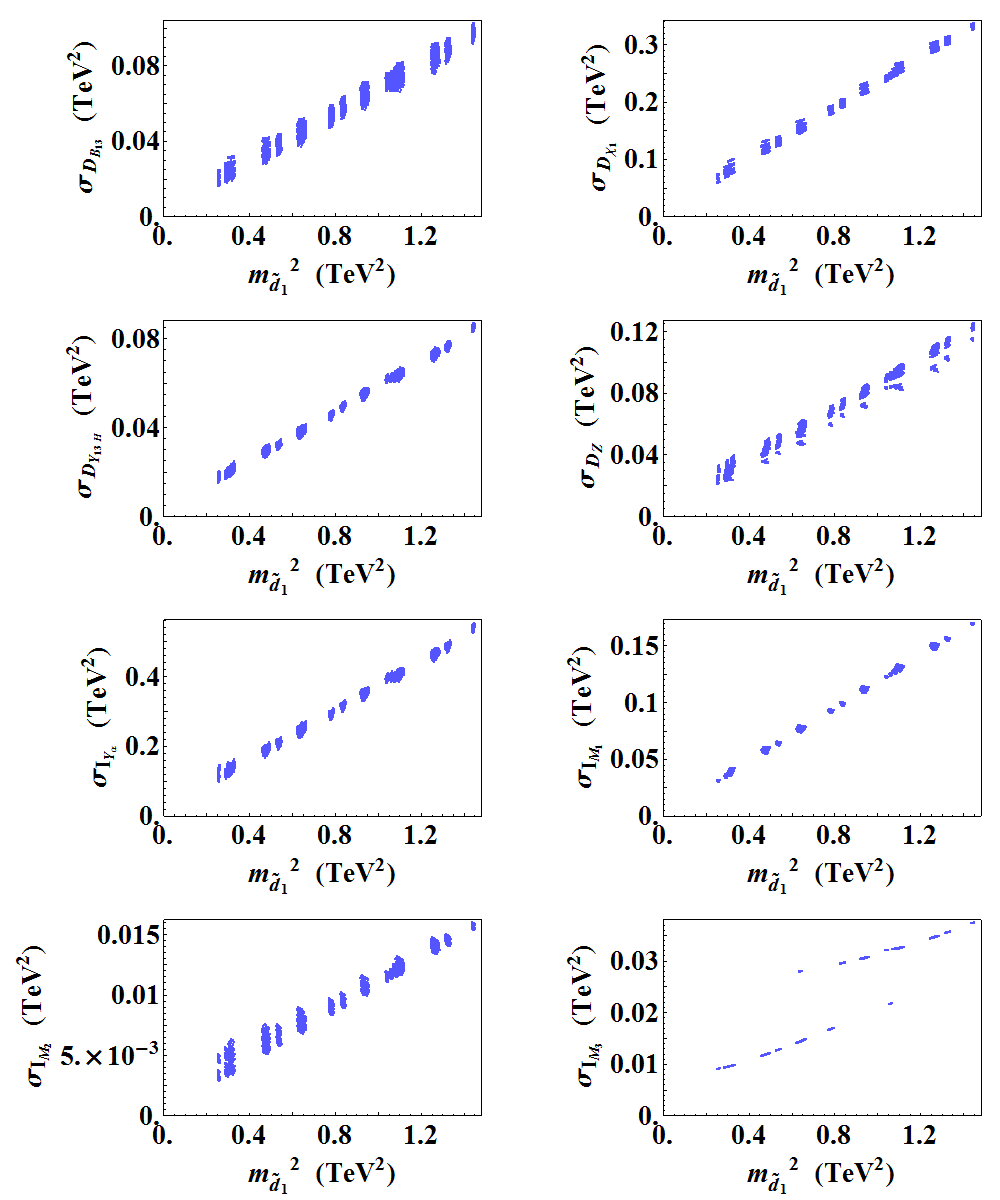}
\caption{Experimental errors in the low scale determination of the RGIs plotted vs. $m_{\tilde{d}_{1}}^2$, assuming 1\% uncertainties in all soft masses and scanning over the high scale parameters as in Eq.~(\ref{scanvals}). The clear correlation seen in these plots demonstrates that although the invariants involve many masses squared, the total uncertainties are dominated almost entirely by the uncertainty in the squark masses. RGIs with small uncertainties are omitted for brevity.}
\label{sigmamdR1}
\end{center}
\end{figure}

\begin{figure}[!htbp]
\begin{center}
\includegraphics[width=0.95\textwidth,trim=1.0in 0in 0 0,clip=true]{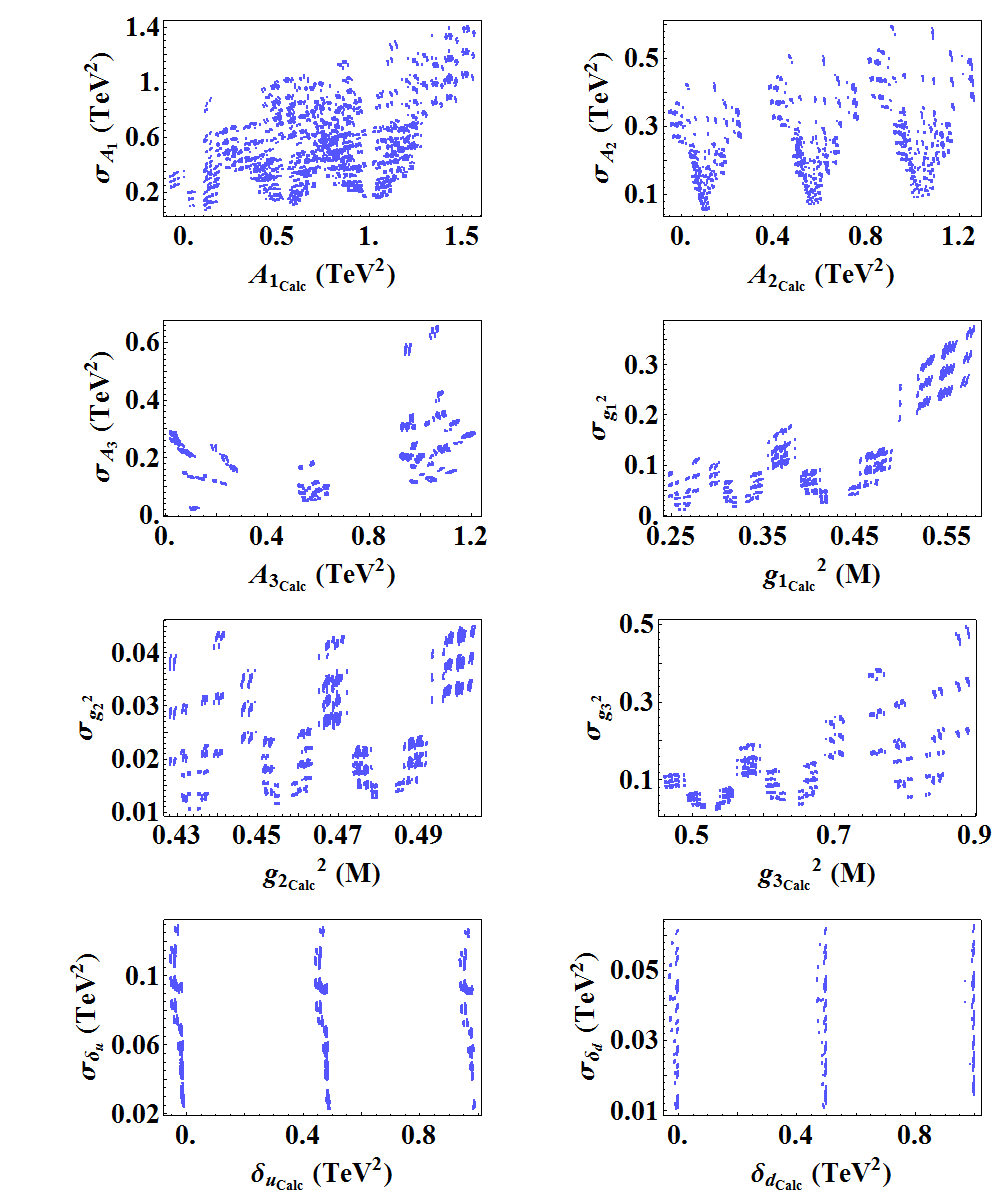}
\caption{Experimental errors in the calculated values of the high scale GGM parameters using the RGI reconstruction method. Parameters with small uncertainties are omitted for brevity. The spread in the calculated $A_r$ parameters is due mainly to 2-loop effects, and can be reduced using the technique discussed in Section 3.3. See Fig.~\ref{GGMshiftreconstruct}.}
\label{GGMreconstruct}
\end{center}
\end{figure}

It is difficult to estimate how well the physical quantities of the MSSM can be measured at the LHC, even if the sparticles are kinematically accessible. Most mass measurement studies in the literature have been performed in mSUGRA/CMSSM benchmark scenarios, under the assumption that sparticles will be produced in long cascade decays from strongly produced squarks and gluinos~(for a detailed study in the context of the LHC and a future linear collider, see~\cite{Weiglein:2004hn}). Endpoint methods and other kinematic variables may then be used to extract the pole masses. As for the stop mixing angle, in Ref.~\cite{Rolbiecki:2009hk} it was shown that the angle and the light stop mass may be inferred from a fit to measurements of ratios of stop branching ratios into charginos and neutralinos~(for other stop mixing studies, see~\cite{Hisano:2003qu,Perelstein:2008zt}). The authors demonstrated that this method can be effective if the decays are kinematically allowed and if the parameters of the neutralino/chargino sector have already been measured. However, it is clear that these methods will only be possible in a portion of the allowed parameter space, even restricting just to the regions compatible with GGM. Therefore, it is challenging to conjecture what can be done at the LHC without reanalyzing the reach point by point in parameter space, an exercise which is beyond the scope of this work.

With this in mind we forgo the application of projected measurement uncertainties to the physical parameters and simply apply a universal uncertainty to the soft masses entering into the RGIs. This allows us to get a general notion of how well the low scale parameters will have to be measured in order to determine the RGIs and the high scale parameters with a certain precision.

We perform a numerical scan over messenger scale inputs, restricting for the purposes of illustration to the parameter space of GGM and taking the following ranges of values:
\begin{eqnarray}
0.1\leq & A_r &\leq 1.0\mbox{ TeV}^2\;;\nonumber\\
0\leq & \delta_{u,d} & \leq 1.0\mbox{ TeV}^2\;;\nonumber\\
0.1\leq & MB_r & \leq 1.0\mbox{ TeV}\;;\nonumber\\
2\leq & \tan\beta &\leq 50\;;\nonumber\\
10^6\lesssim & M & \lesssim 10^{15}\mbox{ GeV}\;.
\label{scanvals}
\end{eqnarray}

For each point in our scanned parameter space of models, we compute the soft spectrum, run the soft masses down to the scale $\mu_i=1 $~TeV, and calculate the resulting central value and uncertainties for the RGIs. We then reconstruct the high scale GGM parameters, assuming a universal 1\% measurement error on the soft parameters. We enforce the conditions that the low energy spectrum is within the reach of the LHC and satisfies $m_{weak}>100\mbox{ GeV}$ and $m_{strong}>300\mbox{ GeV}$, where the $m_{weak}$ are the masses of the weakly-interacting sparticles and $m_{strong}$ are the colored sparticle masses.

As an approximation, we consider the propagation of errors in quadrature, although this is only an estimate since most masses would be obtained in decay chains and therefore their experimental errors are expected to be highly correlated. The quadrature combination also allows the errors from a larger universal soft mass fractional error to be obtained simply by re-scaling the results presented here for 1\%.

Fig.~\ref{sigmamdR1} gives the net experimental error in the RGIs as a function of $m_{\tilde{d}_1}^2$. We omit $D_{L_{13}}$, $I_{B_r}$, and $I_{g_r}$, which have relatively small uncertainties. With the exception of $I_{M_3}$, all the errors reflect a small y-intercept and small variation for a given $m_{\tilde{d}_1}^2$. This demonstrates that although the invariants involve many masses squared, the total uncertainties are dominated almost entirely by the uncertainty in the squark masses, assuming all individual mass errors are of the same order. Since in the GGM spectrum the squark mass splittings are proportional to smaller gauge couplings, they are typically small compared to the masses themselves unless $A_{1,2}\gg A_{3}$. Thus the expected error tends to be controlled essentially by a single parameter, as reflected in the scan. In the case of $I_{M_3}$ the gluino error must also be included.

Going a step further, we can estimate the uncertainties in the reconstructed values for the high scale GGM parameters.  For the $A_r$ this reconstruction relies on the premise that $\delta_u\neq \delta_d$ within experimental errors, which can be determined by testing the consistency of $D_{Y_{13H}}$ with zero. If $\delta_u= \delta_d$ within errors, then the extraction of $g_1$ at the high scale using Eq.~(\ref{g1mess}) leads to potentially meaningless results. Fig.~\ref{GGMreconstruct} gives the errors in the calculated values of the $A_r$, $\delta_{u,d}$, and $g^2_r$ at the messenger scale, for those points in the scan satisfying $|D_{Y_{13H}}| >\sigma_{D_{Y_{13H}}}$. For brevity we omit the $MB_r$, which have small uncertainties controlled entirely by the corresponding gaugino masses.

It is clear from the range of the y-axes in Fig.~\ref{GGMreconstruct} that even a precise determination of the soft masses can lead to moderate uncertainties in the reconstructed $A_r$. This is mostly due to cancelations between large squark mass parameters, which reduce the value of the $I_{M_r}$ while increasing the magnitude of the uncertainties.
This explains why the value of $A_3$ can be determined with relatively good precision, while even for large values of $A_1$, its uncertainty may be as large as its calculated value. In most cases, however, the uncertainties in the $A_r$ are smaller than their calculated values, indicating that useful information can be obtained about the allowed range of these parameters. The considerable spread in the reconstructed values of the $A_r$ is due to 2-loop effects and will be mostly compensated by methods to be discussed in the next section.

As mentioned previously, if measurements suggest $\delta_u= \delta_d$ within errors, one can still obtain constraints on the $A_r$ by making educated guesses for the gauge couplings at the messenger scale. Taking the messenger scale to lie between $10^5$ and $10^{16}$ GeV, the gauge coupling dependence entering into Eq.~(\ref{Ar}) can then be estimated as\footnote{In principle, if the messenger scale is $\lesssim 10^7$ GeV, it may be determined from decays of the NLSP to the gravitino inside the detector. Assuming this is the case one could take the messenger scale between $10^7$ and $10^{16}$ GeV, with $g_1^{-4}\approx 9\pm 5$, $g_2^{-4}\approx 4.5\pm 0.5$, and $g_3^{-4}\approx 2.5\mp 1$.}
\begin{align}
g_1^{-4}&\approx 10\pm 6\;,\nonumber\\
g_2^{-4}&\approx 4.5\pm 0.5\;,\nonumber\\
g_3^{-4}&\approx 2.5\mp 1.5\;.
\end{align}
These errors are not very different from what is obtained by propagating soft mass experimental errors through the $g_1$ reconstruction in the case $\delta_u\neq \delta_d$. Observe that the values of the couplings are correlated and should satisfy the $I_{g_{2,3}}$ RGI constraints.

Therefore, one can always use the RGIs to obtain interesting constraints on the $A_r$, even if $\delta_u= \delta_d$ within uncertainties.

\subsection{2-loop Effects in GGM}

\begin{figure}[!htbp]
\begin{center}
\includegraphics[width=1.02\textwidth,trim=1.4in 85.5in 0 70, clip=true]{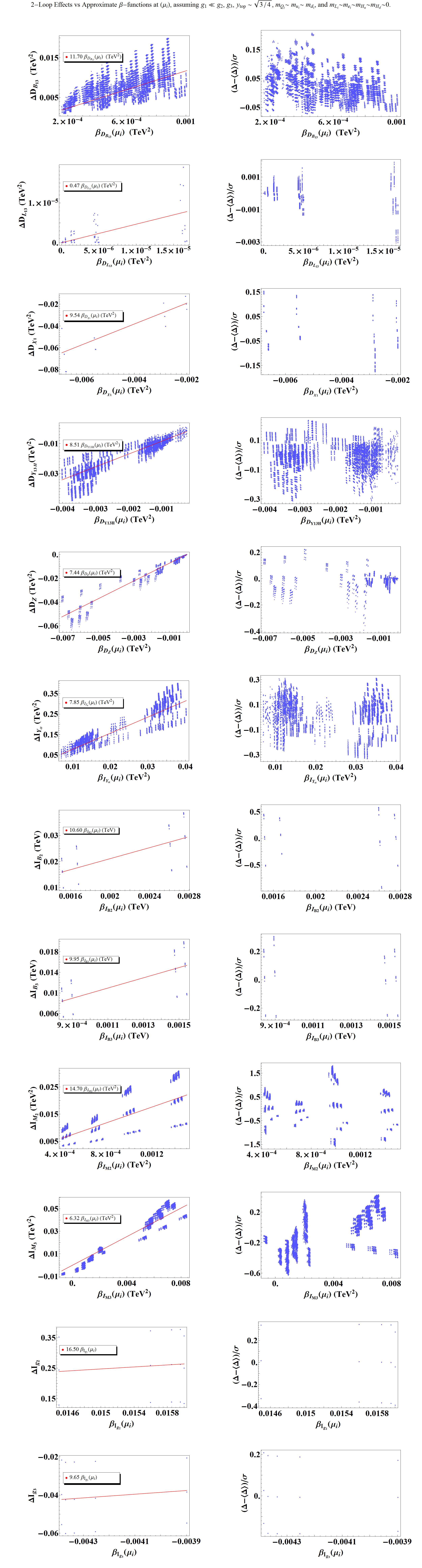}
\caption{\textit{Left:} 2-loop shifts in the low scale values of the RGIs. The x-axis is a simple approximation to the 2-loop $\beta$-function. The red line denotes the best-fit for $\Delta$ in terms of our approximate $\beta$ function. \textit{Right:} The ratio of the residual 2-loop effects, after subtracting the best fit $\beta$ from $\Delta$, to the experimental errors. We assume 1\% uncertainties in all of the soft masses.}
\label{fig2loop}
\end{center}
\end{figure}

\begin{figure}[!htbp]
\begin{center}
\includegraphics[width=1.02\textwidth,trim=1.4in 42in 0 44in, clip=true]{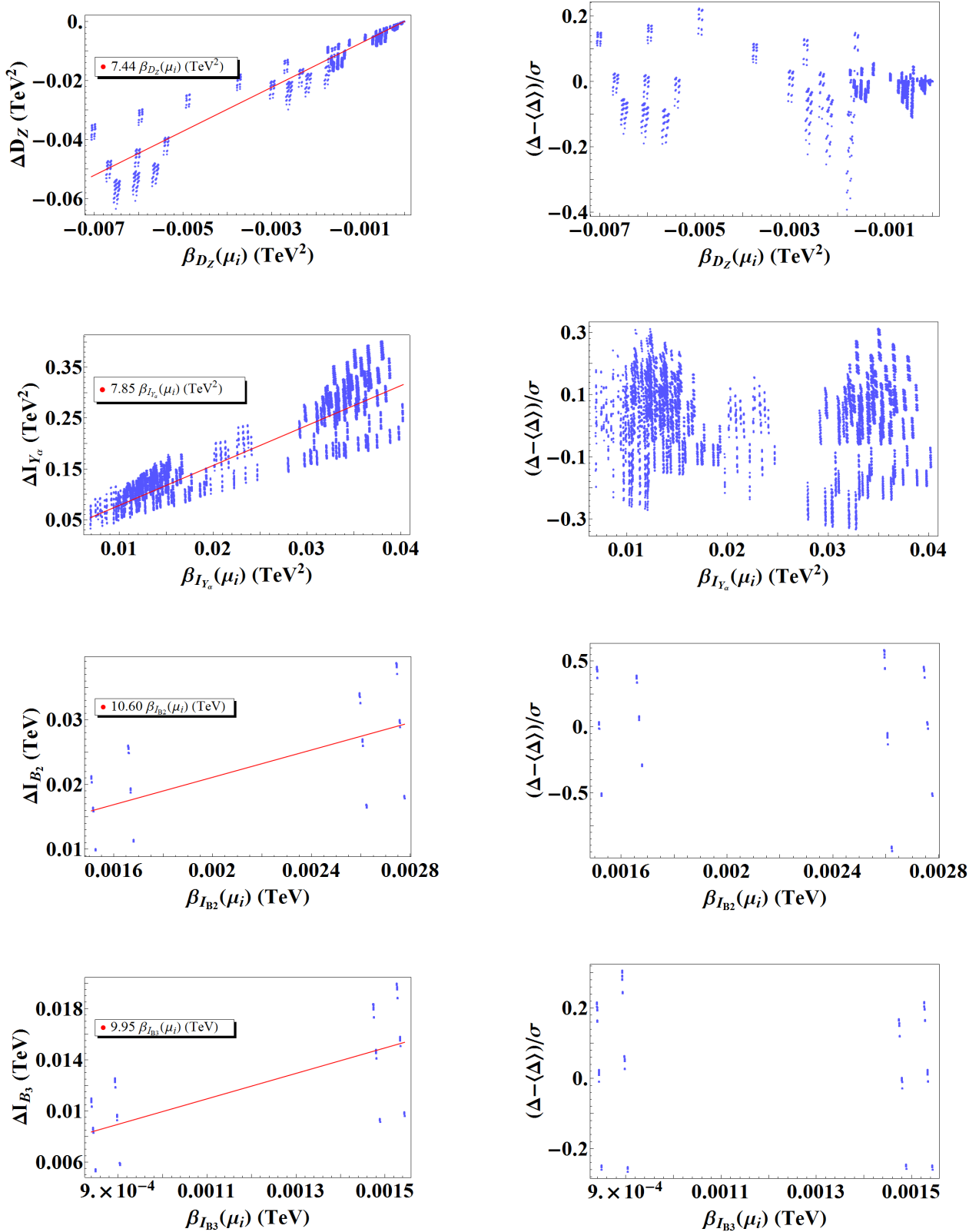}
\caption{\textit{Left:} 2-loop shifts in the low scale values of the RGIs. The x-axis is a simple approximation to the 2-loop $\beta$-function. The red line denotes the best-fit for $\Delta$ in terms of our approximate $\beta$ function. \textit{Right:} The ratio of the residual 2-loop effects, after subtracting the best fit $\beta$ from $\Delta$, to the experimental errors. We assume 1\% uncertainties in all of the soft masses.}
\label{fig2loop2}
\end{center}
\end{figure}

\begin{figure}[!htbp]
\begin{center}
\includegraphics[width=1.02\textwidth,trim=1.4in 0 0 86in, clip=true]{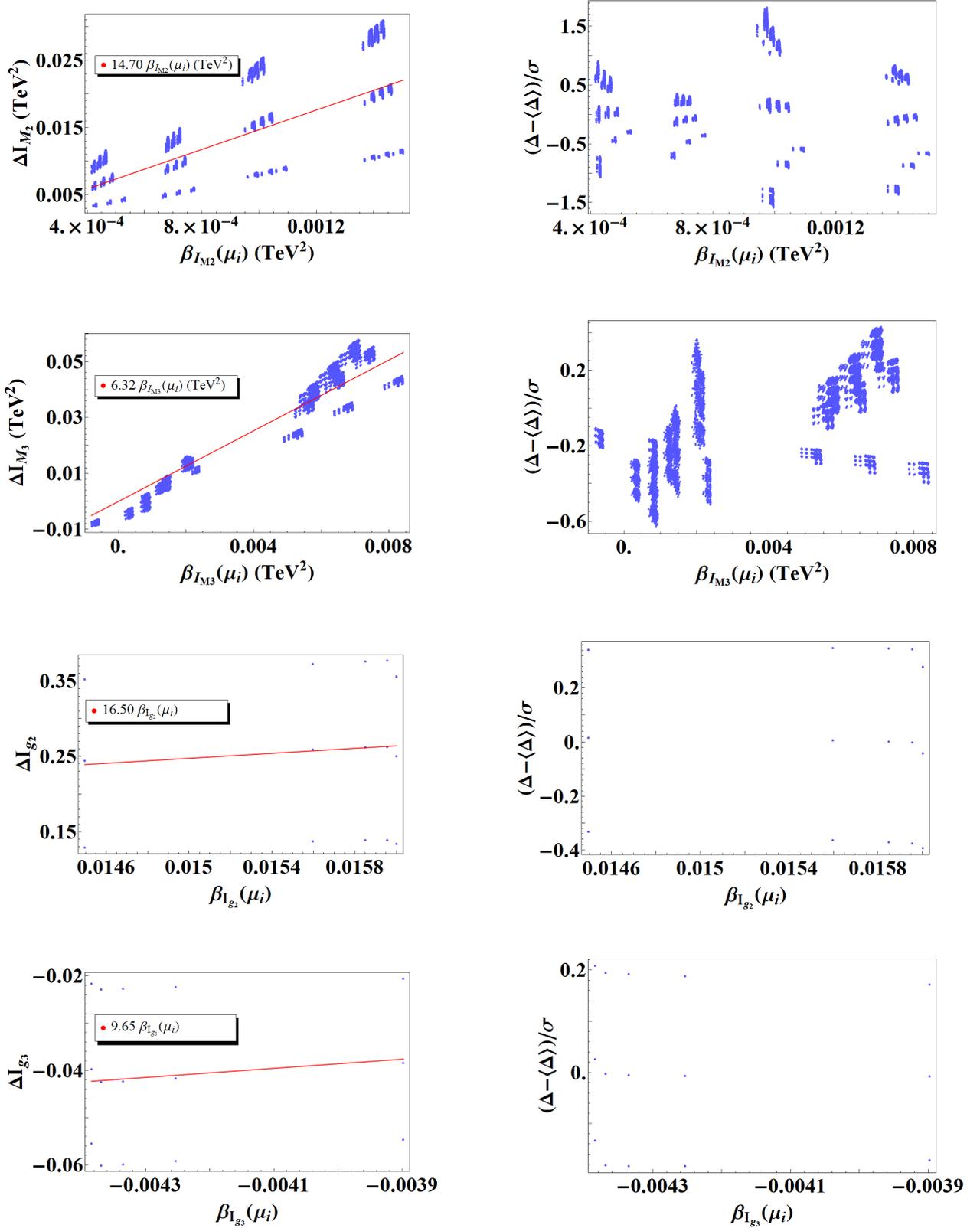}
\caption{\textit{Left:} 2-loop shifts in the low scale values of the RGIs. The x-axis is a simple approximation to the 2-loop $\beta$-function. The red line denotes the best-fit for $\Delta$ in terms of our approximate $\beta$ function. \textit{Right:} The ratio of the residual 2-loop effects, after subtracting the best fit $\beta$ from $\Delta$, to the experimental errors. We assume 1\% uncertainties in all of the soft masses.}
\label{fig2loop3}
\end{center}
\end{figure}

\begin{figure}[!htbp]
\begin{center}
\includegraphics[width=1.02\textwidth,trim=.5in 0in 0 0,clip=true]{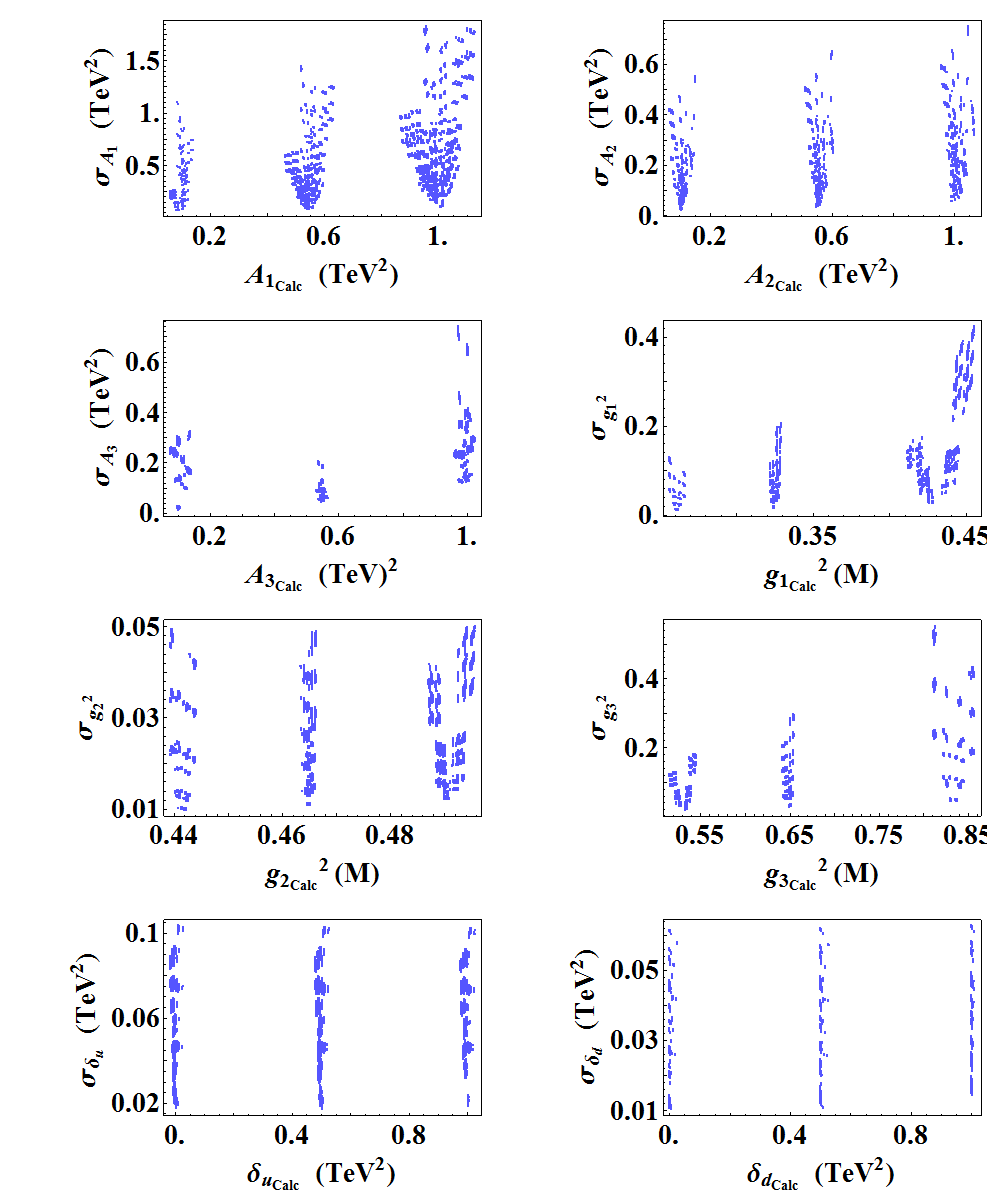}
\caption{Experimental errors in the calculated values of the high scale GGM parameters using the RGI reconstruction method. Parameters with small uncertainties are omitted for brevity. These plots differ from those of Fig.~\ref{GGMreconstruct} in that the bulk of the 2-loop corrections have been accounted for via simple shifts in the RGIs, as detailed in Section 3.3.}
\label{GGMshiftreconstruct}
\end{center}
\end{figure}

\begin{figure}[!htbp]
\begin{center}
\includegraphics[width=1.0\textwidth,trim=0 0in 0 9in, clip=true]{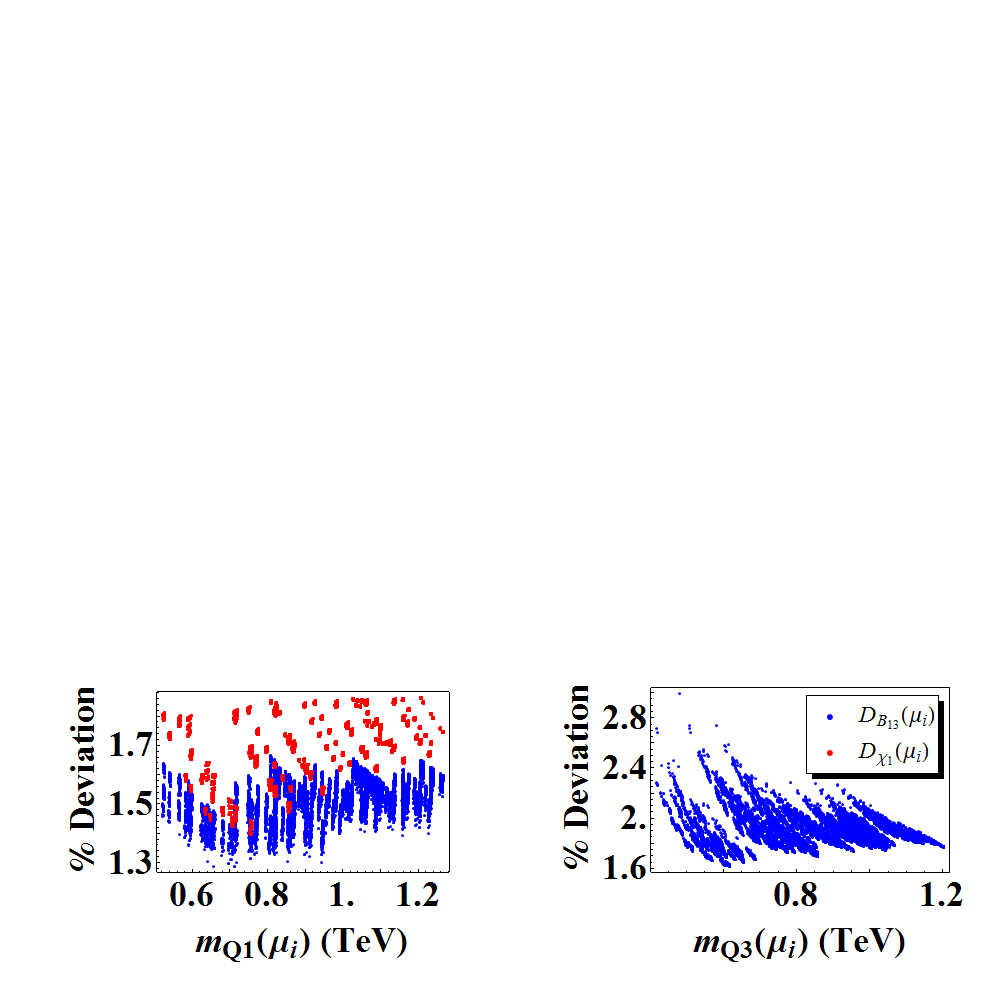}
\caption{Percent deviation in left-handed squark masses necessary to generate values of $D_{B_{13}}$ (blue points) or $D_{\chi_1}$ (red points) more than 1$\sigma$ away from zero, assuming 1\% experimental uncertainties in the soft masses.}
\label{massdev1}
\includegraphics[width=1.0\textwidth,trim=0 0 0 9.in, clip=true]{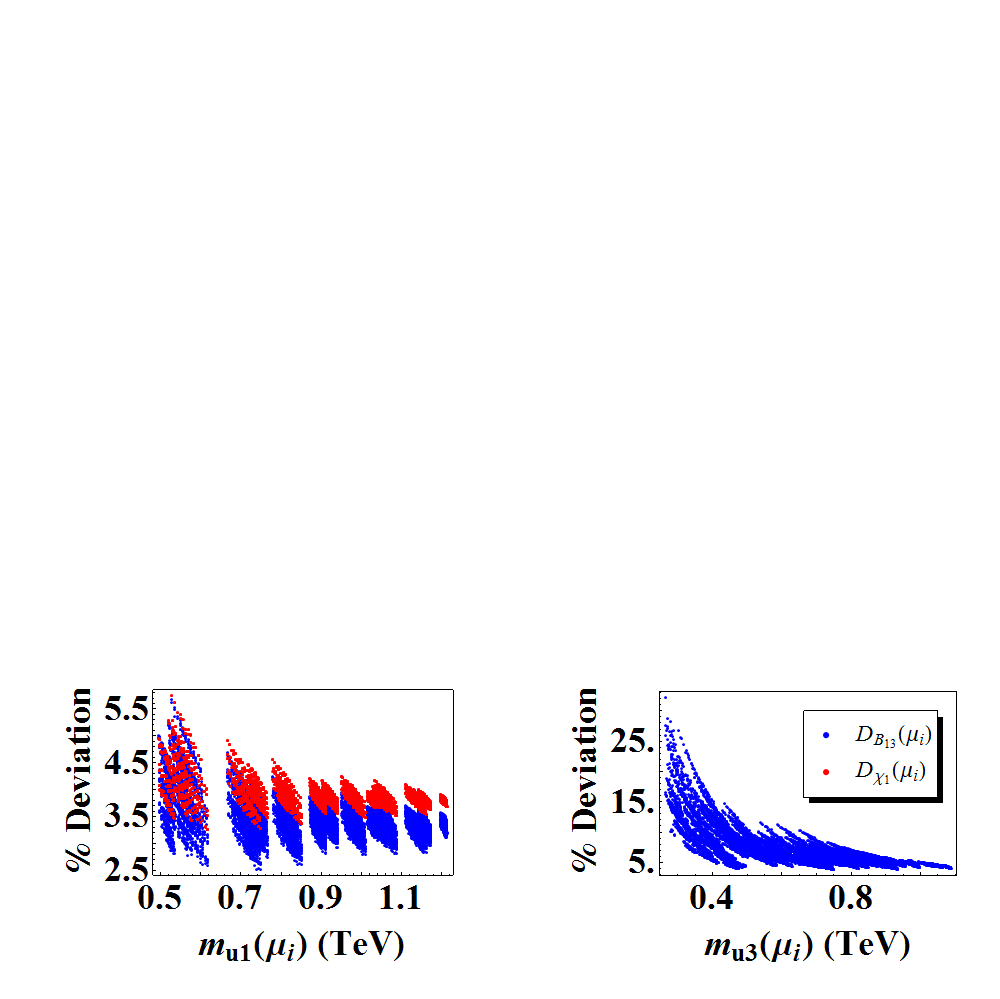}
\caption{Percent deviation in right-handed up-type squark masses necessary to generate values of $D_{B_{13}}$ (blue points) or $D_{\chi_1}$ (red points) more than 1$\sigma$ away from zero, assuming 1\% experimental uncertainties in the soft masses.}
\label{massdev2}
\includegraphics[width=1.0\textwidth,trim=0 0 0 9.in, clip=true]{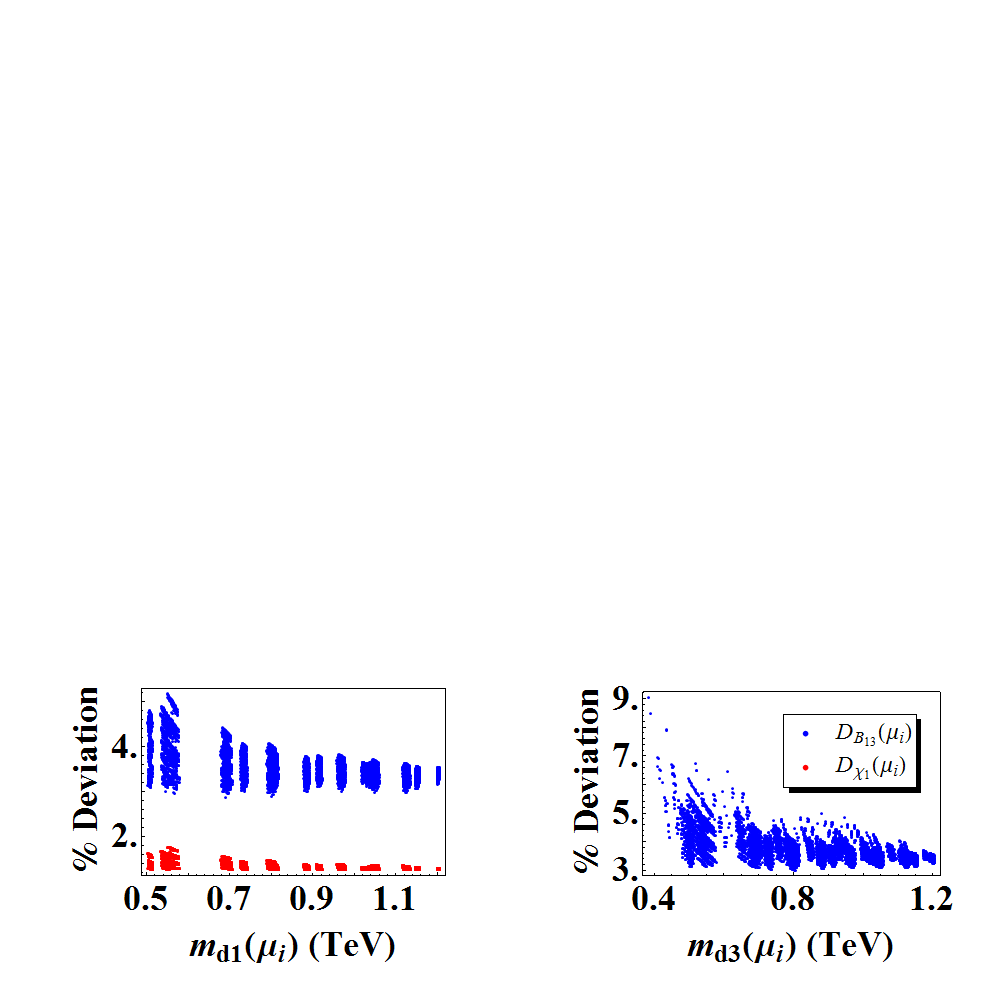}
\caption{Percent deviation in right-handed down-type squark masses necessary to generate values of $D_{B_{13}}$ (blue points) or $D_{\chi_1}$ (red points) more than 1$\sigma$ away from zero, assuming 1\% experimental uncertainties in the soft masses.}
\label{massdev3}
\end{center}
\end{figure}

\begin{figure}[!htb]
\begin{center}
\includegraphics[width=1.0\textwidth,trim=0 0 0 9.in, clip=true]{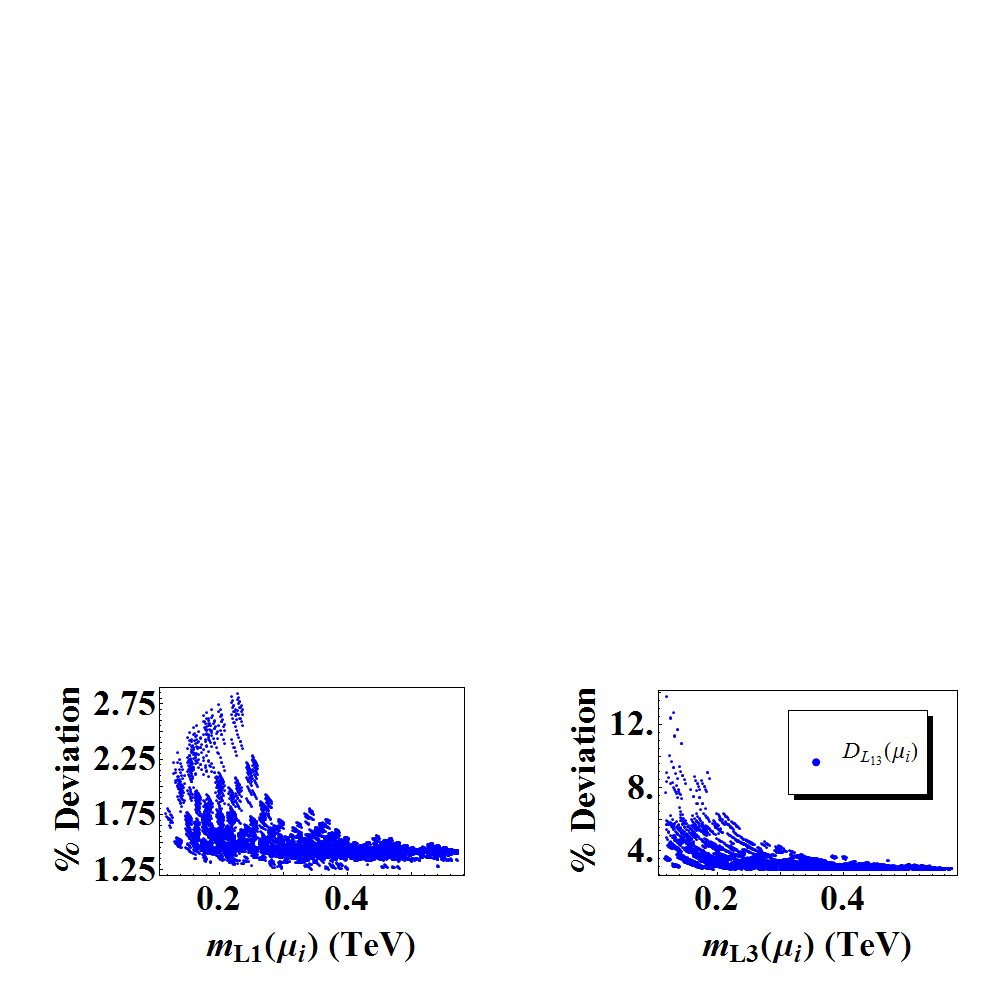}
\caption{Percent deviation in left-handed slepton masses necessary to generate values of $D_{L_{13}}$ more than 1$\sigma$ away from zero, assuming 1\% experimental uncertainties in the soft masses.}
\label{massdev4}
\includegraphics[width=1.0\textwidth,trim=0 0 0 9.in, clip=true]{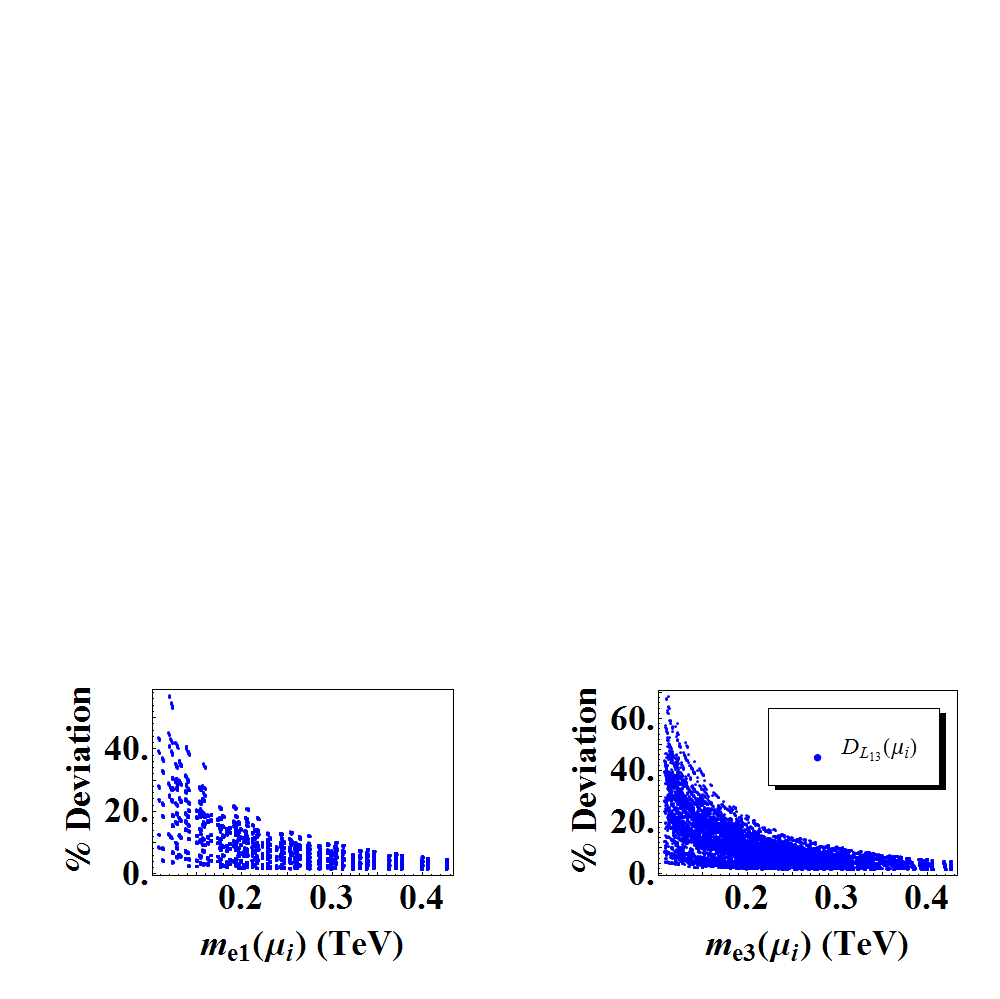}
\caption{Percent deviation in right-handed slepton masses necessary to generate values of $D_{L_{13}}$ more than 1$\sigma$ away from zero, assuming 1\% experimental uncertainties in the soft masses.}
\label{massdev5}
\end{center}
\end{figure}

One can easily check that the RGIs discussed are not strictly preserved at the 2-loop level and therefore it is important to estimate the magnitude of the 2-loop effects. For the scan discussed in the previous section, we implemented full 2-loop RG equations for the soft SUSY-breaking parameters and the gauge and Yukawa couplings~\cite{Martin:1993zk} into Mathematica. We compared the running of the parameters calculated by our code with those obtained from the public program SuSpect~\cite{Djouadi:2002ze}, finding excellent agreement.

The spread in the reconstructed values of the $A_r$ in Fig.~\ref{GGMreconstruct} demonstrates that apart from experimental error, there is also a theoretical uncertainty in the calculated low scale value of invariants since the RGIs defined above have vanishing $\beta$-functions only at 1-loop order. The difference between the low and high scale central values of the RGIs and the difference between the input and reconstructed central values of the GGM parameters gives the 2-loop contributions to these quantities.

In Appendix A we list the 2-loop $\beta$-functions for all the RGIs, ignoring the small contribution from the hypercharge couplings as well as the terms proportional to the trilinear coupling, which however have been included in our numerical work. We see that the contributions to $D_{L_{13}}$ and $I_{M_1}$ vanish in the limit that $y_e\rightarrow 0$, and so in general these functions behave as approximate RGIs at the 2-loop level. Furthermore, the $\beta$-function of $D_Z$ is proportional to the square of the bottom Yukawa coupling and it is therefore only relevant for large values of $\tan\beta$. Finally, the $\beta$-function of $D_{B_{13}}$ does not contain any strong gauge coupling contribution and thus tends to be small for low values of the messenger scale.

To analyze the invariants with larger 2-loop contributions, it is useful to consider a further limit of the $\beta$-functions defined in Appendix A in which we turn off the slepton masses, the bino mass, and the Higgs masses, set the masses of all squarks equal to $m_{\tilde{d}_1}$, and $y_u=\sqrt{3/4}$. This reduces the 2-loop $\beta$-functions to simple functions of $\tan\beta$, $M_2$, $M_3$, and $m_{\tilde{d}_1}$. The 2-loop corrections can then be estimated from a limited number of parameters as the values of these functions multiplied by $\log(M/\mbox{TeV})$.

To demonstrate the utility of these estimates, we analyze numerically the shifts in the RGIs induced by the full 2-loop RGEs. In the left-hand column of Figs.~\ref{fig2loop}-\ref{fig2loop3}, we plot the 2-loop corrections, $\Delta(\mbox{RGI})$, against the approximate 2-loop $\beta$-functions. The best-fit line passing through the origin then gives an estimate for the average value, $\langle\Delta\rangle$, of the 2-loop corrections. The slope of the line corresponds approximately to $\log(M_0/\mbox{TeV})$, where $M_0$ is some intermediate messenger scale. We tabulate the slopes in Table~\ref{BetaShifts}. The deviation from this approximation is due to residual 2-loop running as well as to the different values of the parameters over which we scan. In the right-hand column we subtract off the approximated 2-loop $\beta$-function shifts, $\langle\Delta\rangle$s, from the 2-loop contributions, $\Delta$s, at every point in the scan and plot the ratio of $\Delta-\langle\Delta\rangle$ to the experimental uncertainties against the approximate 2-loop $\beta$-functions. The spread in the y-axis is due to the residual 2-loop effects not accounted for by the $\langle\Delta\rangle$s.

\begin{table}
  \centering
  \begin{tabular}{|| c | c ||c | c ||c | c ||c | c ||}
  \hline
\hline
&&&&&&&\\
~RGI~ & $\langle\Delta\rangle$ & ~RGI~ & $\langle\Delta\rangle$ & ~RGI~ & $\langle\Delta\rangle$& ~RGI~ & $\langle\Delta\rangle$\\ [0.8ex]
\hline
\hline
&&&&&&&\\
$D_{B_{13}}$& $11.7 ~\beta_{D_{B_{13}}}$ & $D_{L_{13}}$& $0.47 ~\beta_{D_{L_{13}}}$ & $D_{\chi_{1}}$& $9.5~ \beta_{D_{\chi_{1}}}$ & $D_{Y_{13H}}$& $8.5 ~\beta_{D_{Y_{13H}}}$\\ [3ex]
\hline
&&&&&&&\\
$D_{Z}$& $7.4 ~\beta_{D_{Z}}$&
$I_{Y_{\alpha}}$& $7.9 ~\beta_{I_{Y_{\alpha}}}$&
$I_{B_{2}}$& $10.6 ~\beta_{I_{B_{2}}}$&
$I_{B_{3}}$& $9.95 ~\beta_{I_{B_{3}}}$\\ [3ex]
\hline
&&&&&&&\\
$I_{M_{2}}$& $14.7 ~\beta_{I_{M_{2}}}$&
$I_{M_{3}}$& $6.3 ~\beta_{I_{M_{3}}}$&
$I_{g_{2}}$& $16.5 ~\beta_{I_{g_{2}}}$&$I_{g_{3}}$& $9.65 ~\beta_{I_{g_{3}}}$\\ [3ex]
\hline
\hline
  \end{tabular}
  \caption{Equations for the best fit line for the 2-loop beta functions, corresponding to Figs.~\ref{fig2loop}-\ref{fig2loop3}.}
  \label{BetaShifts}
\end{table}
%

From the numerical simulation we see that in most cases, even with an optimistic experimental error of 1\% in the soft masses, the experimental error tends to be larger than the residual 2-loop effects on the RGIs once the shift is performed to remove the bulk of the 2-loop corrections. In those cases it is justified to treat the 1-loop RGIs as true invariants in the determination of the parameters of the GGM models. For $I_{M_2}$, the 2-loop corrections can become of the same order as the 1\% experimental errors. One could then in principle combine the uncertainty in $I_{M_2}$ due to the 2-loop effects in quadrature with the experimental errors. Notice that the determination of the messenger scale by any independent method can serve to further reduce most of the uncertainty associated with the 2-loop evolution of the soft parameters.

In Fig.~\ref{GGMshiftreconstruct} we repeat the plots of Fig.~\ref{GGMreconstruct} using the shifted RGIs to compute the GGM input parameters. This further demonstrates the advantage of the simple shifts and the subsequent dominance of experimental errors over residual 2-loop corrections.


Finally, we consider more carefully $D_{B_{13}}$, $D_{L_{13}}$, and $D_{\chi_1}$. The first two are expected to vanish for all flavor-blind models and all three vanish in GGM. To test the power of these invariants as discriminants we calculate the percent deviation in the soft SUSY-breaking parameters that would lead to a departure from zero. For instance, even a 10\% shift in $m_{\tilde{Q}_1}$ entering into the $D_{\chi_1}$ RGI is enough to shift it by more than $5\sigma$ outside the range experimentally consistent with zero. Similar conclusions hold for $D_{B_{13}}$ and $D_{L_{13}}$. For a few of the soft masses we plot in Figs.~\ref{massdev1}-\ref{massdev5} the percent deviation in them that would cause the invariants to take values more than 1$\sigma$ away from zero. $D_{\chi_1}$ and $D_{B_{13}}$ are quite sensitive to the squark masses; they are less sensitive to the generically smaller slepton masses, but for those $D_{L_{13}}$ is an excellent discriminator. As mentioned before, the effect of a larger experimental error in the soft masses can be simply seen as a rescaling of the y-axis of Figs.~\ref{massdev1}-\ref{massdev5}.


\section{Conclusions}

Low energy supersymmetry provides a predictive framework of physics at the weak scale. However, the precise spectrum of the new particles depends on the soft SUSY-breaking parameters, which proceed from the unknown mechanism mediating SUSY-breaking from the hidden sector to the MSSM at high energies. Knowledge of the soft parameters at the messenger scale will greatly contribute to the understanding of UV physics. In principal the RG evolution of the soft parameters will allow the determination of their high scale values as a function of the measured values at the weak scale. The running of each scalar parameter, however, is complicated by the dependence of the $\beta$-function on nearly all other running soft parameters.

In this work we have proposed the use of Renormalization Group invariants to resolve the soft SUSY-breaking parameters at the messenger scale. Of the fourteen RGIs we have considered, two may be used to test high scale flavor universality of the soft parameters, and twelve encode information about the thirteen variables associated with the most general CP-conserving flavor-universal models. Indeed, if the messenger scale can be determined by alternative methods, the whole spectrum at high energies could be established.

Specific models of SUSY-breaking, however, lead to relationships between the different sparticle masses at the messenger scale. For example, in General Gauge Mediation, even assuming a modification of the soft parameters in the Higgs sector, only eleven free parameters remain. Moreover, an additional RGI must vanish and may be used, together with the two RGIs related to the flavor independence, to test consistency of the low scale spectrum with GGM. As we have shown in this work, the deviation from zero of these invariants is a very powerful discriminant for these models. The remaining eleven RGIs can be used to extract most, and in some cases all, of the GGM parameters at the messenger scale. More minimal models, for example the CMSSM~+~NUHM, depend on fewer parameters to define the high energy sparticle spectrum. In those cases the system is over-constrained, leading not only to the determination of all parameters but also to consistency relations between the values of several RGIs.

Although RG invariance holds only at the 1-loop level, we have shown that in general the 2-loop evolution leads to modifications of the RGIs that are smaller than their expected experimental errors obtained from aggressive 1\% uncertainties in the low scale soft masses. Moreover, we have shown that a simple approximation based on the 2-loop $\beta$-functions and dependent on a few low-scale parameters effectively describes the 2-loop corrections to the invariants. These functions can be directly subtracted from the low energy measured values of the RGIs, thereby reducing the theoretical uncertainty. The remaining uncertainty is then primarily due to the unknown messenger scale and can be reduced once further constraints are obtained on the value of this scale.

RGIs offer a simple and modular approach to the reconstruction of messenger scale physics. The dependence of high scale parameters on observed values at the weak scale is reduced from a set of integrals to a set of algebraic equations. Moreover, each RGI depends only on a subset of the low scale masses. Although we have not investigated this possibility in detail, it would be interesting to consider strategies for the use of RGIs if only a limited set of the masses are extracted at the LHC or a future linear collider. In such a situation traditional methods relying on direct integration of the RG equations will fail in the sfermion sector, while if any RGIs depend only on the known masses, they may still be used to provide constraints on the high scale structure. It would also be interesting to examine in greater detail the use and effectiveness of the RGI method in other proposed models for the mediation of SUSY-breaking, as well as in non-minimal models of low scale supersymmetry. Additionally, in this work we have taken a simple quadrature sum in the estimation of experimental errors, neglecting correlations which in many cases may be significant. Finally, a more realistic determination of the expected experimental uncertainty in the extraction of the soft breaking parameters could alter the overall uncertainties in parameter reconstructions. We leave these investigations to future work.

~\\
~\\
{\large{\textbf{Acknowledgements:}}}
\vspace{0.2 cm}

Fermilab is operated by Fermi Research Alliance, LLC under Contract No. DE-AC02-07CH11359 with the U.S. Department of Energy. Work at ANL is supported in part by the U.S. Department of Energy (DOE), Div.~of HEP, Contract DE-AC02-06CH11357. This work was supported in part by the DOE under Task TeV of contract DE-FGO2-96-ER40956.  M.~C., N.~S. and C.~W. would like to thank the Aspen Center for Physics, where part of this work has been done.

\normalsize

\newpage

\appendix

\begin{center}
{\large{\textbf{APPENDIX A}}}
\end{center}

\setcounter{equation}{0}
\begin{center}
\textbf{2-Loop $\beta$-Functions for the 1-Loop RG Invariants}\nonumber
\end{center}
\renewcommand{\theequation}{A.\arabic{equation}}

\noindent In the following set of 2-loop $\beta$-functions, we set the hypercharge gauge coupling $g_1$ and the soft trilinear couplings $A_i$ to zero. The subscripts on the $\beta$ functions are in correspondence with the invariants defined in the text.

~\\
\begin{align}
\beta_{D_{B_{13}}} =\frac{1}{64 \pi ^4}&\bigg(3 g_2  ^2 \bigg(m_{\tilde{d}_3}^2+m_{\tilde{Q}_3}^2+m_{H_d}  ^2+2 M_2 ^2\bigg) y_d  ^2+2 \bigg(m_{\tilde{d}_3}^2+m_{\tilde{Q}_3}^2+m_{H_d}  ^2\bigg) y_d  ^4\nonumber\\
&+\bigg(3 g_2  ^2 \bigg(m_{\tilde{Q}_3}^2+m_{\tilde{u}_3}^2+m_{H_u}  ^2+2 M_2  ^2\bigg)-2 \bigg(m_{\tilde{d}_3}^2+2 m_{\tilde{Q}_3}^2+m_{\tilde{u}_3}^2+m_{H_d} ^2+m_{H_u}  ^2\bigg) y_d  ^2\bigg) y_u  ^2\nonumber\\
&+2 \bigg(m_{\tilde{Q}_3}^2+m_{\tilde{u}_3}^2+m_{H_u}  ^2\bigg) y_u  ^4\bigg) \;,\label{beta_B13} \\
\beta_{D_{L_{13}}} = \frac{1}{64 \pi ^4}&\bigg(3 g_2  ^2 \bigg(m_{\tilde{e}_3}^2+m_{\tilde{L}_3}^2+m_{H_d}  ^2+2 M_2  ^2\bigg) y_e  ^2+2 \bigg(m_{\tilde{e}_3}^2+m_{\tilde{L}_3}^2+m_{H_d}  ^2\bigg) y_e  ^4\bigg)\;,\label{beta_L13} \\
\beta_{D_{\chi_1}} =-\frac{1}{4 \pi ^4}&\bigg(3 g_2  ^2 g_3  ^2 \bigg(M_2  ^2+M_2  M_3 +M_3  ^2\bigg)\bigg)\;, \label{beta_chi} \\
\beta_{D_{Y_{13H}}} =\frac{1}{416 \pi ^4}&\bigg(3 g_2  ^2 \bigg(4 g_3  ^2 \bigg(M_2  ^2+M_2  M_3 +M_3  ^2\bigg)-5 \bigg(m_{\tilde{d}_3}^2+m_{\tilde{Q}_3}^2+m_{H_d}  ^2+2 M_2  ^2\bigg) y_d  ^2\nonumber\\
&-5 \bigg(m_{\tilde{e}_3}^2+m_{\tilde{L}_3}^2+m_{H_d}  ^2+2 M_2  ^2\bigg) y_e  ^2+10 \bigg(m_{\tilde{Q}_3}^2+m_{\tilde{u}_3}^2+m_{H_u}  ^2+2 M_2  ^2\bigg) y_u ^2\bigg)\nonumber\\
&+5 \bigg(8 m_{\tilde{Q}_3}^2 g_3  ^2 y_d  ^2+8 g_3  ^2 m_{H_d}  ^2 y_d  ^2+16 g_3  ^2 M_3  ^2 y_d  ^2+4 m_{\tilde{Q}_3}^2 y_d  ^4+4 m_{H_d}  ^2 y_d  ^4+3 m_{\tilde{e}_3}^2 y_d  ^2 y_e  ^2+3 m_{\tilde{L}_3}^2 y_d  ^2 y_e  ^2\nonumber\\
&+3 m_{\tilde{Q}_3}^2 y_d  ^2 y_e  ^2+6 m_{H_d}  ^2 y_d  ^2 y_e  ^2-2 m_{\tilde{e}_3}^2 y_e  ^4-2 m_{\tilde{L}_3}^2 y_e  ^4-2 m_{H_d}  ^2 y_e  ^4\nonumber\\
&-\bigg(8 g_3  ^2 \bigg(m_{\tilde{Q}_3}^2+m_{\tilde{u}_3}^2+m_{H_u}  ^2+2 M_3  ^2\bigg)+\bigg(2 m_{\tilde{Q}_3}^2+m_{\tilde{u}_3}^2+m_{H_d}  ^2+m_{H_u}  ^2\bigg) y_d  ^2\bigg) y_u  ^2\nonumber\\
&-2 \bigg(m_{\tilde{Q}_3}^2+m_{\tilde{u}_3}^2+m_{H_u}  ^2\bigg) y_u  ^4+m_{\tilde{d}_3}^2 y_d  ^2 \bigg(8 g_3  ^2+4 y_d  ^2+3 y_e  ^2-y_u ^2\bigg)\bigg)\bigg)\;,\label{beta_Y13H}\\
\beta_{D_{Z}} =\frac{1}{64 \pi ^4}&y_d  ^2 \bigg(9 g_2  ^2 m_{H_d}  ^2-16 g_3  ^2 m_{H_d}  ^2+18 g_2  ^2 M_2  ^2-32 g_3  ^2 M_3  ^2-6 m_{H_d}  ^2 y_d  ^2-6 \bigg(m_{\tilde{e}_3}^2+m_{\tilde{L}_3}^2+2 m_{H_d}  ^2\bigg) y_e  ^2\nonumber\\
&+m_{\tilde{d}_3}^2 \bigg(9 g_2  ^2-2 \bigg(8 g_3  ^2+3 \bigg(y_d  ^2+y_e  ^2\bigg)\bigg)\bigg)+m_{\tilde{Q}_3}^2 \bigg(9 g_2  ^2-2 \bigg(8 g_3  ^2+3 \bigg(y_d  ^2+y_e  ^2\bigg)\bigg)\bigg)\bigg)\;,\label{beta_Z}
\end{align}
\begin{align}
\beta_{D_{Y\alpha}} =\frac{1}{64 \pi ^4 g_1  ^2}&\bigg(-8 m_{\tilde{Q}_3}^2 g_3  ^2 y_d  ^2-8 g_3  ^2 m_{H_d}  ^2 y_d  ^2-16 g_3  ^2 M_3  ^2 y_d  ^2-4 m_{\tilde{Q}_3}^2 y_d  ^4-4 m_{H_d}  ^2 y_d  ^4-3 m_{\tilde{e}_3}^2 y_d  ^2 y_e  ^2\nonumber\\
&-3 m_{\tilde{L}_3}^2 y_d  ^2 y_e  ^2-3 m_{\tilde{Q}_3}^2 y_d  ^2 y_e  ^2-6 m_{H_d}  ^2 y_d  ^2 y_e  ^2+2 m_{\tilde{e}_3}^2 y_e ^4+2 m_{\tilde{L}_3}^2 y_e  ^4+2 m_{H_d}  ^2 y_e  ^4\nonumber\\
&+\bigg(8 g_3  ^2 \bigg(m_{\tilde{Q}_3}^2+m_{\tilde{u}_3}^2+m_{H_u}  ^2+2 M_3  ^2\bigg)+\bigg(2 m_{\tilde{Q}_3}^2+m_{\tilde{u}_3}^2+m_{H_d}  ^2+m_{H_u}  ^2\bigg) y_d  ^2\bigg) y_u  ^2\nonumber\\
&+2 \bigg(m_{\tilde{Q}_3}^2+m_{\tilde{u}_3}^2+m_{H_u}  ^2\bigg) y_u  ^4+m_{\tilde{d}_3}^2 y_d  ^2 \bigg(-8 g_3  ^2-4 y_d ^2-3 y_e  ^2+y_u  ^2\bigg)\nonumber\\
&+3 g_2  ^2 \bigg(m_{\tilde{d}_3}^2 y_d  ^2+\bigg(m_{\tilde{e}_3}^2+m_{\tilde{L}_3}^2\bigg) y_e ^2+m_{H_d}  ^2 \bigg(y_d  ^2+y_e  ^2\bigg)-2 \bigg(m_{\tilde{u}_3}^2+m_{H_u}  ^2\bigg) y_u  ^2\nonumber\\
&+m_{\tilde{Q}_3}^2 \bigg(y_d  ^2-2 y_u  ^2\bigg)+2 \bigg(4 g_3  ^2 \bigg(M_2  ^2+M_2  M_3 +M_3  ^2\bigg)+M_2  ^2 \bigg(y_d  ^2+y_e ^2-2 y_u  ^2\bigg)\bigg)\bigg)\bigg)\;,\label{beta_Ya}\\
\beta_{I_{B_1}} =\frac{1}{640 \pi ^4}&\bigg(27 g_2  ^2 M_2 +88 g_3  ^2 M_3 \bigg)\;,\label{beta_B1}\\
\beta_{I_{B_2}} =\frac{1}{128 \pi ^4}&\bigg(25 g_2  ^2 M_2 +24 g_3  ^2 M_3 \bigg)\;,\label{beta_B2}\\
\beta_{I_{B_3}} =\frac{1 }{128 \pi ^4}&\bigg(9 g_2  ^2 M_2 +14 g_3  ^2 M_3\bigg)\;,\label{beta_B3}\\
\beta_{I_{M_1}} =0&\;,\label{beta_M1}\\
\beta_{I_{M_2}} =\frac{1}{128 \pi ^4}&g_2  ^2 \bigg(m_{\tilde{L}_1}^2 g_2  ^2+m_{\tilde{L}_2}^2 g_2  ^2+m_{\tilde{L}_3}^2 g_2 ^2+3 m_{\tilde{Q}_1}^2 g_2  ^2+3 m_{\tilde{Q}_2}^2 g_2  ^2+3 m_{\tilde{Q}_3}^2 g_2  ^2+g_2  ^2 m_{H_d}  ^2+g_2  ^2 m_{H_u} ^2\nonumber\\
&+111 g_2  ^2 M_2  ^2+48 g_3  ^2 M_2  ^2+48 g_3  ^2 M_2  M_3 -12 M_2  ^2 y_d  ^2-4 M_2  ^2 y_e  ^2-12 M_2  ^2 y_u ^2\bigg)\;,\label{beta_M2}\\
\beta_{I_{M_3}} =-\frac{1}{128 \pi ^4}&g_3  ^2 \bigg(3 m_{\tilde{d}_1}^2 g_3  ^2+3 m_{\tilde{d}_2}^2 g_3  ^2+3 m_{\tilde{d}_3}^2 g_3  ^2+6 m_{\tilde{Q}_1}^2 g_3  ^2+6 m_{\tilde{Q}_2}^2 g_3  ^2+6 m_{\tilde{Q}_3}^2 g_3  ^2+3 m_{\tilde{u}_1}^2 g_3  ^2\nonumber\\
&+3 m_{\tilde{u}_2}^2 g_3  ^2+3 m_{\tilde{u}_3}^2 g_3  ^2-80 g_3  ^2 M_3  ^2-18 g_2  ^2 M_3  \bigg(M_2 +M_3 \bigg)+8 M_3  ^2 y_d  ^2+8 M_3  ^2 y_u  ^2\bigg)\;,\label{beta_M3}\\
\beta_{I_{g_2}} =\frac{1}{320 \pi ^4}&\bigg(399 g_2  ^2+352 g_3  ^2-92 y_d  ^2-24 y_e  ^2-86 y_u  ^2\bigg)\;,\label{beta_g2}\\
\beta_{I_{g_3}} =\frac{1}{320 \pi ^4}&\bigg(-63 g_2  ^2-121 g_3  ^2+29 y_d  ^2+9 y_e  ^2+35 y_u  ^2\bigg)\;.\label{beta_g3}
\end{align}

\newpage

\bibliographystyle{apsrev4-1}
\setlength\bibsep{0.0pt}
\setlength{\bibhang}{0ex}
\bibliography{references}

\begin{thebibliography}{100}%
\makeatletter
\providecommand \@ifxundefined [1]{%
 \ifx #1\undefined \expandafter \@firstoftwo
 \else \expandafter \@secondoftwo
\fi
}%
\providecommand \@ifnum [1]{%
 \ifnum #1\expandafter \@firstoftwo
 \else \expandafter \@secondoftwo
\fi
}%
\providecommand \enquote [1]{``#1''}%
\providecommand \bibnamefont  [1]{#1}%
\providecommand \bibfnamefont [1]{#1}%
\providecommand \citenamefont [1]{#1}%
\providecommand\href[0]{\@sanitize\@href}%
\providecommand\@href[1]{\endgroup\@@startlink{#1}\endgroup\@@href}%
\providecommand\@@href[1]{#1\@@endlink}%
\providecommand \@sanitize [0]{\begingroup\catcode`\&12\catcode`\#12\relax}%
\@ifxundefined \pdfoutput {\@firstoftwo}{%
 \@ifnum{\z@=\pdfoutput}{\@firstoftwo}{\@secondoftwo}%
}{%
 \providecommand\@@startlink[1]{\leavevmode\special{html:<a href="#1">}}%
 \providecommand\@@endlink[0]{\special{html:</a>}}%
}{%
 \providecommand\@@startlink[1]{%
  \leavevmode
  \pdfstartlink
   attr{/Border[0 0 1 ]/H/I/C[0 1 1]}%
   user{/Subtype/Link/A<</Type/Action/S/URI/URI(#1)>>}%
  \relax
 }%
 \providecommand\@@endlink[0]{\pdfendlink}%
}%
\providecommand \url  [0]{\begingroup\@sanitize \@url }%
\providecommand \@url [1]{\endgroup\@href {#1}{\urlprefix}}%
\providecommand \urlprefix [0]{URL }%
\providecommand \Eprint[0]{\href }%
\@ifxundefined \urlstyle {%
  \providecommand \doi [1]{doi:\discretionary{}{}{}#1}%
}{%
  \providecommand \doi [0]{doi:\discretionary{}{}{}\begingroup
  \urlstyle{rm}\Url }%
}%
\providecommand \doibase [0]{http://dx.doi.org/}%
\providecommand \Doi[1]{\href{\doibase#1}}%
\providecommand \bibAnnote [3]{%
  \BibitemShut{#1}%
  \begin{quotation}\noindent
    \textsc{Key:}\ #2\\\textsc{Annotation:}\ #3%
  \end{quotation}%
}%
\providecommand \bibAnnoteFile [2]{%
  \IfFileExists{#2}{\bibAnnote {#1} {#2} {\input{#2}}}{}%
}%
\providecommand \typeout [0]{\immediate \write \m@ne }%
\providecommand \selectlanguage [0]{\@gobble}%
\providecommand \bibinfo [0]{\@secondoftwo}%
\providecommand \bibfield [0]{\@secondoftwo}%
\providecommand \translation [1]{[#1]}%
\providecommand \BibitemOpen[0]{}%
\providecommand \bibitemStop [0]{}%
\providecommand \bibitemNoStop [0]{.\EOS\space}%
\providecommand \EOS [0]{\spacefactor3000\relax}%
\providecommand \BibitemShut [1]{\csname bibitem#1\endcsname}%
\bibitem{Nilles:1983ge}%
  \BibitemOpen
  \bibfield{author}{%
  \bibinfo {author} {\bibfnamefont{H.~P.}\ \bibnamefont{Nilles}},\ }%
  \bibfield{journal}{%
  \Doi{10.1016/0370-1573(84)90008-5}{\bibinfo {journal} {Phys. Rept.}}\ }%
  \textbf{\bibinfo {volume} {110}},\ \bibinfo {pages} {1} (\bibinfo {year}
  {1984})%
  \bibAnnoteFile{NoStop}{Nilles:1983ge}%
\bibitem{Haber:1984rc}%
  \BibitemOpen
  \bibfield{author}{%
  \bibinfo {author} {\bibfnamefont{H.~E.}\ \bibnamefont{Haber}}\ and\ \bibinfo
  {author} {\bibfnamefont{G.~L.}\ \bibnamefont{Kane}},\ }%
  \bibfield{journal}{%
  \Doi{10.1016/0370-1573(85)90051-1}{\bibinfo {journal} {Phys. Rept.}}\ }%
  \textbf{\bibinfo {volume} {117}},\ \bibinfo {pages} {75} (\bibinfo {year}
  {1985})%
  \bibAnnoteFile{NoStop}{Haber:1984rc}%
\bibitem{Martin:1997ns}%
  \BibitemOpen
  \bibfield{author}{%
  \bibinfo {author} {\bibfnamefont{S.~P.}\ \bibnamefont{Martin}}}%
   (\bibinfo {year} {1997}),\
  \Eprint{http://arxiv.org/abs/hep-ph/9709356}{arXiv:hep-ph/9709356}%
  \bibAnnoteFile{NoStop}{Martin:1997ns}%
\bibitem{Haber:1990aw}%
  \BibitemOpen
  \bibfield{author}{%
  \bibinfo {author} {\bibfnamefont{H.~E.}\ \bibnamefont{Haber}}\ and\ \bibinfo
  {author} {\bibfnamefont{R.}~\bibnamefont{Hempfling}},\ }%
  \bibfield{journal}{%
  \bibinfo {journal} {Phys. Rev. Lett.}\ }%
  \textbf{\bibinfo {volume} {66}},\ \bibinfo {pages} {1815} (\bibinfo {year}
  {1991})%
  \bibAnnoteFile{NoStop}{Haber:1990aw}%
\bibitem{Okada:1990vk}%
  \BibitemOpen
  \bibfield{author}{%
  \bibinfo {author} {\bibfnamefont{Y.}~\bibnamefont{Okada}}, \bibinfo {author}
  {\bibfnamefont{M.}~\bibnamefont{Yamaguchi}},\ and\ \bibinfo {author}
  {\bibfnamefont{T.}~\bibnamefont{Yanagida}},\ }%
  \bibfield{journal}{%
  \bibinfo {journal} {Prog. Theor. Phys.}\ }%
  \textbf{\bibinfo {volume} {85}},\ \bibinfo {pages} {1} (\bibinfo {year}
  {1991})%
  \bibAnnoteFile{NoStop}{Okada:1990vk}%
\bibitem{Ellis:1991zd}%
  \BibitemOpen
  \bibfield{author}{%
  \bibinfo {author} {\bibfnamefont{J.~R.}\ \bibnamefont{Ellis}}, \bibinfo
  {author} {\bibfnamefont{G.}~\bibnamefont{Ridolfi}},\ and\ \bibinfo {author}
  {\bibfnamefont{F.}~\bibnamefont{Zwirner}},\ }%
  \bibfield{journal}{%
  \bibinfo {journal} {Phys. Lett. B}\ }%
  \textbf{\bibinfo {volume} {262}},\ \bibinfo {pages} {477} (\bibinfo {year}
  {1991})%
  \bibAnnoteFile{NoStop}{Ellis:1991zd}%
\bibitem{Carena:1996}%
  \BibitemOpen
  \bibfield{author}{%
  \bibinfo {author} {\bibfnamefont{M.~S.}\ \bibnamefont{Carena}}, \bibinfo
  {author} {\bibfnamefont{M.}~\bibnamefont{Quiros}},\ and\ \bibinfo {author}
  {\bibfnamefont{C.~E.~M.}\ \bibnamefont{Wagner}},\ }%
  \bibfield{journal}{%
  \bibinfo {journal} {Nucl. Phys. B}\ }%
  \textbf{\bibinfo {volume} {461}},\ \bibinfo {pages} {407} (\bibinfo {year}
  {1996}),\ \Eprint{http://arxiv.org/abs/hep-ph/9508343}{hep-ph/9508343}%
  \bibAnnoteFile{NoStop}{Carena:1996}%
\bibitem{Haber:1996fp}%
  \BibitemOpen
  \bibfield{author}{%
  \bibinfo {author} {\bibfnamefont{H.~E.}\ \bibnamefont{Haber}}, \bibinfo
  {author} {\bibfnamefont{R.}~\bibnamefont{Hempfling}},\ and\ \bibinfo {author}
  {\bibfnamefont{A.~H.}\ \bibnamefont{Hoang}},\ }%
  \bibfield{journal}{%
  \bibinfo {journal} {Z. Phys. C},\ \bibinfo {pages} {539}}%
   (\bibinfo {year} {1997})%
  \bibAnnoteFile{NoStop}{Haber:1996fp}%
\bibitem{Heinemeyer:1998np}%
  \BibitemOpen
  \bibfield{author}{%
  \bibinfo {author} {\bibfnamefont{S.}~\bibnamefont{Heinemeyer}}, \bibinfo
  {author} {\bibfnamefont{W.}~\bibnamefont{Hollik}},\ and\ \bibinfo {author}
  {\bibfnamefont{G.}~\bibnamefont{Weiglein}},\ }%
  \bibfield{journal}{%
  \bibinfo {journal} {Eur. Phys. J. C}\ }%
  \textbf{\bibinfo {volume} {9}},\ \bibinfo {pages} {343} (\bibinfo {year}
  {1999})%
  \bibAnnoteFile{NoStop}{Heinemeyer:1998np}%
\bibitem{Espinosa:2000df}%
  \BibitemOpen
  \bibfield{author}{%
  \bibinfo {author} {\bibfnamefont{J.~R.}\ \bibnamefont{Espinosa}}\ and\
  \bibinfo {author} {\bibfnamefont{R.~J.}\ \bibnamefont{Zhang}},\ }%
  \bibfield{journal}{%
  \bibinfo {journal} {Nucl. Phys. B}\ }%
  \textbf{\bibinfo {volume} {586}},\ \bibinfo {pages} {3} (\bibinfo {year}
  {2000})%
  \bibAnnoteFile{NoStop}{Espinosa:2000df}%
\bibitem{Carena:2000npb}%
  \BibitemOpen
  \bibfield{author}{%
  \bibinfo {author} {\bibfnamefont{M.}~\bibnamefont{Carena}}, \bibinfo {author}
  {\bibfnamefont{H.~E.}\ \bibnamefont{Haber}}, \bibinfo {author}
  {\bibfnamefont{S.}~\bibnamefont{Heinemeyer}}, \bibinfo {author}
  {\bibfnamefont{W.}~\bibnamefont{Hollik}}, \bibinfo {author}
  {\bibfnamefont{C.~E.~M.}\ \bibnamefont{Wagner}},\ and\ \bibinfo {author}
  {\bibfnamefont{G.}~\bibnamefont{Weiglein}},\ }%
  \bibfield{journal}{%
  \bibinfo {journal} {Nucl. Phys. B}\ }%
  \textbf{\bibinfo {volume} {580}},\ \bibinfo {pages} {29} (\bibinfo {year}
  {2000})%
  \bibAnnoteFile{NoStop}{Carena:2000npb}%
\bibitem{Degrassi2}%
  \BibitemOpen
  \bibfield{author}{%
  \bibinfo {author} {\bibfnamefont{A.}~\bibnamefont{Brignole}}, \bibinfo
  {author} {\bibfnamefont{G.}~\bibnamefont{Degrassi}}, \bibinfo {author}
  {\bibfnamefont{P.}~\bibnamefont{Slavich}},\ and\ \bibinfo {author}
  {\bibfnamefont{F.}~\bibnamefont{Zwirner}},\ }%
  \bibfield{journal}{%
  \bibinfo {journal} {Nucl. Phys. B}\ }%
  \textbf{\bibinfo {volume} {631}},\ \bibinfo {pages} {195} (\bibinfo {year}
  {2002})%
  \bibAnnoteFile{NoStop}{Degrassi2}%
\bibitem{Martin:2003}%
  \BibitemOpen
  \bibfield{author}{%
  \bibinfo {author} {\bibfnamefont{S.~P.}\ \bibnamefont{Martin}},\ }%
  \bibfield{journal}{%
  \bibinfo {journal} {Phys. Rev. D}\ }%
  \textbf{\bibinfo {volume} {67}},\ \bibinfo {pages} {095012} (\bibinfo {year}
  {2003})%
  \bibAnnoteFile{NoStop}{Martin:2003}%
\bibitem{Abazov:2008eb}%
  \BibitemOpen
  \bibfield{author}{%
  \bibinfo {author} {\bibfnamefont{V.~M.}\ \bibnamefont{Abazov}} \emph{et~al.}
  (\bibinfo {collaboration} {D0}),\ }%
  \bibfield{journal}{%
  \Doi{10.1103/PhysRevLett.102.051803}{\bibinfo {journal} {Phys. Rev. Lett.}}\
  }%
  \textbf{\bibinfo {volume} {102}},\ \bibinfo {pages} {051803} (\bibinfo {year}
  {2009}),\ \Eprint{http://arxiv.org/abs/0808.1970}{arXiv:0808.1970 [hep-ex]}%
  \bibAnnoteFile{NoStop}{Abazov:2008eb}%
\bibitem{Aaltonen:2009dh}%
  \BibitemOpen
  \bibfield{author}{%
  \bibinfo {author} {\bibfnamefont{T.}~\bibnamefont{Aaltonen}} \emph{et~al.}
  (\bibinfo {collaboration} {CDF}),\ }%
  \bibfield{journal}{%
  \Doi{10.1103/PhysRevLett.103.101802}{\bibinfo {journal} {Phys. Rev. Lett.}}\
  }%
  \textbf{\bibinfo {volume} {103}},\ \bibinfo {pages} {101802} (\bibinfo {year}
  {2009}),\ \Eprint{http://arxiv.org/abs/0906.5613}{arXiv:0906.5613 [hep-ex]}%
  \bibAnnoteFile{NoStop}{Aaltonen:2009dh}%
\bibitem{:2009je}%
  \BibitemOpen
  \bibfield{author}{%
  \bibinfo {author} {\bibnamefont{CDF}}\ and\ \bibinfo {author}
  {\bibnamefont{D0}}}%
   (\bibinfo {year} {2009}),\
  \Eprint{http://arxiv.org/abs/0911.3930}{arXiv:0911.3930 [hep-ex]}%
  \bibAnnoteFile{NoStop}{:2009je}%
\bibitem{Aad:2009wy}%
  \BibitemOpen
  \bibfield{author}{%
  \bibinfo {author} {\bibfnamefont{G.}~\bibnamefont{Aad}} \emph{et~al.}
  (\bibinfo {collaboration} {ATLAS})}%
   (\bibinfo {year} {2009}),\
  \Eprint{http://arxiv.org/abs/0901.0512}{arXiv:0901.0512 [hep-ex]}%
  \bibAnnoteFile{NoStop}{Aad:2009wy}%
\bibitem{Ball:2007zza}%
  \BibitemOpen
  \bibfield{author}{%
  \bibinfo {author} {\bibfnamefont{G.~L.}\ \bibnamefont{Bayatian}}
  \emph{et~al.} (\bibinfo {collaboration} {CMS}),\ }%
  \bibfield{journal}{%
  \Doi{10.1088/0954-3899/34/6/S01}{\bibinfo {journal} {J. Phys.}}\ }%
  \textbf{\bibinfo {volume} {G34}},\ \bibinfo {pages} {995} (\bibinfo {year}
  {2007})%
  \bibAnnoteFile{NoStop}{Ball:2007zza}%
\bibitem{Stange:1993ya}%
  \BibitemOpen
  \bibfield{author}{%
  \bibinfo {author} {\bibfnamefont{A.}~\bibnamefont{Stange}}, \bibinfo {author}
  {\bibfnamefont{W.~J.}\ \bibnamefont{Marciano}},\ and\ \bibinfo {author}
  {\bibfnamefont{S.}~\bibnamefont{Willenbrock}},\ }%
  \bibfield{journal}{%
  \Doi{10.1103/PhysRevD.49.1354}{\bibinfo {journal} {Phys. Rev.}}\ }%
  \textbf{\bibinfo {volume} {D49}},\ \bibinfo {pages} {1354} (\bibinfo {year}
  {1994}),\ \Eprint{http://arxiv.org/abs/hep-ph/9309294}{arXiv:hep-ph/9309294}%
  \bibAnnoteFile{NoStop}{Stange:1993ya}%
\bibitem{Spira:1998wh}%
  \BibitemOpen
  \bibfield{author}{%
  \bibinfo {author} {\bibfnamefont{M.}~\bibnamefont{Spira}}}%
   (\bibinfo {year} {1998}),\
  \Eprint{http://arxiv.org/abs/hep-ph/9810289}{arXiv:hep-ph/9810289}%
  \bibAnnoteFile{NoStop}{Spira:1998wh}%
\bibitem{Hahn:2006my}%
  \BibitemOpen
  \bibfield{author}{%
  \bibinfo {author} {\bibfnamefont{T.}~\bibnamefont{Hahn}}, \bibinfo {author}
  {\bibfnamefont{S.}~\bibnamefont{Heinemeyer}}, \bibinfo {author}
  {\bibfnamefont{F.}~\bibnamefont{Maltoni}}, \bibinfo {author}
  {\bibfnamefont{G.}~\bibnamefont{Weiglein}},\ and\ \bibinfo {author}
  {\bibfnamefont{S.}~\bibnamefont{Willenbrock}}}%
   (\bibinfo {year} {2006}),\
  \Eprint{http://arxiv.org/abs/hep-ph/0607308}{arXiv:hep-ph/0607308}%
  \bibAnnoteFile{NoStop}{Hahn:2006my}%
\bibitem{Berger:2003pd}%
  \BibitemOpen
  \bibfield{author}{%
  \bibinfo {author} {\bibfnamefont{E.~L.}\ \bibnamefont{Berger}}\ and\ \bibinfo
  {author} {\bibfnamefont{J.-w.}\ \bibnamefont{Qiu}},\ }%
  \bibfield{journal}{%
  \Doi{10.1103/PhysRevLett.91.222003}{\bibinfo {journal} {Phys. Rev. Lett.}}\
  }%
  \textbf{\bibinfo {volume} {91}},\ \bibinfo {pages} {222003} (\bibinfo {year}
  {2003}),\ \Eprint{http://arxiv.org/abs/hep-ph/0304267}{arXiv:hep-ph/0304267}%
  \bibAnnoteFile{NoStop}{Berger:2003pd}%
\bibitem{Dittmaier:2003ej}%
  \BibitemOpen
  \bibfield{author}{%
  \bibinfo {author} {\bibfnamefont{S.}~\bibnamefont{Dittmaier}}, \bibinfo
  {author} {\bibfnamefont{M.}~\bibnamefont{Kramer}},\ and\ \bibinfo {author}
  {\bibfnamefont{M.}~\bibnamefont{Spira}},\ }%
  \bibfield{journal}{%
  \Doi{10.1103/PhysRevD.70.074010}{\bibinfo {journal} {Phys. Rev.}}\ }%
  \textbf{\bibinfo {volume} {D70}},\ \bibinfo {pages} {074010} (\bibinfo {year}
  {2004}),\ \Eprint{http://arxiv.org/abs/hep-ph/0309204}{arXiv:hep-ph/0309204}%
  \bibAnnoteFile{NoStop}{Dittmaier:2003ej}%
\bibitem{Dawson:2005vi}%
  \BibitemOpen
  \bibfield{author}{%
  \bibinfo {author} {\bibfnamefont{S.}~\bibnamefont{Dawson}}, \bibinfo {author}
  {\bibfnamefont{C.~B.}\ \bibnamefont{Jackson}}, \bibinfo {author}
  {\bibfnamefont{L.}~\bibnamefont{Reina}},\ and\ \bibinfo {author}
  {\bibfnamefont{D.}~\bibnamefont{Wackeroth}},\ }%
  \bibfield{journal}{%
  \Doi{10.1142/S0217732306019256}{\bibinfo {journal} {Mod. Phys. Lett.}}\ }%
  \textbf{\bibinfo {volume} {A21}},\ \bibinfo {pages} {89} (\bibinfo {year}
  {2006}),\ \Eprint{http://arxiv.org/abs/hep-ph/0508293}{arXiv:hep-ph/0508293}%
  \bibAnnoteFile{NoStop}{Dawson:2005vi}%
\bibitem{Balazs:1998nt}%
  \BibitemOpen
  \bibfield{author}{%
  \bibinfo {author} {\bibfnamefont{C.}~\bibnamefont{Balazs}}, \bibinfo {author}
  {\bibfnamefont{J.~L.}\ \bibnamefont{Diaz-Cruz}}, \bibinfo {author}
  {\bibfnamefont{H.~J.}\ \bibnamefont{He}}, \bibinfo {author}
  {\bibfnamefont{T.~M.~P.}\ \bibnamefont{Tait}},\ and\ \bibinfo {author}
  {\bibfnamefont{C.~P.}\ \bibnamefont{Yuan}},\ }%
  \bibfield{journal}{%
  \Doi{10.1103/PhysRevD.59.055016}{\bibinfo {journal} {Phys. Rev.}}\ }%
  \textbf{\bibinfo {volume} {D59}},\ \bibinfo {pages} {055016} (\bibinfo {year}
  {1999}),\ \Eprint{http://arxiv.org/abs/hep-ph/9807349}{arXiv:hep-ph/9807349}%
  \bibAnnoteFile{NoStop}{Balazs:1998nt}%
\bibitem{Carena:1998gk}%
  \BibitemOpen
  \bibfield{author}{%
  \bibinfo {author} {\bibfnamefont{M.~S.}\ \bibnamefont{Carena}}, \bibinfo
  {author} {\bibfnamefont{S.}~\bibnamefont{Mrenna}},\ and\ \bibinfo {author}
  {\bibfnamefont{C.~E.~M.}\ \bibnamefont{Wagner}},\ }%
  \bibfield{journal}{%
  \Doi{10.1103/PhysRevD.60.075010}{\bibinfo {journal} {Phys. Rev.}}\ }%
  \textbf{\bibinfo {volume} {D60}},\ \bibinfo {pages} {075010} (\bibinfo {year}
  {1999}),\ \Eprint{http://arxiv.org/abs/hep-ph/9808312}{arXiv:hep-ph/9808312}%
  \bibAnnoteFile{NoStop}{Carena:1998gk}%
\bibitem{Carena:1999bh}%
  \BibitemOpen
  \bibfield{author}{%
  \bibinfo {author} {\bibfnamefont{M.~S.}\ \bibnamefont{Carena}}, \bibinfo
  {author} {\bibfnamefont{S.}~\bibnamefont{Mrenna}},\ and\ \bibinfo {author}
  {\bibfnamefont{C.~E.~M.}\ \bibnamefont{Wagner}},\ }%
  \bibfield{journal}{%
  \Doi{10.1103/PhysRevD.62.055008}{\bibinfo {journal} {Phys. Rev.}}\ }%
  \textbf{\bibinfo {volume} {D62}},\ \bibinfo {pages} {055008} (\bibinfo {year}
  {2000}),\ \Eprint{http://arxiv.org/abs/hep-ph/9907422}{arXiv:hep-ph/9907422}%
  \bibAnnoteFile{NoStop}{Carena:1999bh}%
\bibitem{Plehn:1999nw}%
  \BibitemOpen
  \bibfield{author}{%
  \bibinfo {author} {\bibfnamefont{T.}~\bibnamefont{Plehn}}, \bibinfo {author}
  {\bibfnamefont{D.~L.}\ \bibnamefont{Rainwater}},\ and\ \bibinfo {author}
  {\bibfnamefont{D.}~\bibnamefont{Zeppenfeld}},\ }%
  \bibfield{journal}{%
  \Doi{10.1016/S0370-2693(99)00408-6}{\bibinfo {journal} {Phys. Lett.}}\ }%
  \textbf{\bibinfo {volume} {B454}},\ \bibinfo {pages} {297} (\bibinfo {year}
  {1999}),\ \Eprint{http://arxiv.org/abs/hep-ph/9902434}{arXiv:hep-ph/9902434}%
  \bibAnnoteFile{NoStop}{Plehn:1999nw}%
\bibitem{Draper:2009fh}%
  \BibitemOpen
  \bibfield{author}{%
  \bibinfo {author} {\bibfnamefont{P.}~\bibnamefont{Draper}}, \bibinfo {author}
  {\bibfnamefont{T.}~\bibnamefont{Liu}},\ and\ \bibinfo {author}
  {\bibfnamefont{C.~E.~M.}\ \bibnamefont{Wagner}},\ }%
  \bibfield{journal}{%
  \Doi{10.1103/PhysRevD.80.035025}{\bibinfo {journal} {Phys. Rev.}}\ }%
  \textbf{\bibinfo {volume} {D80}},\ \bibinfo {pages} {035025} (\bibinfo {year}
  {2009}),\ \Eprint{http://arxiv.org/abs/0905.4721}{arXiv:0905.4721 [hep-ph]}%
  \bibAnnoteFile{NoStop}{Draper:2009fh}%
\bibitem{Draper:2009au}%
  \BibitemOpen
  \bibfield{author}{%
  \bibinfo {author} {\bibfnamefont{P.}~\bibnamefont{Draper}}, \bibinfo {author}
  {\bibfnamefont{T.}~\bibnamefont{Liu}},\ and\ \bibinfo {author}
  {\bibfnamefont{C.~E.~M.}\ \bibnamefont{Wagner}},\ }%
  \bibfield{journal}{%
  \Doi{10.1103/PhysRevD.81.015014}{\bibinfo {journal} {Phys. Rev.}}\ }%
  \textbf{\bibinfo {volume} {D81}},\ \bibinfo {pages} {015014} (\bibinfo {year}
  {2010}),\ \Eprint{http://arxiv.org/abs/0911.0034}{arXiv:0911.0034 [hep-ph]}%
  \bibAnnoteFile{NoStop}{Draper:2009au}%
\bibitem{Barbieri:1989dc}%
  \BibitemOpen
  \bibfield{author}{%
  \bibinfo {author} {\bibfnamefont{R.}~\bibnamefont{Barbieri}}, \bibinfo
  {author} {\bibfnamefont{M.}~\bibnamefont{Frigeni}}, \bibinfo {author}
  {\bibfnamefont{F.}~\bibnamefont{Giuliani}},\ and\ \bibinfo {author}
  {\bibfnamefont{H.~E.}\ \bibnamefont{Haber}},\ }%
  \bibfield{journal}{%
  \Doi{10.1016/0550-3213(90)90181-C}{\bibinfo {journal} {Nucl. Phys.}}\ }%
  \textbf{\bibinfo {volume} {B341}},\ \bibinfo {pages} {309} (\bibinfo {year}
  {1990})%
  \bibAnnoteFile{NoStop}{Barbieri:1989dc}%
\bibitem{Chankowski:1993eu}%
  \BibitemOpen
  \bibfield{author}{%
  \bibinfo {author} {\bibfnamefont{P.~H.}\ \bibnamefont{Chankowski}}
  \emph{et~al.},\ }%
  \bibfield{journal}{%
  \Doi{10.1016/0550-3213(94)90539-8}{\bibinfo {journal} {Nucl. Phys.}}\ }%
  \textbf{\bibinfo {volume} {B417}},\ \bibinfo {pages} {101} (\bibinfo {year}
  {1994})%
  \bibAnnoteFile{NoStop}{Chankowski:1993eu}%
\bibitem{Erler:1998ur}%
  \BibitemOpen
  \bibfield{author}{%
  \bibinfo {author} {\bibfnamefont{J.}~\bibnamefont{Erler}}\ and\ \bibinfo
  {author} {\bibfnamefont{D.~M.}\ \bibnamefont{Pierce}},\ }%
  \bibfield{journal}{%
  \Doi{10.1016/S0550-3213(98)00359-9}{\bibinfo {journal} {Nucl. Phys.}}\ }%
  \textbf{\bibinfo {volume} {B526}},\ \bibinfo {pages} {53} (\bibinfo {year}
  {1998}),\ \Eprint{http://arxiv.org/abs/hep-ph/9801238}{arXiv:hep-ph/9801238}%
  \bibAnnoteFile{NoStop}{Erler:1998ur}%
\bibitem{Baer:2003wx}%
  \BibitemOpen
  \bibfield{author}{%
  \bibinfo {author} {\bibfnamefont{H.}~\bibnamefont{Baer}}, \bibinfo {author}
  {\bibfnamefont{C.}~\bibnamefont{Balazs}}, \bibinfo {author}
  {\bibfnamefont{A.}~\bibnamefont{Belyaev}}, \bibinfo {author}
  {\bibfnamefont{T.}~\bibnamefont{Krupovnickas}},\ and\ \bibinfo {author}
  {\bibfnamefont{X.}~\bibnamefont{Tata}},\ }%
  \bibfield{journal}{%
  \bibinfo {journal} {JHEP}\ }%
  \textbf{\bibinfo {volume} {06}},\ \bibinfo {pages} {054} (\bibinfo {year}
  {2003}),\ \Eprint{http://arxiv.org/abs/hep-ph/0304303}{arXiv:hep-ph/0304303}%
  \bibAnnoteFile{NoStop}{Baer:2003wx}%
\bibitem{Chattopadhyay:2003xi}%
  \BibitemOpen
  \bibfield{author}{%
  \bibinfo {author} {\bibfnamefont{U.}~\bibnamefont{Chattopadhyay}}, \bibinfo
  {author} {\bibfnamefont{A.}~\bibnamefont{Corsetti}},\ and\ \bibinfo {author}
  {\bibfnamefont{P.}~\bibnamefont{Nath}},\ }%
  \bibfield{journal}{%
  \Doi{10.1103/PhysRevD.68.035005}{\bibinfo {journal} {Phys. Rev.}}\ }%
  \textbf{\bibinfo {volume} {D68}},\ \bibinfo {pages} {035005} (\bibinfo {year}
  {2003}),\ \Eprint{http://arxiv.org/abs/hep-ph/0303201}{arXiv:hep-ph/0303201}%
  \bibAnnoteFile{NoStop}{Chattopadhyay:2003xi}%
\bibitem{Ellis:2004bx}%
  \BibitemOpen
  \bibfield{author}{%
  \bibinfo {author} {\bibfnamefont{J.~R.}\ \bibnamefont{Ellis}}, \bibinfo
  {author} {\bibfnamefont{K.~A.}\ \bibnamefont{Olive}}, \bibinfo {author}
  {\bibfnamefont{Y.}~\bibnamefont{Santoso}},\ and\ \bibinfo {author}
  {\bibfnamefont{V.~C.}\ \bibnamefont{Spanos}},\ }%
  \bibfield{journal}{%
  \Doi{10.1016/j.physletb.2004.09.073}{\bibinfo {journal} {Phys. Lett.}}\ }%
  \textbf{\bibinfo {volume} {B603}},\ \bibinfo {pages} {51} (\bibinfo {year}
  {2004}),\ \Eprint{http://arxiv.org/abs/hep-ph/0408118}{arXiv:hep-ph/0408118}%
  \bibAnnoteFile{NoStop}{Ellis:2004bx}%
\bibitem{Battaglia:2003ab}%
  \BibitemOpen
  \bibfield{author}{%
  \bibinfo {author} {\bibfnamefont{M.}~\bibnamefont{Battaglia}} \emph{et~al.},\
  }%
  \bibfield{journal}{%
  \Doi{10.1140/epjc/s2003-01575-7}{\bibinfo {journal} {Eur. Phys. J.}}\ }%
  \textbf{\bibinfo {volume} {C33}},\ \bibinfo {pages} {273} (\bibinfo {year}
  {2004}),\ \Eprint{http://arxiv.org/abs/hep-ph/0306219}{arXiv:hep-ph/0306219}%
  \bibAnnoteFile{NoStop}{Battaglia:2003ab}%
\bibitem{Djouadi:2006be}%
  \BibitemOpen
  \bibfield{author}{%
  \bibinfo {author} {\bibfnamefont{A.}~\bibnamefont{Djouadi}}, \bibinfo
  {author} {\bibfnamefont{M.}~\bibnamefont{Drees}},\ and\ \bibinfo {author}
  {\bibfnamefont{J.-L.}\ \bibnamefont{Kneur}},\ }%
  \bibfield{journal}{%
  \bibinfo {journal} {JHEP}\ }%
  \textbf{\bibinfo {volume} {03}},\ \bibinfo {pages} {033} (\bibinfo {year}
  {2006}),\ \Eprint{http://arxiv.org/abs/hep-ph/0602001}{arXiv:hep-ph/0602001}%
  \bibAnnoteFile{NoStop}{Djouadi:2006be}%
\bibitem{Baltz:2006fm}%
  \BibitemOpen
  \bibfield{author}{%
  \bibinfo {author} {\bibfnamefont{E.~A.}\ \bibnamefont{Baltz}}, \bibinfo
  {author} {\bibfnamefont{M.}~\bibnamefont{Battaglia}}, \bibinfo {author}
  {\bibfnamefont{M.~E.}\ \bibnamefont{Peskin}},\ and\ \bibinfo {author}
  {\bibfnamefont{T.}~\bibnamefont{Wizansky}},\ }%
  \bibfield{journal}{%
  \Doi{10.1103/PhysRevD.74.103521}{\bibinfo {journal} {Phys. Rev.}}\ }%
  \textbf{\bibinfo {volume} {D74}},\ \bibinfo {pages} {103521} (\bibinfo {year}
  {2006}),\ \Eprint{http://arxiv.org/abs/hep-ph/0602187}{arXiv:hep-ph/0602187}%
  \bibAnnoteFile{NoStop}{Baltz:2006fm}%
\bibitem{Allanach:2007qk}%
  \BibitemOpen
  \bibfield{author}{%
  \bibinfo {author} {\bibfnamefont{B.~C.}\ \bibnamefont{Allanach}}, \bibinfo
  {author} {\bibfnamefont{K.}~\bibnamefont{Cranmer}}, \bibinfo {author}
  {\bibfnamefont{C.~G.}\ \bibnamefont{Lester}},\ and\ \bibinfo {author}
  {\bibfnamefont{A.~M.}\ \bibnamefont{Weber}},\ }%
  \bibfield{journal}{%
  \Doi{10.1088/1126-6708/2007/08/023}{\bibinfo {journal} {JHEP}}\ }%
  \textbf{\bibinfo {volume} {08}},\ \bibinfo {pages} {023} (\bibinfo {year}
  {2007}),\ \Eprint{http://arxiv.org/abs/0705.0487}{arXiv:0705.0487 [hep-ph]}%
  \bibAnnoteFile{NoStop}{Allanach:2007qk}%
\bibitem{Berger:2008cq}%
  \BibitemOpen
  \bibfield{author}{%
  \bibinfo {author} {\bibfnamefont{C.~F.}\ \bibnamefont{Berger}}, \bibinfo
  {author} {\bibfnamefont{J.~S.}\ \bibnamefont{Gainer}}, \bibinfo {author}
  {\bibfnamefont{J.~L.}\ \bibnamefont{Hewett}},\ and\ \bibinfo {author}
  {\bibfnamefont{T.~G.}\ \bibnamefont{Rizzo}},\ }%
  \bibfield{journal}{%
  \Doi{10.1088/1126-6708/2009/02/023}{\bibinfo {journal} {JHEP}}\ }%
  \textbf{\bibinfo {volume} {02}},\ \bibinfo {pages} {023} (\bibinfo {year}
  {2009}),\ \Eprint{http://arxiv.org/abs/0812.0980}{arXiv:0812.0980 [hep-ph]}%
  \bibAnnoteFile{NoStop}{Berger:2008cq}%
\bibitem{Lafaye:2004cn}%
  \BibitemOpen
  \bibfield{author}{%
  \bibinfo {author} {\bibfnamefont{R.}~\bibnamefont{Lafaye}}, \bibinfo {author}
  {\bibfnamefont{T.}~\bibnamefont{Plehn}},\ and\ \bibinfo {author}
  {\bibfnamefont{D.}~\bibnamefont{Zerwas}}}%
   (\bibinfo {year} {2004}),\
  \Eprint{http://arxiv.org/abs/hep-ph/0404282}{arXiv:hep-ph/0404282}%
  \bibAnnoteFile{NoStop}{Lafaye:2004cn}%
\bibitem{deltamb1}%
  \BibitemOpen
  \bibfield{author}{%
  \bibinfo {author} {\bibfnamefont{R.}~\bibnamefont{Hempfling}},\ }%
  \bibfield{journal}{%
  \bibinfo {journal} {Phys. Rev.}\ }%
  \textbf{\bibinfo {volume} {D49}},\ \bibinfo {pages} {6168} (\bibinfo {year}
  {1994})%
  \bibAnnoteFile{NoStop}{deltamb1}%
\bibitem{deltamb2}%
  \BibitemOpen
  \bibfield{author}{%
  \bibinfo {author} {\bibfnamefont{L.~J.}\ \bibnamefont{Hall}}, \bibinfo
  {author} {\bibfnamefont{R.}~\bibnamefont{Rattazzi}},\ and\ \bibinfo {author}
  {\bibfnamefont{U.}~\bibnamefont{Sarid}},\ }%
  \bibfield{journal}{%
  \bibinfo {journal} {Phys. Rev.}\ }%
  \textbf{\bibinfo {volume} {D 50}},\ \bibinfo {pages} {7048} (\bibinfo {year}
  {1994}),\ \Eprint{http://arxiv.org/abs/hep-ph/9306309}{hep-ph/9306309}%
  \bibAnnoteFile{NoStop}{deltamb2}%
\bibitem{deltamb3}%
  \BibitemOpen
  \bibfield{author}{%
  \bibinfo {author} {\bibfnamefont{M.}~\bibnamefont{Carena}}, \bibinfo {author}
  {\bibfnamefont{M.}~\bibnamefont{Olechowski}}, \bibinfo {author}
  {\bibfnamefont{S.}~\bibnamefont{Pokorski}},\ and\ \bibinfo {author}
  {\bibfnamefont{C.}~\bibnamefont{Wagner}},\ }%
  \bibfield{journal}{%
  \bibinfo {journal} {Nucl. Phys.}\ }%
  \textbf{\bibinfo {volume} {B 426}} (\bibinfo {year} {1994}),\
  \Eprint{http://arxiv.org/abs/hep-ph/9402253}{hep-ph/9402253}%
  \bibAnnoteFile{NoStop}{deltamb3}%
\bibitem{AguilarSaavedra:2005pw}%
  \BibitemOpen
  \bibfield{author}{%
  \bibinfo {author} {\bibfnamefont{J.~A.}\ \bibnamefont{Aguilar-Saavedra}}
  \emph{et~al.},\ }%
  \bibfield{journal}{%
  \Doi{10.1140/epjc/s2005-02460-1}{\bibinfo {journal} {Eur. Phys. J.}}\ }%
  \textbf{\bibinfo {volume} {C46}},\ \bibinfo {pages} {43} (\bibinfo {year}
  {2006}),\ \Eprint{http://arxiv.org/abs/hep-ph/0511344}{arXiv:hep-ph/0511344}%
  \bibAnnoteFile{NoStop}{AguilarSaavedra:2005pw}%
\bibitem{Kneur:2008ur}%
  \BibitemOpen
  \bibfield{author}{%
  \bibinfo {author} {\bibfnamefont{J.~L.}\ \bibnamefont{Kneur}}\ and\ \bibinfo
  {author} {\bibfnamefont{N.}~\bibnamefont{Sahoury}},\ }%
  \bibfield{journal}{%
  \Doi{10.1103/PhysRevD.79.075010}{\bibinfo {journal} {Phys. Rev.}}\ }%
  \textbf{\bibinfo {volume} {D79}},\ \bibinfo {pages} {075010} (\bibinfo {year}
  {2009}),\ \Eprint{http://arxiv.org/abs/0808.0144}{arXiv:0808.0144 [hep-ph]}%
  \bibAnnoteFile{NoStop}{Kneur:2008ur}%
\bibitem{Kneur:1998gy}%
  \BibitemOpen
  \bibfield{author}{%
  \bibinfo {author} {\bibfnamefont{J.~L.}\ \bibnamefont{Kneur}}\ and\ \bibinfo
  {author} {\bibfnamefont{G.}~\bibnamefont{Moultaka}},\ }%
  \bibfield{journal}{%
  \Doi{10.1103/PhysRevD.59.015005}{\bibinfo {journal} {Phys. Rev.}}\ }%
  \textbf{\bibinfo {volume} {D59}},\ \bibinfo {pages} {015005} (\bibinfo {year}
  {1998}),\ \Eprint{http://arxiv.org/abs/hep-ph/9807336}{arXiv:hep-ph/9807336}%
  \bibAnnoteFile{NoStop}{Kneur:1998gy}%
\bibitem{Blair:2002pg}%
  \BibitemOpen
  \bibfield{author}{%
  \bibinfo {author} {\bibfnamefont{G.~A.}\ \bibnamefont{Blair}}, \bibinfo
  {author} {\bibfnamefont{W.}~\bibnamefont{Porod}},\ and\ \bibinfo {author}
  {\bibfnamefont{P.~M.}\ \bibnamefont{Zerwas}},\ }%
  \bibfield{journal}{%
  \Doi{10.1140/epjc/s2002-01117-y}{\bibinfo {journal} {Eur. Phys. J.}}\ }%
  \textbf{\bibinfo {volume} {C27}},\ \bibinfo {pages} {263} (\bibinfo {year}
  {2003}),\ \Eprint{http://arxiv.org/abs/hep-ph/0210058}{arXiv:hep-ph/0210058}%
  \bibAnnoteFile{NoStop}{Blair:2002pg}%
\bibitem{Blair:2005ui}%
  \BibitemOpen
  \bibfield{author}{%
  \bibinfo {author} {\bibfnamefont{G.~A.}\ \bibnamefont{Blair}} \emph{et~al.},\
  }%
  \bibfield{journal}{%
  \bibinfo {journal} {Acta Phys. Polon.}\ }%
  \textbf{\bibinfo {volume} {B36}},\ \bibinfo {pages} {3445} (\bibinfo {year}
  {2005}),\ \Eprint{http://arxiv.org/abs/hep-ph/0512084}{arXiv:hep-ph/0512084}%
  \bibAnnoteFile{NoStop}{Blair:2005ui}%
\bibitem{Carena:1996km}%
  \BibitemOpen
  \bibfield{author}{%
  \bibinfo {author} {\bibfnamefont{M.~S.}\ \bibnamefont{Carena}}, \bibinfo
  {author} {\bibfnamefont{P.~H.}\ \bibnamefont{Chankowski}}, \bibinfo {author}
  {\bibfnamefont{M.}~\bibnamefont{Olechowski}}, \bibinfo {author}
  {\bibfnamefont{S.}~\bibnamefont{Pokorski}},\ and\ \bibinfo {author}
  {\bibfnamefont{C.~E.~M.}\ \bibnamefont{Wagner}},\ }%
  \bibfield{journal}{%
  \Doi{10.1016/S0550-3213(97)00109-0}{\bibinfo {journal} {Nucl. Phys.}}\ }%
  \textbf{\bibinfo {volume} {B491}},\ \bibinfo {pages} {103} (\bibinfo {year}
  {1997}),\ \Eprint{http://arxiv.org/abs/hep-ph/9612261}{arXiv:hep-ph/9612261}%
  \bibAnnoteFile{NoStop}{Carena:1996km}%
\bibitem{Kane:2006hd}%
  \BibitemOpen
  \bibfield{author}{%
  \bibinfo {author} {\bibfnamefont{G.~L.}\ \bibnamefont{Kane}}, \bibinfo
  {author} {\bibfnamefont{P.}~\bibnamefont{Kumar}}, \bibinfo {author}
  {\bibfnamefont{D.~E.}\ \bibnamefont{Morrissey}},\ and\ \bibinfo {author}
  {\bibfnamefont{M.}~\bibnamefont{Toharia}},\ }%
  \bibfield{journal}{%
  \Doi{10.1103/PhysRevD.75.115018}{\bibinfo {journal} {Phys. Rev.}}\ }%
  \textbf{\bibinfo {volume} {D75}},\ \bibinfo {pages} {115018} (\bibinfo {year}
  {2007}),\ \Eprint{http://arxiv.org/abs/hep-ph/0612287}{arXiv:hep-ph/0612287}%
  \bibAnnoteFile{NoStop}{Kane:2006hd}%
\bibitem{Cohen:2006qc}%
  \BibitemOpen
  \bibfield{author}{%
  \bibinfo {author} {\bibfnamefont{A.~G.}\ \bibnamefont{Cohen}}, \bibinfo
  {author} {\bibfnamefont{T.~S.}\ \bibnamefont{Roy}},\ and\ \bibinfo {author}
  {\bibfnamefont{M.}~\bibnamefont{Schmaltz}},\ }%
  \bibfield{journal}{%
  \bibinfo {journal} {JHEP}\ }%
  \textbf{\bibinfo {volume} {02}},\ \bibinfo {pages} {027} (\bibinfo {year}
  {2007}),\ \Eprint{http://arxiv.org/abs/hep-ph/0612100}{arXiv:hep-ph/0612100}%
  \bibAnnoteFile{NoStop}{Cohen:2006qc}%
\bibitem{Meade:2008wd}%
  \BibitemOpen
  \bibfield{author}{%
  \bibinfo {author} {\bibfnamefont{P.}~\bibnamefont{Meade}}, \bibinfo {author}
  {\bibfnamefont{N.}~\bibnamefont{Seiberg}},\ and\ \bibinfo {author}
  {\bibfnamefont{D.}~\bibnamefont{Shih}},\ }%
  \bibfield{journal}{%
  \Doi{10.1143/PTPS.177.143}{\bibinfo {journal} {Prog. Theor. Phys. Suppl.}}\
  }%
  \textbf{\bibinfo {volume} {177}},\ \bibinfo {pages} {143} (\bibinfo {year}
  {2009}),\ \Eprint{http://arxiv.org/abs/0801.3278}{arXiv:0801.3278 [hep-ph]}%
  \bibAnnoteFile{NoStop}{Meade:2008wd}%
\bibitem{Demir:2004aq}%
  \BibitemOpen
  \bibfield{author}{%
  \bibinfo {author} {\bibfnamefont{D.~A.}\ \bibnamefont{Demir}},\ }%
  \bibfield{journal}{%
  \bibinfo {journal} {JHEP}\ }%
  \textbf{\bibinfo {volume} {11}},\ \bibinfo {pages} {003} (\bibinfo {year}
  {2005}),\ \Eprint{http://arxiv.org/abs/hep-ph/0408043}{arXiv:hep-ph/0408043}%
  \bibAnnoteFile{NoStop}{Demir:2004aq}%
\bibitem{Martin:1993ft}%
  \BibitemOpen
  \bibfield{author}{%
  \bibinfo {author} {\bibfnamefont{S.~P.}\ \bibnamefont{Martin}}\ and\ \bibinfo
  {author} {\bibfnamefont{P.}~\bibnamefont{Ramond}},\ }%
  \bibfield{journal}{%
  \Doi{10.1103/PhysRevD.48.5365}{\bibinfo {journal} {Phys. Rev.}}\ }%
  \textbf{\bibinfo {volume} {D48}},\ \bibinfo {pages} {5365} (\bibinfo {year}
  {1993}),\ \Eprint{http://arxiv.org/abs/hep-ph/9306314}{arXiv:hep-ph/9306314}%
  \bibAnnoteFile{NoStop}{Martin:1993ft}%
\bibitem{Ananthanarayan:2007fj}%
  \BibitemOpen
  \bibfield{author}{%
  \bibinfo {author} {\bibfnamefont{B.}~\bibnamefont{Ananthanarayan}}\ and\
  \bibinfo {author} {\bibfnamefont{P.~N.}\ \bibnamefont{Pandita}},\ }%
  \bibfield{journal}{%
  \Doi{10.1142/S0217751X07036889}{\bibinfo {journal} {Int. J. Mod. Phys.}}\ }%
  \textbf{\bibinfo {volume} {A22}},\ \bibinfo {pages} {3229} (\bibinfo {year}
  {2007}),\ \Eprint{http://arxiv.org/abs/0706.2560}{arXiv:0706.2560 [hep-ph]}%
  \bibAnnoteFile{NoStop}{Ananthanarayan:2007fj}%
\bibitem{Ananthanarayan:2003ca}%
  \BibitemOpen
  \bibfield{author}{%
  \bibinfo {author} {\bibfnamefont{B.}~\bibnamefont{Ananthanarayan}}\ and\
  \bibinfo {author} {\bibfnamefont{P.~N.}\ \bibnamefont{Pandita}},\ }%
  \bibfield{journal}{%
  \Doi{10.1142/S0217732399002522}{\bibinfo {journal} {Mod. Phys. Lett.}}\ }%
  \textbf{\bibinfo {volume} {A19}},\ \bibinfo {pages} {467} (\bibinfo {year}
  {2004}),\ \Eprint{http://arxiv.org/abs/hep-ph/0312361}{arXiv:hep-ph/0312361}%
  \bibAnnoteFile{NoStop}{Ananthanarayan:2003ca}%
\bibitem{Ananthanarayan:2004dm}%
  \BibitemOpen
  \bibfield{author}{%
  \bibinfo {author} {\bibfnamefont{B.}~\bibnamefont{Ananthanarayan}}\ and\
  \bibinfo {author} {\bibfnamefont{P.~N.}\ \bibnamefont{Pandita}},\ }%
  \bibfield{journal}{%
  \Doi{10.1142/S0217751X05022445}{\bibinfo {journal} {Int. J. Mod. Phys.}}\ }%
  \textbf{\bibinfo {volume} {A20}},\ \bibinfo {pages} {4241} (\bibinfo {year}
  {2005}),\ \Eprint{http://arxiv.org/abs/hep-ph/0412125}{arXiv:hep-ph/0412125}%
  \bibAnnoteFile{NoStop}{Ananthanarayan:2004dm}%
\bibitem{Balazs:2010ha}%
  \BibitemOpen
  \bibfield{author}{%
  \bibinfo {author} {\bibfnamefont{C.}~\bibnamefont{Balazs}}, \bibinfo {author}
  {\bibfnamefont{T.}~\bibnamefont{Li}}, \bibinfo {author}
  {\bibfnamefont{D.~V.}\ \bibnamefont{Nanopoulos}},\ and\ \bibinfo {author}
  {\bibfnamefont{F.}~\bibnamefont{Wang}}}%
   (\bibinfo {year} {2010}),\
  \Eprint{http://arxiv.org/abs/1006.5559}{arXiv:1006.5559 [hep-ph]}%
  \bibAnnoteFile{NoStop}{Balazs:2010ha}%
\bibitem{Everett:2009cn}%
  \BibitemOpen
  \bibfield{author}{%
  \bibinfo {author} {\bibfnamefont{L.~L.}\ \bibnamefont{Everett}}, \bibinfo
  {author} {\bibfnamefont{J.}~\bibnamefont{Jiang}}, \bibinfo {author}
  {\bibfnamefont{P.~G.}\ \bibnamefont{Langacker}},\ and\ \bibinfo {author}
  {\bibfnamefont{T.}~\bibnamefont{Liu}}}%
   (\bibinfo {year} {2009}),\
  \Eprint{http://arxiv.org/abs/0911.5349}{arXiv:0911.5349 [hep-ph]}%
  \bibAnnoteFile{NoStop}{Everett:2009cn}%
\bibitem{Ellis:1981ts}%
  \BibitemOpen
  \bibfield{author}{%
  \bibinfo {author} {\bibfnamefont{J.~R.}\ \bibnamefont{Ellis}}\ and\ \bibinfo
  {author} {\bibfnamefont{D.~V.}\ \bibnamefont{Nanopoulos}},\ }%
  \bibfield{journal}{%
  \Doi{10.1016/0370-2693(82)90948-0}{\bibinfo {journal} {Phys. Lett.}}\ }%
  \textbf{\bibinfo {volume} {B110}},\ \bibinfo {pages} {44} (\bibinfo {year}
  {1982})%
  \bibAnnoteFile{NoStop}{Ellis:1981ts}%
\bibitem{Bertolini:1990if}%
  \BibitemOpen
  \bibfield{author}{%
  \bibinfo {author} {\bibfnamefont{S.}~\bibnamefont{Bertolini}}, \bibinfo
  {author} {\bibfnamefont{F.}~\bibnamefont{Borzumati}}, \bibinfo {author}
  {\bibfnamefont{A.}~\bibnamefont{Masiero}},\ and\ \bibinfo {author}
  {\bibfnamefont{G.}~\bibnamefont{Ridolfi}},\ }%
  \bibfield{journal}{%
  \Doi{10.1016/0550-3213(91)90320-W}{\bibinfo {journal} {Nucl. Phys.}}\ }%
  \textbf{\bibinfo {volume} {B353}},\ \bibinfo {pages} {591} (\bibinfo {year}
  {1991})%
  \bibAnnoteFile{NoStop}{Bertolini:1990if}%
\bibitem{Isidori:2001fv}%
  \BibitemOpen
  \bibfield{author}{%
  \bibinfo {author} {\bibfnamefont{G.}~\bibnamefont{Isidori}}\ and\ \bibinfo
  {author} {\bibfnamefont{A.}~\bibnamefont{Retico}},\ }%
  \bibfield{journal}{%
  \bibinfo {journal} {JHEP}\ }%
  \textbf{\bibinfo {volume} {11}},\ \bibinfo {pages} {001} (\bibinfo {year}
  {2001}),\ \Eprint{http://arxiv.org/abs/hep-ph/0110121}{arXiv:hep-ph/0110121}%
  \bibAnnoteFile{NoStop}{Isidori:2001fv}%
\bibitem{Buras:2002vd}%
  \BibitemOpen
  \bibfield{author}{%
  \bibinfo {author} {\bibfnamefont{A.~J.}\ \bibnamefont{Buras}}, \bibinfo
  {author} {\bibfnamefont{P.~H.}\ \bibnamefont{Chankowski}}, \bibinfo {author}
  {\bibfnamefont{J.}~\bibnamefont{Rosiek}},\ and\ \bibinfo {author}
  {\bibfnamefont{L.}~\bibnamefont{Slawianowska}},\ }%
  \bibfield{journal}{%
  \Doi{10.1016/S0550-3213(03)00190-1}{\bibinfo {journal} {Nucl. Phys.}}\ }%
  \textbf{\bibinfo {volume} {B659}},\ \bibinfo {pages} {3} (\bibinfo {year}
  {2003}),\ \Eprint{http://arxiv.org/abs/hep-ph/0210145}{arXiv:hep-ph/0210145}%
  \bibAnnoteFile{NoStop}{Buras:2002vd}%
\bibitem{Babu:1999hn}%
  \BibitemOpen
  \bibfield{author}{%
  \bibinfo {author} {\bibfnamefont{K.~S.}\ \bibnamefont{Babu}}\ and\ \bibinfo
  {author} {\bibfnamefont{C.~F.}\ \bibnamefont{Kolda}},\ }%
  \bibfield{journal}{%
  \Doi{10.1103/PhysRevLett.84.228}{\bibinfo {journal} {Phys. Rev. Lett.}}\ }%
  \textbf{\bibinfo {volume} {84}},\ \bibinfo {pages} {228} (\bibinfo {year}
  {2000}),\ \Eprint{http://arxiv.org/abs/hep-ph/9909476}{arXiv:hep-ph/9909476}%
  \bibAnnoteFile{NoStop}{Babu:1999hn}%
\bibitem{Dedes:2002er}%
  \BibitemOpen
  \bibfield{author}{%
  \bibinfo {author} {\bibfnamefont{A.}~\bibnamefont{Dedes}}\ and\ \bibinfo
  {author} {\bibfnamefont{A.}~\bibnamefont{Pilaftsis}},\ }%
  \bibfield{journal}{%
  \Doi{10.1103/PhysRevD.67.015012}{\bibinfo {journal} {Phys. Rev.}}\ }%
  \textbf{\bibinfo {volume} {D67}},\ \bibinfo {pages} {015012} (\bibinfo {year}
  {2003}),\ \Eprint{http://arxiv.org/abs/hep-ph/0209306}{arXiv:hep-ph/0209306}%
  \bibAnnoteFile{NoStop}{Dedes:2002er}%
\bibitem{Demir:2003bv}%
  \BibitemOpen
  \bibfield{author}{%
  \bibinfo {author} {\bibfnamefont{D.~A.}\ \bibnamefont{Demir}},\ }%
  \bibfield{journal}{%
  \Doi{10.1016/j.physletb.2003.07.069}{\bibinfo {journal} {Phys. Lett.}}\ }%
  \textbf{\bibinfo {volume} {B571}},\ \bibinfo {pages} {193} (\bibinfo {year}
  {2003}),\ \Eprint{http://arxiv.org/abs/hep-ph/0303249}{arXiv:hep-ph/0303249}%
  \bibAnnoteFile{NoStop}{Demir:2003bv}%
\bibitem{Carena:2006ai}%
  \BibitemOpen
  \bibfield{author}{%
  \bibinfo {author} {\bibfnamefont{M.~S.}\ \bibnamefont{Carena}}, \bibinfo
  {author} {\bibfnamefont{A.}~\bibnamefont{Menon}}, \bibinfo {author}
  {\bibfnamefont{R.}~\bibnamefont{Noriega-Papaqui}}, \bibinfo {author}
  {\bibfnamefont{A.}~\bibnamefont{Szynkman}},\ and\ \bibinfo {author}
  {\bibfnamefont{C.~E.~M.}\ \bibnamefont{Wagner}},\ }%
  \bibfield{journal}{%
  \Doi{10.1103/PhysRevD.74.015009}{\bibinfo {journal} {Phys. Rev.}}\ }%
  \textbf{\bibinfo {volume} {D74}},\ \bibinfo {pages} {015009} (\bibinfo {year}
  {2006}),\ \Eprint{http://arxiv.org/abs/hep-ph/0603106}{arXiv:hep-ph/0603106}%
  \bibAnnoteFile{NoStop}{Carena:2006ai}%
\bibitem{Lunghi:2006uf}%
  \BibitemOpen
  \bibfield{author}{%
  \bibinfo {author} {\bibfnamefont{E.}~\bibnamefont{Lunghi}}, \bibinfo {author}
  {\bibfnamefont{W.}~\bibnamefont{Porod}},\ and\ \bibinfo {author}
  {\bibfnamefont{O.}~\bibnamefont{Vives}},\ }%
  \bibfield{journal}{%
  \Doi{10.1103/PhysRevD.74.075003}{\bibinfo {journal} {Phys. Rev.}}\ }%
  \textbf{\bibinfo {volume} {D74}},\ \bibinfo {pages} {075003} (\bibinfo {year}
  {2006}),\ \Eprint{http://arxiv.org/abs/hep-ph/0605177}{arXiv:hep-ph/0605177}%
  \bibAnnoteFile{NoStop}{Lunghi:2006uf}%
\bibitem{Carena:2007aq}%
  \BibitemOpen
  \bibfield{author}{%
  \bibinfo {author} {\bibfnamefont{M.~S.}\ \bibnamefont{Carena}}, \bibinfo
  {author} {\bibfnamefont{A.}~\bibnamefont{Menon}},\ and\ \bibinfo {author}
  {\bibfnamefont{C.~E.~M.}\ \bibnamefont{Wagner}},\ }%
  \bibfield{journal}{%
  \Doi{10.1103/PhysRevD.76.035004}{\bibinfo {journal} {Phys. Rev.}}\ }%
  \textbf{\bibinfo {volume} {D76}},\ \bibinfo {pages} {035004} (\bibinfo {year}
  {2007}),\ \Eprint{http://arxiv.org/abs/0704.1143}{arXiv:0704.1143 [hep-ph]}%
  \bibAnnoteFile{NoStop}{Carena:2007aq}%
\bibitem{Ellis:2006jy}%
  \BibitemOpen
  \bibfield{author}{%
  \bibinfo {author} {\bibfnamefont{J.~R.}\ \bibnamefont{Ellis}}, \bibinfo
  {author} {\bibfnamefont{K.~A.}\ \bibnamefont{Olive}}, \bibinfo {author}
  {\bibfnamefont{Y.}~\bibnamefont{Santoso}},\ and\ \bibinfo {author}
  {\bibfnamefont{V.~C.}\ \bibnamefont{Spanos}},\ }%
  \bibfield{journal}{%
  \bibinfo {journal} {JHEP}\ }%
  \textbf{\bibinfo {volume} {05}},\ \bibinfo {pages} {063} (\bibinfo {year}
  {2006}),\ \Eprint{http://arxiv.org/abs/hep-ph/0603136}{arXiv:hep-ph/0603136}%
  \bibAnnoteFile{NoStop}{Ellis:2006jy}%
\bibitem{Ellis:2007ss}%
  \BibitemOpen
  \bibfield{author}{%
  \bibinfo {author} {\bibfnamefont{J.~R.}\ \bibnamefont{Ellis}}, \bibinfo
  {author} {\bibfnamefont{S.}~\bibnamefont{Heinemeyer}}, \bibinfo {author}
  {\bibfnamefont{K.~A.}\ \bibnamefont{Olive}},\ and\ \bibinfo {author}
  {\bibfnamefont{G.}~\bibnamefont{Weiglein}},\ }%
  \bibfield{journal}{%
  \Doi{10.1016/j.physletb.2007.07.056}{\bibinfo {journal} {Phys. Lett.}}\ }%
  \textbf{\bibinfo {volume} {B653}},\ \bibinfo {pages} {292} (\bibinfo {year}
  {2007}),\ \Eprint{http://arxiv.org/abs/0706.0977}{arXiv:0706.0977 [hep-ph]}%
  \bibAnnoteFile{NoStop}{Ellis:2007ss}%
\bibitem{Isidori:2007jw}%
  \BibitemOpen
  \bibfield{author}{%
  \bibinfo {author} {\bibfnamefont{G.}~\bibnamefont{Isidori}}, \bibinfo
  {author} {\bibfnamefont{F.}~\bibnamefont{Mescia}}, \bibinfo {author}
  {\bibfnamefont{P.}~\bibnamefont{Paradisi}},\ and\ \bibinfo {author}
  {\bibfnamefont{D.}~\bibnamefont{Temes}},\ }%
  \bibfield{journal}{%
  \Doi{10.1103/PhysRevD.75.115019}{\bibinfo {journal} {Phys. Rev.}}\ }%
  \textbf{\bibinfo {volume} {D75}},\ \bibinfo {pages} {115019} (\bibinfo {year}
  {2007}),\ \Eprint{http://arxiv.org/abs/hep-ph/0703035}{arXiv:hep-ph/0703035}%
  \bibAnnoteFile{NoStop}{Isidori:2007jw}%
\bibitem{Barenboim:2007sk}%
  \BibitemOpen
  \bibfield{author}{%
  \bibinfo {author} {\bibfnamefont{G.}~\bibnamefont{Barenboim}}, \bibinfo
  {author} {\bibfnamefont{P.}~\bibnamefont{Paradisi}}, \bibinfo {author}
  {\bibfnamefont{O.}~\bibnamefont{Vives}}, \bibinfo {author}
  {\bibfnamefont{E.}~\bibnamefont{Lunghi}},\ and\ \bibinfo {author}
  {\bibfnamefont{W.}~\bibnamefont{Porod}},\ }%
  \bibfield{journal}{%
  \Doi{10.1088/1126-6708/2008/04/079}{\bibinfo {journal} {JHEP}}\ }%
  \textbf{\bibinfo {volume} {04}},\ \bibinfo {pages} {079} (\bibinfo {year}
  {2008}),\ \Eprint{http://arxiv.org/abs/0712.3559}{arXiv:0712.3559 [hep-ph]}%
  \bibAnnoteFile{NoStop}{Barenboim:2007sk}%
\bibitem{Paradisi:2008qh}%
  \BibitemOpen
  \bibfield{author}{%
  \bibinfo {author} {\bibfnamefont{P.}~\bibnamefont{Paradisi}}, \bibinfo
  {author} {\bibfnamefont{M.}~\bibnamefont{Ratz}}, \bibinfo {author}
  {\bibfnamefont{R.}~\bibnamefont{Schieren}},\ and\ \bibinfo {author}
  {\bibfnamefont{C.}~\bibnamefont{Simonetto}},\ }%
  \bibfield{journal}{%
  \Doi{10.1016/j.physletb.2008.08.025}{\bibinfo {journal} {Phys. Lett.}}\ }%
  \textbf{\bibinfo {volume} {B668}},\ \bibinfo {pages} {202} (\bibinfo {year}
  {2008}),\ \Eprint{http://arxiv.org/abs/0805.3989}{arXiv:0805.3989 [hep-ph]}%
  \bibAnnoteFile{NoStop}{Paradisi:2008qh}%
\bibitem{Ellis:2007kb}%
  \BibitemOpen
  \bibfield{author}{%
  \bibinfo {author} {\bibfnamefont{J.~R.}\ \bibnamefont{Ellis}}, \bibinfo
  {author} {\bibfnamefont{J.~S.}\ \bibnamefont{Lee}},\ and\ \bibinfo {author}
  {\bibfnamefont{A.}~\bibnamefont{Pilaftsis}},\ }%
  \bibfield{journal}{%
  \Doi{10.1103/PhysRevD.76.115011}{\bibinfo {journal} {Phys. Rev.}}\ }%
  \textbf{\bibinfo {volume} {D76}},\ \bibinfo {pages} {115011} (\bibinfo {year}
  {2007}),\ \Eprint{http://arxiv.org/abs/0708.2079}{arXiv:0708.2079 [hep-ph]}%
  \bibAnnoteFile{NoStop}{Ellis:2007kb}%
\bibitem{Carena:2008ue}%
  \BibitemOpen
  \bibfield{author}{%
  \bibinfo {author} {\bibfnamefont{M.}~\bibnamefont{Carena}}, \bibinfo {author}
  {\bibfnamefont{A.}~\bibnamefont{Menon}},\ and\ \bibinfo {author}
  {\bibfnamefont{C.~E.~M.}\ \bibnamefont{Wagner}},\ }%
  \bibfield{journal}{%
  \Doi{10.1103/PhysRevD.79.075025}{\bibinfo {journal} {Phys. Rev.}}\ }%
  \textbf{\bibinfo {volume} {D79}},\ \bibinfo {pages} {075025} (\bibinfo {year}
  {2009}),\ \Eprint{http://arxiv.org/abs/0812.3594}{arXiv:0812.3594 [hep-ph]}%
  \bibAnnoteFile{NoStop}{Carena:2008ue}%
\bibitem{Foster:2005wb}%
  \BibitemOpen
  \bibfield{author}{%
  \bibinfo {author} {\bibfnamefont{J.}~\bibnamefont{Foster}}, \bibinfo {author}
  {\bibfnamefont{K.-i.}\ \bibnamefont{Okumura}},\ and\ \bibinfo {author}
  {\bibfnamefont{L.}~\bibnamefont{Roszkowski}},\ }%
  \bibfield{journal}{%
  \bibinfo {journal} {JHEP}\ }%
  \textbf{\bibinfo {volume} {08}},\ \bibinfo {pages} {094} (\bibinfo {year}
  {2005}),\ \Eprint{http://arxiv.org/abs/hep-ph/0506146}{arXiv:hep-ph/0506146}%
  \bibAnnoteFile{NoStop}{Foster:2005wb}%
\bibitem{Giudice:1998bp}%
  \BibitemOpen
  \bibfield{author}{%
  \bibinfo {author} {\bibfnamefont{G.~F.}\ \bibnamefont{Giudice}}\ and\
  \bibinfo {author} {\bibfnamefont{R.}~\bibnamefont{Rattazzi}},\ }%
  \bibfield{journal}{%
  \Doi{10.1016/S0370-1573(99)00042-3}{\bibinfo {journal} {Phys. Rept.}}\ }%
  \textbf{\bibinfo {volume} {322}},\ \bibinfo {pages} {419} (\bibinfo {year}
  {1999}),\ \Eprint{http://arxiv.org/abs/hep-ph/9801271}{arXiv:hep-ph/9801271}%
  \bibAnnoteFile{NoStop}{Giudice:1998bp}%
\bibitem{Affleck:1984xz}%
  \BibitemOpen
  \bibfield{author}{%
  \bibinfo {author} {\bibfnamefont{I.}~\bibnamefont{Affleck}}, \bibinfo
  {author} {\bibfnamefont{M.}~\bibnamefont{Dine}},\ and\ \bibinfo {author}
  {\bibfnamefont{N.}~\bibnamefont{Seiberg}},\ }%
  \bibfield{journal}{%
  \Doi{10.1016/0550-3213(85)90408-0}{\bibinfo {journal} {Nucl. Phys.}}\ }%
  \textbf{\bibinfo {volume} {B256}},\ \bibinfo {pages} {557} (\bibinfo {year}
  {1985})%
  \bibAnnoteFile{NoStop}{Affleck:1984xz}%
\bibitem{Dine:1995ag}%
  \BibitemOpen
  \bibfield{author}{%
  \bibinfo {author} {\bibfnamefont{M.}~\bibnamefont{Dine}}, \bibinfo {author}
  {\bibfnamefont{A.~E.}\ \bibnamefont{Nelson}}, \bibinfo {author}
  {\bibfnamefont{Y.}~\bibnamefont{Nir}},\ and\ \bibinfo {author}
  {\bibfnamefont{Y.}~\bibnamefont{Shirman}},\ }%
  \bibfield{journal}{%
  \Doi{10.1103/PhysRevD.53.2658}{\bibinfo {journal} {Phys. Rev.}}\ }%
  \textbf{\bibinfo {volume} {D53}},\ \bibinfo {pages} {2658} (\bibinfo {year}
  {1996}),\ \Eprint{http://arxiv.org/abs/hep-ph/9507378}{arXiv:hep-ph/9507378}%
  \bibAnnoteFile{NoStop}{Dine:1995ag}%
\bibitem{Martin:1996zb}%
  \BibitemOpen
  \bibfield{author}{%
  \bibinfo {author} {\bibfnamefont{S.~P.}\ \bibnamefont{Martin}},\ }%
  \bibfield{journal}{%
  \Doi{10.1103/PhysRevD.55.3177}{\bibinfo {journal} {Phys. Rev.}}\ }%
  \textbf{\bibinfo {volume} {D55}},\ \bibinfo {pages} {3177} (\bibinfo {year}
  {1997}),\ \Eprint{http://arxiv.org/abs/hep-ph/9608224}{arXiv:hep-ph/9608224}%
  \bibAnnoteFile{NoStop}{Martin:1996zb}%
\bibitem{Wagner:1998vd}%
  \BibitemOpen
  \bibfield{author}{%
  \bibinfo {author} {\bibfnamefont{C.~E.~M.}\ \bibnamefont{Wagner}},\ }%
  \bibfield{journal}{%
  \Doi{10.1016/S0550-3213(98)00473-8}{\bibinfo {journal} {Nucl. Phys.}}\ }%
  \textbf{\bibinfo {volume} {B528}},\ \bibinfo {pages} {3} (\bibinfo {year}
  {1998}),\ \Eprint{http://arxiv.org/abs/hep-ph/9801376}{arXiv:hep-ph/9801376}%
  \bibAnnoteFile{NoStop}{Wagner:1998vd}%
\bibitem{McGarrie:2010kh}%
  \BibitemOpen
  \bibfield{author}{%
  \bibinfo {author} {\bibfnamefont{M.}~\bibnamefont{McGarrie}}\ and\ \bibinfo
  {author} {\bibfnamefont{R.}~\bibnamefont{Russo}}}%
   (\bibinfo {year} {2010}),\
  \Eprint{http://arxiv.org/abs/1004.3305}{arXiv:1004.3305 [hep-ph]}%
  \bibAnnoteFile{NoStop}{McGarrie:2010kh}%
\bibitem{Komargodski:2009pc}%
  \BibitemOpen
  \bibfield{author}{%
  \bibinfo {author} {\bibfnamefont{Z.}~\bibnamefont{Komargodski}}\ and\
  \bibinfo {author} {\bibfnamefont{N.}~\bibnamefont{Seiberg}},\ }%
  \bibfield{journal}{%
  \Doi{10.1088/1126-6708/2009/06/007}{\bibinfo {journal} {JHEP}}\ }%
  \textbf{\bibinfo {volume} {06}},\ \bibinfo {pages} {007} (\bibinfo {year}
  {2009}),\ \Eprint{http://arxiv.org/abs/0904.1159}{arXiv:0904.1159 [hep-th]}%
  \bibAnnoteFile{NoStop}{Komargodski:2009pc}%
\bibitem{Dienes:2009td}%
  \BibitemOpen
  \bibfield{author}{%
  \bibinfo {author} {\bibfnamefont{K.~R.}\ \bibnamefont{Dienes}}\ and\ \bibinfo
  {author} {\bibfnamefont{B.}~\bibnamefont{Thomas}},\ }%
  \bibfield{journal}{%
  \Doi{10.1103/PhysRevD.81.065023}{\bibinfo {journal} {Phys. Rev.}}\ }%
  \textbf{\bibinfo {volume} {D81}},\ \bibinfo {pages} {065023} (\bibinfo {year}
  {2010}),\ \Eprint{http://arxiv.org/abs/0911.0677}{arXiv:0911.0677 [hep-th]}%
  \bibAnnoteFile{NoStop}{Dienes:2009td}%
\bibitem{Matalliotakis:1994ft}%
  \BibitemOpen
  \bibfield{author}{%
  \bibinfo {author} {\bibfnamefont{D.}~\bibnamefont{Matalliotakis}}\ and\
  \bibinfo {author} {\bibfnamefont{H.~P.}\ \bibnamefont{Nilles}},\ }%
  \bibfield{journal}{%
  \Doi{10.1016/0550-3213(94)00487-Y}{\bibinfo {journal} {Nucl. Phys.}}\ }%
  \textbf{\bibinfo {volume} {B435}},\ \bibinfo {pages} {115} (\bibinfo {year}
  {1995}),\ \Eprint{http://arxiv.org/abs/hep-ph/9407251}{arXiv:hep-ph/9407251}%
  \bibAnnoteFile{NoStop}{Matalliotakis:1994ft}%
\bibitem{Olechowski:1994gm}%
  \BibitemOpen
  \bibfield{author}{%
  \bibinfo {author} {\bibfnamefont{M.}~\bibnamefont{Olechowski}}\ and\ \bibinfo
  {author} {\bibfnamefont{S.}~\bibnamefont{Pokorski}},\ }%
  \bibfield{journal}{%
  \Doi{10.1016/0370-2693(94)01571-S}{\bibinfo {journal} {Phys. Lett.}}\ }%
  \textbf{\bibinfo {volume} {B344}},\ \bibinfo {pages} {201} (\bibinfo {year}
  {1995}),\ \Eprint{http://arxiv.org/abs/hep-ph/9407404}{arXiv:hep-ph/9407404}%
  \bibAnnoteFile{NoStop}{Olechowski:1994gm}%
\bibitem{Berezinsky:1995cj}%
  \BibitemOpen
  \bibfield{author}{%
  \bibinfo {author} {\bibfnamefont{V.}~\bibnamefont{Berezinsky}}
  \emph{et~al.},\ }%
  \bibfield{journal}{%
  \Doi{10.1016/0927-6505(95)00048-8}{\bibinfo {journal} {Astropart. Phys.}}\ }%
  \textbf{\bibinfo {volume} {5}},\ \bibinfo {pages} {1} (\bibinfo {year}
  {1996}),\ \Eprint{http://arxiv.org/abs/hep-ph/9508249}{arXiv:hep-ph/9508249}%
  \bibAnnoteFile{NoStop}{Berezinsky:1995cj}%
\bibitem{Drees:2000he}%
  \BibitemOpen
  \bibfield{author}{%
  \bibinfo {author} {\bibfnamefont{M.}~\bibnamefont{Drees}} \emph{et~al.},\ }%
  \bibfield{journal}{%
  \Doi{10.1103/PhysRevD.63.035008}{\bibinfo {journal} {Phys. Rev.}}\ }%
  \textbf{\bibinfo {volume} {D63}},\ \bibinfo {pages} {035008} (\bibinfo {year}
  {2001}),\ \Eprint{http://arxiv.org/abs/hep-ph/0007202}{arXiv:hep-ph/0007202}%
  \bibAnnoteFile{NoStop}{Drees:2000he}%
\bibitem{Nath:1997qm}%
  \BibitemOpen
  \bibfield{author}{%
  \bibinfo {author} {\bibfnamefont{P.}~\bibnamefont{Nath}}\ and\ \bibinfo
  {author} {\bibfnamefont{R.~L.}\ \bibnamefont{Arnowitt}},\ }%
  \bibfield{journal}{%
  \Doi{10.1103/PhysRevD.56.2820}{\bibinfo {journal} {Phys. Rev.}}\ }%
  \textbf{\bibinfo {volume} {D56}},\ \bibinfo {pages} {2820} (\bibinfo {year}
  {1997}),\ \Eprint{http://arxiv.org/abs/hep-ph/9701301}{arXiv:hep-ph/9701301}%
  \bibAnnoteFile{NoStop}{Nath:1997qm}%
\bibitem{Ellis:2002iu}%
  \BibitemOpen
  \bibfield{author}{%
  \bibinfo {author} {\bibfnamefont{J.~R.}\ \bibnamefont{Ellis}}, \bibinfo
  {author} {\bibfnamefont{T.}~\bibnamefont{Falk}}, \bibinfo {author}
  {\bibfnamefont{K.~A.}\ \bibnamefont{Olive}},\ and\ \bibinfo {author}
  {\bibfnamefont{Y.}~\bibnamefont{Santoso}},\ }%
  \bibfield{journal}{%
  \Doi{10.1016/S0550-3213(02)01144-6}{\bibinfo {journal} {Nucl. Phys.}}\ }%
  \textbf{\bibinfo {volume} {B652}},\ \bibinfo {pages} {259} (\bibinfo {year}
  {2003}),\ \Eprint{http://arxiv.org/abs/hep-ph/0210205}{arXiv:hep-ph/0210205}%
  \bibAnnoteFile{NoStop}{Ellis:2002iu}%
\bibitem{Ellis:2008eu}%
  \BibitemOpen
  \bibfield{author}{%
  \bibinfo {author} {\bibfnamefont{J.~R.}\ \bibnamefont{Ellis}}, \bibinfo
  {author} {\bibfnamefont{K.~A.}\ \bibnamefont{Olive}},\ and\ \bibinfo {author}
  {\bibfnamefont{P.}~\bibnamefont{Sandick}},\ }%
  \bibfield{journal}{%
  \Doi{10.1103/PhysRevD.78.075012}{\bibinfo {journal} {Phys. Rev.}}\ }%
  \textbf{\bibinfo {volume} {D78}},\ \bibinfo {pages} {075012} (\bibinfo {year}
  {2008}),\ \Eprint{http://arxiv.org/abs/0805.2343}{arXiv:0805.2343 [hep-ph]}%
  \bibAnnoteFile{NoStop}{Ellis:2008eu}%
\bibitem{Choi:2004sx}%
  \BibitemOpen
  \bibfield{author}{%
  \bibinfo {author} {\bibfnamefont{K.}~\bibnamefont{Choi}}, \bibinfo {author}
  {\bibfnamefont{A.}~\bibnamefont{Falkowski}}, \bibinfo {author}
  {\bibfnamefont{H.~P.}\ \bibnamefont{Nilles}}, \bibinfo {author}
  {\bibfnamefont{M.}~\bibnamefont{Olechowski}},\ and\ \bibinfo {author}
  {\bibfnamefont{S.}~\bibnamefont{Pokorski}},\ }%
  \bibfield{journal}{%
  \Doi{10.1088/1126-6708/2004/11/076}{\bibinfo {journal} {JHEP}}\ }%
  \textbf{\bibinfo {volume} {11}},\ \bibinfo {pages} {076} (\bibinfo {year}
  {2004}),\ \Eprint{http://arxiv.org/abs/hep-th/0411066}{arXiv:hep-th/0411066}%
  \bibAnnoteFile{NoStop}{Choi:2004sx}%
\bibitem{Choi:2005uz}%
  \BibitemOpen
  \bibfield{author}{%
  \bibinfo {author} {\bibfnamefont{K.}~\bibnamefont{Choi}}, \bibinfo {author}
  {\bibfnamefont{K.~S.}\ \bibnamefont{Jeong}},\ and\ \bibinfo {author}
  {\bibfnamefont{K.-i.}\ \bibnamefont{Okumura}},\ }%
  \bibfield{journal}{%
  \bibinfo {journal} {JHEP}\ }%
  \textbf{\bibinfo {volume} {09}},\ \bibinfo {pages} {039} (\bibinfo {year}
  {2005}),\ \Eprint{http://arxiv.org/abs/hep-ph/0504037}{arXiv:hep-ph/0504037}%
  \bibAnnoteFile{NoStop}{Choi:2005uz}%
\bibitem{Endo:2005uy}%
  \BibitemOpen
  \bibfield{author}{%
  \bibinfo {author} {\bibfnamefont{M.}~\bibnamefont{Endo}}, \bibinfo {author}
  {\bibfnamefont{M.}~\bibnamefont{Yamaguchi}},\ and\ \bibinfo {author}
  {\bibfnamefont{K.}~\bibnamefont{Yoshioka}},\ }%
  \bibfield{journal}{%
  \Doi{10.1103/PhysRevD.72.015004}{\bibinfo {journal} {Phys. Rev.}}\ }%
  \textbf{\bibinfo {volume} {D72}},\ \bibinfo {pages} {015004} (\bibinfo {year}
  {2005}),\ \Eprint{http://arxiv.org/abs/hep-ph/0504036}{arXiv:hep-ph/0504036}%
  \bibAnnoteFile{NoStop}{Endo:2005uy}%
\bibitem{Randall:1998uk}%
  \BibitemOpen
  \bibfield{author}{%
  \bibinfo {author} {\bibfnamefont{L.}~\bibnamefont{Randall}}\ and\ \bibinfo
  {author} {\bibfnamefont{R.}~\bibnamefont{Sundrum}},\ }%
  \bibfield{journal}{%
  \Doi{10.1016/S0550-3213(99)00359-4}{\bibinfo {journal} {Nucl. Phys.}}\ }%
  \textbf{\bibinfo {volume} {B557}},\ \bibinfo {pages} {79} (\bibinfo {year}
  {1999}),\ \Eprint{http://arxiv.org/abs/hep-th/9810155}{arXiv:hep-th/9810155}%
  \bibAnnoteFile{NoStop}{Randall:1998uk}%
\bibitem{Giudice:1998xp}%
  \BibitemOpen
  \bibfield{author}{%
  \bibinfo {author} {\bibfnamefont{G.~F.}\ \bibnamefont{Giudice}}, \bibinfo
  {author} {\bibfnamefont{M.~A.}\ \bibnamefont{Luty}}, \bibinfo {author}
  {\bibfnamefont{H.}~\bibnamefont{Murayama}},\ and\ \bibinfo {author}
  {\bibfnamefont{R.}~\bibnamefont{Rattazzi}},\ }%
  \bibfield{journal}{%
  \bibinfo {journal} {JHEP}\ }%
  \textbf{\bibinfo {volume} {12}},\ \bibinfo {pages} {027} (\bibinfo {year}
  {1998}),\ \Eprint{http://arxiv.org/abs/hep-ph/9810442}{arXiv:hep-ph/9810442}%
  \bibAnnoteFile{NoStop}{Giudice:1998xp}%
\bibitem{Weiglein:2004hn}%
  \BibitemOpen
  \bibfield{author}{%
  \bibinfo {author} {\bibfnamefont{G.}~\bibnamefont{Weiglein}} \emph{et~al.}
  (\bibinfo {collaboration} {LHC/LC Study Group}),\ }%
  \bibfield{journal}{%
  \Doi{10.1016/j.physrep.2005.12.003}{\bibinfo {journal} {Phys. Rept.}}\ }%
  \textbf{\bibinfo {volume} {426}},\ \bibinfo {pages} {47} (\bibinfo {year}
  {2006}),\ \Eprint{http://arxiv.org/abs/hep-ph/0410364}{arXiv:hep-ph/0410364}%
  \bibAnnoteFile{NoStop}{Weiglein:2004hn}%
\bibitem{Rolbiecki:2009hk}%
  \BibitemOpen
  \bibfield{author}{%
  \bibinfo {author} {\bibfnamefont{K.}~\bibnamefont{Rolbiecki}}, \bibinfo
  {author} {\bibfnamefont{J.}~\bibnamefont{Tattersall}},\ and\ \bibinfo
  {author} {\bibfnamefont{G.}~\bibnamefont{Moortgat-Pick}}}%
   (\bibinfo {year} {2009}),\
  \Eprint{http://arxiv.org/abs/0909.3196}{arXiv:0909.3196 [hep-ph]}%
  \bibAnnoteFile{NoStop}{Rolbiecki:2009hk}%
\bibitem{Hisano:2003qu}%
  \BibitemOpen
  \bibfield{author}{%
  \bibinfo {author} {\bibfnamefont{J.}~\bibnamefont{Hisano}}, \bibinfo {author}
  {\bibfnamefont{K.}~\bibnamefont{Kawagoe}},\ and\ \bibinfo {author}
  {\bibfnamefont{M.~M.}\ \bibnamefont{Nojiri}},\ }%
  \bibfield{journal}{%
  \Doi{10.1103/PhysRevD.68.035007}{\bibinfo {journal} {Phys. Rev.}}\ }%
  \textbf{\bibinfo {volume} {D68}},\ \bibinfo {pages} {035007} (\bibinfo {year}
  {2003}),\ \Eprint{http://arxiv.org/abs/hep-ph/0304214}{arXiv:hep-ph/0304214}%
  \bibAnnoteFile{NoStop}{Hisano:2003qu}%
\bibitem{Perelstein:2008zt}%
  \BibitemOpen
  \bibfield{author}{%
  \bibinfo {author} {\bibfnamefont{M.}~\bibnamefont{Perelstein}}\ and\ \bibinfo
  {author} {\bibfnamefont{A.}~\bibnamefont{Weiler}},\ }%
  \bibfield{journal}{%
  \Doi{10.1088/1126-6708/2009/03/141}{\bibinfo {journal} {JHEP}}\ }%
  \textbf{\bibinfo {volume} {03}},\ \bibinfo {pages} {141} (\bibinfo {year}
  {2009}),\ \Eprint{http://arxiv.org/abs/0811.1024}{arXiv:0811.1024 [hep-ph]}%
  \bibAnnoteFile{NoStop}{Perelstein:2008zt}%
\bibitem{Martin:1993zk}%
  \BibitemOpen
  \bibfield{author}{%
  \bibinfo {author} {\bibfnamefont{S.~P.}\ \bibnamefont{Martin}}\ and\ \bibinfo
  {author} {\bibfnamefont{M.~T.}\ \bibnamefont{Vaughn}},\ }%
  \bibfield{journal}{%
  \Doi{10.1103/PhysRevD.50.2282}{\bibinfo {journal} {Phys. Rev.}}\ }%
  \textbf{\bibinfo {volume} {D50}},\ \bibinfo {pages} {2282} (\bibinfo {year}
  {1994}),\ \Eprint{http://arxiv.org/abs/hep-ph/9311340}{arXiv:hep-ph/9311340}%
  \bibAnnoteFile{NoStop}{Martin:1993zk}%
\bibitem{Djouadi:2002ze}%
  \BibitemOpen
  \bibfield{author}{%
  \bibinfo {author} {\bibfnamefont{A.}~\bibnamefont{Djouadi}}, \bibinfo
  {author} {\bibfnamefont{J.-L.}\ \bibnamefont{Kneur}},\ and\ \bibinfo {author}
  {\bibfnamefont{G.}~\bibnamefont{Moultaka}},\ }%
  \bibfield{journal}{%
  \Doi{10.1016/j.cpc.2006.11.009}{\bibinfo {journal} {Comput. Phys. Commun.}}\
  }%
  \textbf{\bibinfo {volume} {176}},\ \bibinfo {pages} {426} (\bibinfo {year}
  {2007}),\ \Eprint{http://arxiv.org/abs/hep-ph/0211331}{arXiv:hep-ph/0211331}%
  \bibAnnoteFile{NoStop}{Djouadi:2002ze}%
\end{thebibliography}%

\end{document}